\documentclass[12pt]{article}
\usepackage{latexsym,graphicx,a4,epsfig,psfrag,here}    

\newcommand{\Pint}{-\hspace{-2.3ex}\int}
\newcommand{\bdm}{\begin{displaymath}}
\newcommand{\edm}{\end{displaymath}}
\newcommand{\no}{\nonumber \\}
\newcommand{\sos}{\Delta_{\mbox{\tiny{Roy}}}^2}
\renewcommand{\theequation}{\arabic{section}.\arabic{equation}}
\renewcommand{\Re}{{\rm Re}\,}
\renewcommand{\Im}{{\rm Im}\,}
\newcommand{\be}{\begin{equation}}
\newcommand{\ee}{\end{equation}}
\newcommand{\bea}{\begin{eqnarray}}
\newcommand{\eea}{\end{eqnarray}}
\newcommand{\fs}{\; \; .}
\newcommand{\co}{\; \; ,}

\newcommand{\scs}{\co \;}

\newcommand{\per}{ \; .}

\newcommand{\al}{&\!\!\!\!}
\newcommand{\ind}{\scriptscriptstyle}

\newcommand{\SP}{\hspace{-0.03em}\rule[-0.2em]{0em}{0em}_{\mbox{\tiny\it SP}}}

\newcommand{\cnnnl}{\co\nonumber\\}

\newcommand{\gsim}{\,\raisebox{-0.3em}{$\stackrel{\raisebox{-0.1em}
{$>$}}{\sim}$}\,}

\newcommand{\Ezero}{\sqrt{\rule[0.1em]{0em}{0.5em}s_0}}
\newcommand{\Eone}{\sqrt{\rule[0.1em]{0em}{0.5em}s_1}}
\newcommand{\Etwo}{\sqrt{\rule[0.1em]{0em}{0.5em}s_2}}

\begin{document}

\begin{titlepage}

\begin{flushright}
IISc-CTS-12/99\\
ZU-TH 10/00\\
BUTP-99/33

\end{flushright}

\begin{center}
\vspace{0.5cm}

{\Large\bf Roy equation analysis of $\pi\pi$ scattering}

\vspace{0.5cm}
May 30, 2000

\vspace{0.5cm}

B.~Ananthanarayan$^a$, G.~Colangelo$^b$,
J.~Gasser$^c$ and  
H.~Leutwyler$^c$ 

\vspace{2em}
\footnotesize{\begin{tabular}{c}
$^a\,$Centre for Theoretical Studies, 
Indian Institute of Science\\ Bangalore, 560 012 India \\
$^b\,$Institute for Theoretical Physics, University of 
Z\"urich\\
Winterthurerstr. 190, CH-8057 Z\"urich, Switzerland\\
$^c\,$Institute for Theoretical Physics, University of 
Bern\\   
Sidlerstr. 5, CH-3012 Bern, Switzerland
\end{tabular}  }

\vspace{.5cm}

\begin{abstract}

We analyze the Roy equations for the lowest partial waves
of elastic $\pi\pi$ scattering. In the first part of the paper, 
we review the mathematical properties of these equations as well as 
their phenomenological applications.
In particular, the experimental situation concerning the 
contributions from intermediate energies and the evaluation
of the driving terms are discussed in detail. We then demonstrate that the two 
$S$-wave scattering lengths $a_0^0$ and $a_0^2$ are the essential parameters
in the low energy region: Once these are known, the 
available experimental information 
determines the behaviour near threshold to within remarkably small
uncertainties. An explicit
numerical representation for the energy dependence of the $S$- and $P$-waves
is given and it is shown that the threshold parameters of the $D$- 
and $F$-waves are also fixed very sharply in terms of $a_0^0$ and $a^2_0$.
In agreement with earlier work, which is reviewed in some detail,
we find that the Roy equations admit physically acceptable solutions
only within a band of the ($a_0^0$,$a_0^2$) plane. We show that 
the data on the reactions $e^+e^-\rightarrow \pi\,\pi$ and 
$\tau\rightarrow\pi\,\pi\,\nu$ reduce the width of this band quite
significantly. Furthermore, we discuss the relevance of the decay 
$K\rightarrow\pi\,\pi\, e\,\nu$
in restricting the allowed range of $a_0^0$,
preparing the grounds for an analysis of the forthcoming precision data
on this decay and on pionic atoms. We expect these to reduce the
uncertainties in the two basic low energy parameters very substantially, 
so that a meaningful test of the chiral perturbation theory predictions
will become possible.
\end{abstract}

\footnotesize{\begin{tabular}{ll}
{\bf{Pacs:}}$\!\!\!\!$& 11.30.Rd, 11.55.Fv, 11.80.Et, 13.75.Lb\\
{\bf{Keywords:}}$\!\!\!\!$& Roy equations, Dispersion relations, 
Partial wave analysis,\\&
Meson-meson interactions, Pion-pion scattering, Chiral symmetries
\end{tabular}}
\end{center}

\end{titlepage}
\setcounter{page}{2}
\tableofcontents
\setcounter{page}{3}

\section{Introduction}
\label{intro}

The present paper deals with the properties of the $\pi\pi$ scattering
amplitude in the low energy region. 
Our analysis relies on a set of dispersion relations for the partial wave
amplitudes due to Roy \cite{Roy}. These equations involve two subtraction
constants, which may be identified with the $S$-wave scattering lengths, 
$a_0^0$ and $a_0^2$. We demonstrate that the subtraction constants represent
the essential parameters in the low energy region --
once these are known, the Roy equations allow
us to calculate the partial waves in terms of the available data, to within
small uncertainties.
 Given the strong dominance of the two $S$-waves and of the $P$-wave, 
it makes sense to solve  the equations only for these,
using experimental as well as theoretical information 
to determine the contributions from higher energies and from the higher partial
waves. More specifically, we solve the relevant integral 
 equations on the interval
$2M_\pi< \sqrt{s}< 0.8\,\mbox{GeV}$. 
One of the main results of this work is an accurate numerical representation
of the $S$-  and $P$-waves for a given pair of
 scattering lengths $a_0^0$ and $a_0^2$.

Before describing the outline of the present paper, we review
previous work concerning the  Roy equations.
Roy's representation \cite{Roy} for the partial wave amplitudes
 $t^I_l$ of elastic  $\pi\pi$ scattering reads
\be\label{eq:req1} t_\ell^I(s)= k_\ell^I(s)+
\sum_{I'=0}^2\sum_{\ell'=0}^\infty \int_{4M_\pi^2}^\infty
ds'\,K_{\ell\ell'}^{I I'}(s,s')\,\mbox{Im} \, t_{\ell'}^{I'}(s')\, ,
\ee
where $I$ and $\ell$ denote isospin and angular
momentum, respectively and $k^I_\ell(s)$ is the partial wave 
projection of the subtraction term. 
It  shows up only in the $S$- and $P$-waves,
\bea 
\label{eq:subconst1}
\hspace{-6mm}
k^I_\ell(s)\al=\al a_0^I \,\delta_\ell^0 +\frac{s-4M_\pi^2}{4M_\pi^2}\,
(2a_0^0-5a_0^2)\,\left(\frac{1}{3}\,\delta_0^I\,\delta_\ell^0
+\frac{1}{18}\,\delta_1^I\,
\delta_\ell^1-\frac{1}{6}\,\delta_2^I\,\delta^0_\ell\right).
\eea
The kernels $K_{\ell\ell'}^{I I'}(s,s')$ are explicitly known functions 
(see appendix A). They contain a diagonal, singular Cauchy kernel that
generates the right hand cut in the partial wave amplitudes, as well as a
logarithmically singular piece that accounts for the left hand cut. 
The validity of these equations has rigorously been established on the
interval $-\,4 M_\pi^2<s< 60 M_\pi^2$.

The relations (\ref{eq:req1})  are consequences of the
analyticity properties of the $\pi\pi$ scattering amplitude, of the
Froissart bound and of crossing symmetry.
Combined with unitarity, the Roy equations amount
to an infinite system of coupled, singular integral equations for
the phase shifts. The integration 
is split into a low energy interval $4M_\pi^2<s'<s_0$
and a remainder, $s_0<s'<\infty$.
We refer to $s_0$ as the {\em matching point}, which is chosen
somewhere in the range where the Roy equations are valid.
The two $S$-wave scattering lengths, the elasticity parameters below 
the matching point and the imaginary parts above that point are
treated as an externally assigned input.
The mathematical problem consists in solving Roy's integral equations with
this input.

Soon after the original article of Roy \cite{Roy} had appeared, extensive
phenomenological applications were performed \cite{Basdevant Guillou
Navelet}--\cite{Froggatt Petersen}, resulting in a detailed analysis and
exploitation of the then available experimental data on $\pi\pi$
scattering.  For a recent review of those results, we refer the reader to
the article by Morgan and Pennington \cite{Morgan Pennington handbook}.
Parallel to these phenomenological applications, the very structure of the
Roy equations was investigated.  In \cite{mahoux}, a family of partial wave
equations was derived, on the basis of manifestly crossing symmetric
dispersion relations in the variables $s\,t+t\,u+u\,s$ and $s\,t\,u$. Each
set in this family is valid in an interval $s_0<s<s_1$, and the union of
these intervals covers the domain $-28M_\pi^2\leq {\mbox {Re}}\,\,s \leq
125.3 M_\pi^2$ (for a recent application of these dispersion relations, see
\cite{Anant}).  Using hyperbolae in the plane of the above variables, Auberson
and Epele \cite{Auberson Epele} proved the existence of partial wave
equations up to ${\mbox {Re}}\,\, s=165 M_\pi^2$. Furthermore, the manifold
of solutions of Roy's equations was investigated, in the single channel
\cite{pomponiuw}--\cite{slim} as well as in the coupled channel case
\cite{epelew}.  In the late seventies, Pool \cite{proofpool} provided a
proof that the original, infinite set of integral equations does have at
least one solution for $\Ezero < 4.8\,M_\pi$, provided that the driving
terms are not too large, see also \cite{proofheemskerk}.  Heemskerk and
Pool also examined numerically the solutions of the Roy equations, both by
solving the $N$ equation \cite{proofheemskerk} and by using an iterative
method \cite{heemskerk}.

It emerged from these investigations that -- for a given input of $S$-wave
scattering lengths, elasticity parameters and imaginary parts -- there are
in general many possible solutions to the Roy equations. This
non-uniqueness is due to the singular Cauchy kernel on the right hand side
of (\ref{eq:req1}).  In order to investigate the uniqueness properties of
the Roy system, one may -- in a first step -- keep only this part of the
kernels, so that the integral equations decouple: one is left with a single
channel problem, that is a single partial wave, which, moreover, does not have
a left hand cut. This mathematical problem was examined by Pomponiu and
Wanders, who also studied the effects due to the presence of a left hand
cut \cite{pomponiuw}.  Investigating the infinitesimal neighbourhood of a
given solution, they found that the multiplicity of the solution increases
by one whenever the value of the phase shift at the matching point goes
through a multiple of $\pi/2$.  Note that the situation for the usual
partial wave equation is different: There, the number of parameters in
general increases by two whenever the phase shift at infinity passes
through a positive integer multiple of $\pi$, see
for instance \cite{lovelace,brander} and references cited therein. 

After 1980, interest in the Roy equations waned, until recently. For
instance, in refs.~\cite{buettiker}
these equations are used to analyze
the threshold parameters for the higher partial waves, relying on the
approach of Basdevant, Froggatt and Petersen \cite{BFP1,BFP2}.  The
uncertainties in the values of $a_0^0$ and $a_0^2$ are reexamined in
refs.~\cite{patarakin}. In recent years, it has become increasingly clear,
however, that a new analysis of the $\pi\pi$ scattering amplitude at low
energies is urgently needed.  New $K_{e4}$ experiments and a measurement of
the combination $a_0^0-a_0^2$ based on the decay of pionic atoms are under
way \cite{mainz}--\cite{dirac}.  It is expected that these will
significantly reduce the uncertainties inherent in the data underlying
previous Roy equation studies, provided the structure of these equations
can be brought under firm control.  For this reason, the one-channel
problem has been revisited in great detail in a recent publication
\cite{Gasser Wanders}, while the role of the input in Roy's equations is
discussed in ref.~\cite{wamultich}.

The main reason for performing an improved determination of the $\pi\pi$
scattering amplitude is that this will allow us to test one of the basic
properties of QCD, namely the occurrence of an approximate, spontaneously
broken symmetry: The symmetry leads to a sharp prediction for the two
$S$-wave scattering lengths \cite{Weinberg 1966}--\cite{pipi6}. The
prediction relies on the standard hypothesis, according to which the quark
condensate is the leading order parameter of the spontaneously broken
symmetry. Hence an accurate test of the prediction would allow us to verify
or falsify that hypothesis \cite{gchpt}.  First steps in this program have
already been performed \cite{Borges}--\cite{pocanic}. However, in the present
paper, we do not discuss this issue. We follow the
phenomenological path and ignore the constraints imposed by 
chiral symmetry altogether, in order not
to bias the data analysis with theoretical pre\-judice. In a future
publication, we intend to match the chiral perturbation theory
representation of the scattering amplitude to two loops \cite{pipi6} with the
phenomenological one obtained in the present work.

Finally, we describe the content of the present paper. Our notation is
specified in section 2. Sections 3 and 4
contain a discussion of the background amplitude and of the driving 
terms, which account for the contributions from the higher partial waves and
from the high-energy region. As is recalled in section 5, unitarity 
leads to a set of three
singular integral equations for the two $S$-waves and for the $P$-wave.
The uniqueness
properties of the solutions to these  equations are 
discussed in section 6, while section 7 
contains a description
of the experimental input used for energies between 0.8 and 2 GeV. In
particular we also discuss the information concerning the $P$-wave phase 
shift, obtained on the basis of
the $e^+e^-\rightarrow\pi\pi$ and $\tau\rightarrow \pi\pi\nu$
data.  In section 8,
we describe the  method used to solve the integral equations for a
given input. The resulting universal band in
the ($a_0^0$,$a_0^2$) plane is discussed in section 9, where we show that, 
in the region below $0.8\,\mbox{GeV}$,
any point in this band leads to a decent numerical solution for 
the three lowest partial waves. As discussed
in section 10, however, the behaviour of the solutions above that energy
is consistent with the input used for the imaginary parts only in
part of the universal band -- approximately the same region of the
($a_0^0$,$a_0^2$) plane, where the Olsson sum rule
is obeyed (section 11). The solutions are compared with available
experimental data in section 12, and in
section 13, we draw our conclusions concerning the allowed  
range of $a_0^0$ and $a_0^2$. The other threshold parameters can be determined
quite accurately in terms of these two. The outcome of our numerical 
evaluation of the scattering lengths and effective ranges of the lowest 
six partial waves as functions of $a_0^0$ and $a_0^2$ 
is given in section 14, while in section 15,  
we describe our results for the values of the phase shifts relevant
for $K\rightarrow \pi\pi$.  
Section 16 contains a comparison with earlier work. A
summary and concluding remarks are given in section 17. 

In appendix A we describe some properties of the Roy kernels, 
which are extensively used in
our work.  The background from the higher partial waves and from the
high energy tail of the dispersion integrals is discussed in detail in
appendix B. In particular, 
we show that the constraints imposed by crossing symmetry 
reduce the uncertainties in the background, so that the driving terms
can be evaluated in a reliable manner.  
In appendix C we discuss sum rules connected with the asymptotic
behaviour of the amplitude and show that these relate the imaginary part
of the $P$-wave to the one of the higher partial waves, thereby
offering a sensitive test of our framework. 
Explicit numerical solutions of the Roy equations 
are given in appendix D and, in appendix E,
we recall the main features of the well-known
Lovelace-Shapiro-Veneziano model, which provides a useful guide
for the analysis of the asymptotic contributions.

\setcounter{equation}{0}
\section{Scattering amplitude}
\label{sec:roy}
We consider elastic $\pi\pi$ scattering in the framework of QCD  
 and  restrict our analysis to the isospin symmetry limit,  where 
the masses of the up and down quarks are taken equal and the e.m.~interaction
 is ignored\footnote{In our numerical work, we 
identify the value of $M_\pi$ with the mass of the charged pion.}. In this
 case, the scattering  process
is described by a single Lorentz invariant amplitude $A(s,t,u)$,
\bea
\al\rule{0em}{0em}\hspace{-16em}\langle\pi^d(p_4)\pi^c(p_3)\,
\mbox{out}|\pi^a(p_1)\pi^b(p_2) \,\mbox{in}\rangle=\delta_{\!fi}+
\no\al\hspace{2em }
(2\pi)^4i\,\delta^4(P_f-P_i)\{\delta^{ab}\delta^{cd} A(s,t,u)
+\delta^{ac}\delta^{bd} A(t,u,s)
+\delta^{ad}\delta^{bc}A(u,s,t)\}\, .
\nonumber\eea
The amplitude only depends on the Mandelstam variables $s$, $t$, $u$, which
 are constrained by $s+t+u=4M_\pi^2$. Moreover, crossing symmetry implies
\bea
A(s,t,u)=A(s,u,t)\per\nonumber\eea
The $s$-channel isospin components of the  amplitude are given by
\bea
T^0(s,t)\al=\al 3A(s,t,u)+A(t,u,s)+A(u,s,t)\cnnnl
T^1(s,t)\al=\al A(t,u,s)-A(u,s,t)\co\\
T^2(s,t)\al=\al A(t,u,s)+A(u,s,t)\fs\nonumber
\eea
In our normalization, the partial wave decomposition reads
\bea T^I(s,t)\al=\al 32\,\pi \sum_\ell\, (2\ell + 1)\,P_\ell
\left(1+\frac{2t}{s-4M_\pi^2}\right) t^I_\ell(s)\co \nonumber\\
\label{eq:tIl}
t_\ell^I(s)\al =\al\frac{1}{2i\sigma(s)}
\left\{\eta_\ell^I(s)\,e^{2i\delta_\ell^I(s)}-1\right\}\co\\  
\sigma(s)\al=\al\sqrt{1-\frac{4M_\pi^2}{s}}\per
\nonumber\eea
The threshold parameters are the coefficients of the expansion
\be \mbox{Re}\,t^I_\ell(s)= q^{2\ell}\,\{a_\ell^I + q^2\,b_\ell^I+q^4\,c^I_\ell
+\ldots\,\}\co \label{eq:threxp}
\ee
with $s=4(M_\pi^2+q^2)$.

The isospin amplitudes  $\vec{T}=(T^0,T^1,T^2)$ 
obey  fixed-$t$ dispersion relations, valid in the interval 
$-28M_\pi^2 < t < 4M_\pi^2$ \cite{Martin}. 
As shown by Roy \cite{Roy}, these
can be written in the form\footnote{For an explicit 
representation of the kernels
 $g_2(s,t,s')$, $g_3(s,t,s')$ and of the crossing
matrices $C_{st}$, $C_{su}$, we
 refer to appendix A.} 
\bea
\label{fixedt} 
\vec{T }(s,t)\al=\al(4M_\pi^2)^{-1}\,(s\,{\bf 1} + t\,
C_{st} + u\, C_{su})\,\vec{T}(4M_\pi^2,0) \\
\al\al+\int_{4M_\pi^2}^\infty
\!ds'\,g_2(s,t,s')\,\mbox{Im}\,\vec{T}(s',0)
+\int_{4M_\pi^2}^\infty \!ds'\,g_3(s,t,s')\,\mbox{Im}\,\vec{T}(s',t)
\,. \nonumber
\eea
The subtraction term is fixed by the $S$-wave scattering lengths: 
\bea
\vec{T}(4M_\pi^2,0)=32\,\pi\,(a_0^0,0,a_0^2)\nonumber\,.
\eea 

The Roy equations (\ref{eq:req1}) represent the partial wave projections of
eq.~(\ref{fixedt}). Since the partial wave expansion of the absorptive
parts converges in the large Lehmann--Martin ellipse, these equations are
rigorously valid in the interval $-4M_\pi^2 < s < 60M_\pi^2$. If the
scattering amplitude obeys Mandelstam analyti\-city, the
fixed-$t$ dispersion relations can be shown to hold for $-32M_\pi^2 < t <
4M_\pi^2$ and the Roy equations are then also valid in a larger domain:
$-4M_\pi^2<s<68M_\pi^2$ (for a review, see \cite{Roy HPA}). In fact, as we
mentioned in the introduction, the range of validity can be extended even
further \cite{mahoux, Auberson Epele}, so that Roy equations could be used
to study the behaviour of the partial waves above
$\sqrt{68}\,M_\pi=1.15\,\mbox{GeV}$, where the uncertainties in the data
are still considerable. In the following, however, we focus on the low
energy region. We assume Mandelstam analyticity and analyze the Roy
equations in the interval from threshold to \bdm
s_1=68M_\pi^2\co\hspace{2em}\Eone=1.15\,\mbox{GeV}\fs\edm

\setcounter{equation}{0}
\section{Background amplitude}\label{section background}
The dispersion relation (\ref{fixedt}) shows that, at low 
energies, the scattering amplitude
is fully determined by the imaginary parts of the partial waves in the
physical region, except for the two subtraction constants $a_0^0,a_0^2$.
In view of the two subtractions, the dispersion integrals converge rapidly.
In the region between 0.8 and 
2 GeV, the available phase shift analyses provide a rather detailed description
of the imaginary parts of the various partial waves. 
Our analysis of the Roy equations allows us to 
extend this description down to threshold. 
For small values of $s$ and $t$, the contributions to the dispersion integrals 
from the region above 2 GeV are very small. We will rely on Regge asymptotics
to estimate these. In the following, we split the interval of integration 
into a low energy part 
($4M_\pi^2\leq s'\leq s_2$) and a high energy tail ($s_2\leq s'<\infty$), with
\bdm \Etwo=2\,\mbox{GeV}\co\hspace{2em}s_2=205.3\,M_\pi^2\fs\edm

For small values of $s$ and $t$, the scattering amplitude $\vec{T}(s,t)$
is dominated by the contributions from the subtraction constants and 
from the low energy part of the dispersion integral 
over the imaginary parts of the $S$- and $P$-waves. 
We denote this part of the
amplitude by $\vec{T}(s,t)\SP$. The  
corresponding contribution to the partial waves is given by 
\bea t^I_\ell(s)\SP\al =\al k_\ell^I(s)+
\sum_{I'=0}^2\sum_{\ell'=0}^1  \int_{4M_\pi^2}^{s_2}
ds'\,K_{\ell\ell'}^{I I'}(s,s')\,\mbox{Im} \, t_{\ell'}^{I'}(s')\fs\eea
\newpage\noindent
The remainder of the partial wave amplitude,
\bea\label{dt} d_\ell^I(s)\al=\al\sum_{I'=0}^2\sum_{\ell'=2}^\infty  
\int_{4M_\pi^2}^{s_2}
ds'\,K_{\ell\ell'}^{I I'}(s,s')\,\mbox{Im} \, t_{\ell'}^{I'}(s')\\
\al\al\hspace{4em}+\sum_{I'=0}^2\sum_{\ell'=0}^\infty  \int_{s_2}^{\infty}
ds'\,K_{\ell\ell'}^{I I'}(s,s')\,\mbox{Im} \, t_{\ell'}^{I'}(s')\co\nonumber
\eea
is called the {\em driving term}. It accounts for those contributions 
to the r.h.s.~of the Roy equations that arise from the imaginary parts of the 
waves with $\ell=2,3,\ldots$
and in addition also contains
those generated by the imaginary parts of the 
$S$- and $P$-waves above 2 GeV. 
By construction, we have
\bea t^I_\ell(s)= t^I_\ell(s)\SP +d^I_\ell(s)\fs\eea
For the scattering amplitude, the corresponding decomposition reads
\bea \vec{T}(s,t)=\vec{T}(s,t)\SP+\vec{T}(s,t)_d\fs\eea
We refer to $\vec{T}(s,t)_d$ as the {\em background amplitude}.

The contribution from the imaginary parts of the $S$- and $P$-waves turns 
out to
be crossing symmetric by itself. In this sense, crossing
symmetry does not 
constrain the imaginary parts of the $S$- and $P$-waves\footnote{The
asymptotic behaviour of the scattering amplitude does tie the imaginary part 
of the $P$-wave to the contributions from the higher partial waves, 
see appendix \ref{C1}.}.
The symmetry can be exhibited explicitly by
representing the three components of the vector 
$\vec{T}(s,t)\SP$ as the isospin projections of a single amplitude
$A(s,t,u)\SP$ that is even with respect to the exchange of $t$ and $u$.
The explicit expression involves three functions of a single
variable \cite{mahoux,pipigchpt}: 
\bea\label{ASP} A(s,t,u)\SP 
\al=\al32\pi\left\{\mbox{$\frac{1}{3}$}W^0(s)+
\mbox{$\frac{3}{2}$}(s-u)W^1(t)
+\mbox{$\frac{3}{2}$}(s-t)W^1(u)\right.\no
\al\al \left.+\mbox{$\frac{1}{2}$}W^2(t)+\mbox{$\frac{1}{2}$}W^2(u)
-\mbox{$\frac{1}{3}$} W^2(s) \right\}\fs\eea
These are determined by the imaginary parts of the $S$- and $P$-waves
and by the two subtraction constants $a_0^0,a_0^2$:
\bea\label{W} W^0(s)\al=\al\frac{a_0^0\, s}{4M_\pi^2} +
\frac{s(s-4M_\pi^2)}{\pi}\int_{4M_\pi^2}^{s_2}
\frac{ds'\;\mbox{Im}\, t_0^0(s')}{s'(s'-4M_\pi^2)(s'-s)}\co \no
      W^1(s) \al=\al \frac{s}{\pi}\int_{4M_\pi^2}^{s_2}
\frac{ds'\;\mbox{Im}\, t_1^1(s')}{s'(s'-4M_\pi^2)(s'-s)}\co\\
       W^2(s)\al=\al\frac{a_0^2\, s}{4M_\pi^2}  +
\frac{s(s-4M_\pi^2)}{\pi}\int_{4M_\pi^2}^{s_2}
\frac{ds'\;\mbox{Im}\, t_0^2(s')}{s'(s'-4M_\pi^2)(s'-s)}\fs \nonumber\eea
The representation
\bea A(s,t,u)=A(s,t,u)\SP+A(s,t,u)_d\eea
yields a manifestly crossing symmetric decomposition of the scattering
amplitude into a leading term generated by the imaginary parts of the
$S$- and $P$-waves at energies below $s_2$ and a background, arising from
the imaginary parts of the higher
partial waves and from the high energy tail of the 
dispersion integrals.

\setcounter{equation}{0}
\section{Driving terms}\label{section driving terms}
In the present paper,
we restrict ourselves to an analysis of the Roy
equations for the $S$- and $P$- waves, which dominate the behaviour at
low energies. The background amplitude only generates small corrections,
which can be worked out on the basis of the available experimental information.
The calculation is described in detail in appendix
\ref{background}. In particular, we show that crossing symmetry
implies a strong constraint on the asymptotic contributions.

\begin{figure}[H]
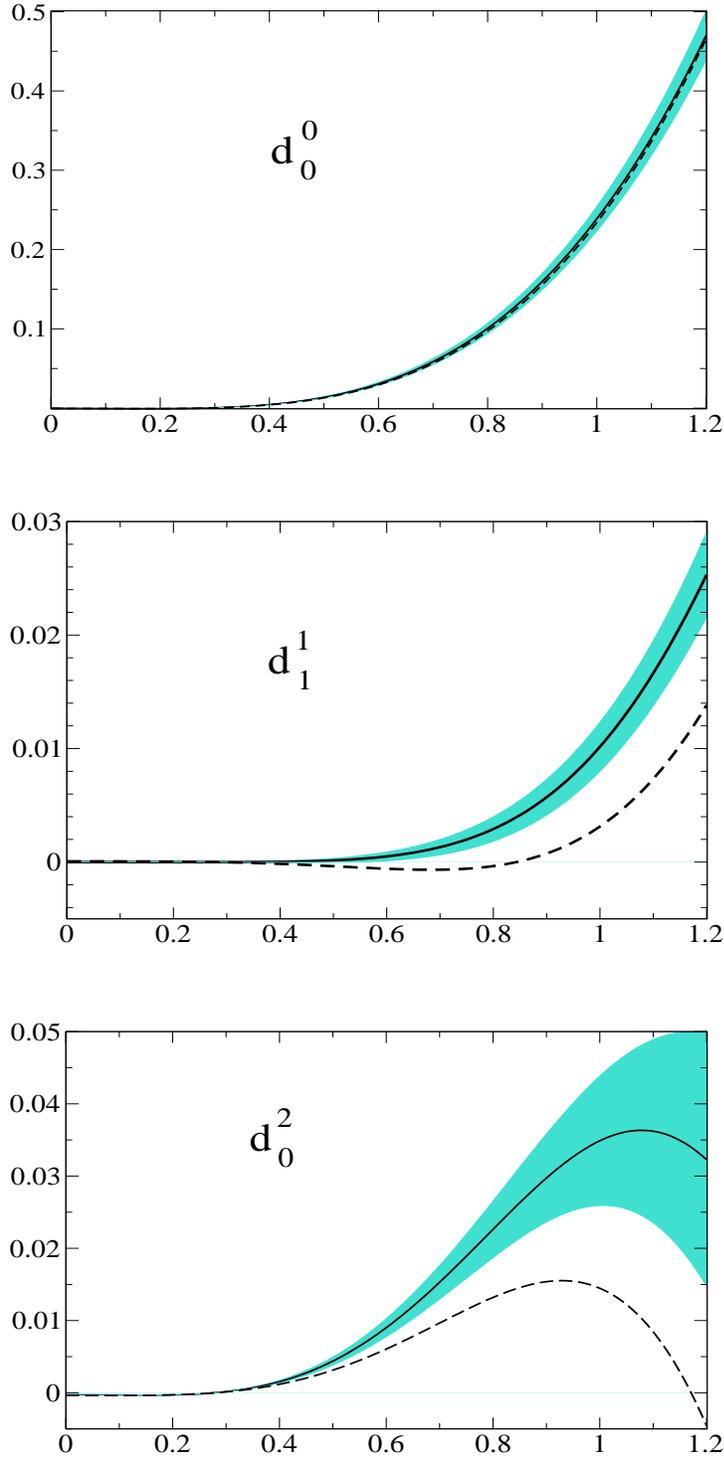

\centering
\begin{tabular}{c}
\raisebox{0.9em}
{\includegraphics[width=9.5cm,height=5.75cm]{f1_driv_d00.eps}}\\
\\
\raisebox{0.9em}
{\includegraphics[width=9.5cm,height=5.75cm]{f1_driv_d11.eps}}\\
\\
{\includegraphics[width=9.5cm,height=5.75cm]{f1_driv_d20.eps}}
\end{tabular}
\caption{\label{fig:driving} Driving terms versus energy in GeV. 
The full lines 
show the result of the calculation described in appendix B.
The shaded regions indicate the uncertainties associated with the input of
that calculation.  The dashed curves represent the contributions from the
$D$- and $F$-waves below 2 GeV. }
\end{figure}

The resulting numerical 
values for the driving terms are well described by
polynomials in $s$, or, equivalently, in the square of the center of mass 
momentum $q^2=\frac{1}{4}(s-4M_\pi^2)$. By definition, the driving
terms vanish at threshold, so that the polynomials do not contain 
$q$-independent terms. In view of their relevance in the evaluation of the
threshold parameters, we fix the coefficients of the terms proportional to
$q^2$ with the derivatives at threshold and also pin down the term
of order $q^4$ in the $P$-wave, such that it correctly
accounts for the background contribution to the effective range of this
partial wave. The remaining coefficients of the polynomial are obtained from
a fit on the interval from threshold to $s_1$. The explicit result
reads
\bea\label{numerical driving terms} 
d^0_0(s)\al =\al 0.116\,  q^2 + 4.79\,  q^4 - 4.09\,  q^6 + 2.69\, q^8
\co\no
d^1_1(s)\al=\al 0.00021\, q^2 + 0.038\, q^4 + 0.94\, q^6 - 1.21\,  q^8 
\label{dttot}\co\\
d^2_0(s)\al =\al 
0.0447\, q^2 + 1.59\, q^4 - 6.26\, q^6 + 5.94\,  q^8 \co 
\nonumber\eea
where $q$ is taken in GeV units (the range
$4M_\pi^2<s<68M_\pi^2$ corresponds to
$0<q<0.56\,\mbox{GeV}$). 
The driving term of the $I=0$ $S$-wave is larger than the other two
by an order of magnitude. It is dominated almost entirely by the
contribution from the $D$-wave with $I=0$. In 
$d^1_1(s)$, the $D$- and $F$-waves nearly cancel, so that
the main contributions arise from the region above 2 GeV. The term $d^2_0(s)$
picks up small contributions both from low energies and from the asymptotic
domain.  
The above polynomials are shown as full 
lines in fig.~\ref{fig:driving}. 
The shaded regions represent the uncertainties of the result, which 
may be represented as 
$d^I_\ell(s)\pm e^I_\ell(s)$, with
\bea\label{errors in driving terms} 
e^0_0(s)\al =\al  0.008\, q^2 + 0.31\, q^4 - 0.33\, q^6 + 0.41\, q^8\co\no
e^1_1(s)\al=\al 0.002\, q^2 + 0.06\, q^4 - 0.17\,  q^6 + 0.21\, q^8\co\\
e^2_0(s)\al =\al 0.005\, q^2 + 0.20\,  q^4 - 0.32\,q^6 + 0.39\,q^8\fs
\nonumber\eea
Above threshold, the error bars in $d_0^0(s)$, $d^1_1(s)$ and $d^2_0(s)$ 
roughly correspond to 6\%, 1\% and 4\% of $d_0^0(s)$, respectively.

As far as $d^0_0(s)$ is concerned,
our result roughly agrees with earlier calculations \cite{PP1,BFP2}. 
Our values for $d^1_1(s)$ and $d^2_0(s)$, however, are much smaller.  
The bulk of the difference is of purely kinematic origin:
The values taken for $s_2$ are different. 
While we are working with $\Etwo=
2\,\mbox{GeV}$, the values used in refs.~\cite{PP1} and \cite{BFP2} are 
$\sqrt{53}\,M_\pi\simeq 1\,\mbox{GeV}$ and 
$\sqrt{110}\,M_\pi\simeq 1.5\,\mbox{GeV}$, respectively. 
The value of $s_2$ enters the definition of the driving terms in
eq.~(3.2) as the lower limit of the integration 
over the imaginary parts of the $S$- and $P$-waves. 
We have checked that, once this difference in the range of integration is
accounted for, the driving terms given in these references are 
consistent with the above representation.
Note however, that our uncertainties are considerably smaller, and we do rely 
on this accuracy in the following.
It then matters that not only the range of integration, but also 
the integrands used in
\cite{PP1,BFP2} differ from ours: 
In these references, it is assumed that, above the value taken for $s_2$,
the behaviour of
the $S$- and $P$-wave imaginary parts is 
adequately described
by a Regge representation. 

The difference between such a picture and our representation
for the background amplitude is best illustrated with the simple model
used in the early literature, where
the asymptotic region is described by a Pomeron term with
$\sigma_{tot}=20\,\mbox{mb}$ and a contribution from the 
$\rho\,$-$f$-trajectory, taken from the Lovelace-Shapiro-Veneziano model 
(appendix \ref{veneziano}). 
As discussed in detail in appendix \ref{B4}, the assumption that 
an asymptotic behaviour of this type sets in early is
in conflict with crossing symmetry \cite{Pennington Annals}. In particular, 
the model overestimates the contribution
to the driving terms from the region above 1.5 GeV, roughly by a factor of
two.  Either the value of $\sigma_{tot}$
or the residue of the leading Regge trajectory or both must be
reduced in order for the model not to violate the sum rule (\ref{SRL}). 
The manner in which the asymptotic contribution
is split into one from the Pomeron and one from the leading Regge trajectory 
is not crucial. For any reasonable partition that obeys 
the sum rule (\ref{SRL}), the outcome for
the driving terms is approximately the same. The result for 
$d^1_1(s)$ and $d^2_0(s)$ is considerably smaller than what is expected from
the above model. The leading term $d^0_0(s)$, on the other hand, 
is dominated by the resonance
$f_2(1275)$ and is therefore not sensitive to the behaviour of the
imaginary parts 
in the region above $1.5\,\mbox{GeV}$.

\setcounter{equation}{0}
\section{Roy equations as integral equations}
Once the driving terms are pinned down, the Roy equations for the 
$S$- and $P$- waves express the real parts of the partial
waves in terms of the $S$-wave scattering lengths and of a principal value
integral over their imaginary parts from $4M_\pi^2$ to $s_2$.
Unitarity implies that, in the elastic domain $4M_\pi^2<s<16M_\pi^2$, 
the real and imaginary parts of the partial wave amplitudes are determined by
a single real parameter, the phase shift. If we were to restrict ourselves to
the elastic region, setting $s_2=16 M_\pi^2$,
the Roy equations would amount to a set of coupled,
nonlinear singular integral equations for the phase shifts. We may extend 
this range, provided the elasticity parameters $\eta^I_\ell(s)$ are 
known. On the other hand, since the Roy equations 
do not constrain the behaviour of the partial waves for $s>68M_\pi^2$,
the integrals occurring on the r.h.s.~of these equations 
can be evaluated only if the 
imaginary parts in that region are known, together with
the subtraction constants $a_0^0$, $a_0^2$, which also represent parameters 
to be assigned externally.

In the present paper, we do not solve the Roy equations in their full domain
of validity, but use a smaller interval, $4 M_\pi^2<s<s_0$. 
The reason why it is advantageous to use a value of $s_0$ below the 
mathematical upper limit, $s_0<s_1$, is 
that the Roy equations in general admit more than one
solution. As will be discussed in detail in section \ref{sec:unique},
the solution does become unique if the value of $s_0$ is chosen 
between the $\rho$ mass
and the energy where the $I=0$ $S$-wave phase passes through 
$\pi/2$ -- this happens around
$0.86$ GeV. In the following, we use 
\bdm \Ezero=0.8\,\mbox{GeV}\co\hspace{2em} s_0=32.9\,M_\pi^2\fs\edm
In the variable $s$, our matching point is nearly at the 
center of the interval between threshold and $s_1=68\,M_\pi^2$. 
We are thus solving the Roy equations on the lower half of
their range of validity, using the upper half to check the consistency of the
solutions so obtained (section \ref{sec:math}). 
Our results are not sensitive to the precise value 
taken for $s_0$ (section \ref{sec:UB}).

The Roy equations for the $S$- and $P$-waves may be rewritten in 
the form
\begin{eqnarray}\label{eq:rieq1}
\mbox{Re}\, t_\ell^I(s) \al=\al k_\ell^I(s)
\hspace{-1mm}+ \Pint_{4M_\pi^2}^{s_0} ds'
K_{\ell\, 0}^{I\, 0} (s,s')\, \mbox{Im}\, t_0^0(s')\hspace{-1mm} +
\Pint_{4M_\pi^2}^{s_0}
ds' K_{\ell\, 1}^{I\, 1} (s,s')\, \mbox{Im}\, t_1^1(s') \nonumber \\ 
\al\al+
\Pint_{4M_\pi^2}^{s_0} ds' K_{\ell \, 0}^{I\, 2} (s,s')\, \mbox{Im}\,
t_0^2(s') + f_\ell^I(s) + d_\ell^{I}(s) \; \; ,
\end{eqnarray}
where $I$ and $\ell$ take only the values ($I, \ell$) =(0,0), (1,1)
and (2,0). The bar across the integral sign denotes the principal value
integral. 
The functions $f_\ell^I(s)$ contain the part of the dispersive integrals
over the three lowest partial waves that comes from the region between $s_0$
and $s_2$, where we are using experimental data as input.
They are defined as
\begin{eqnarray}\label{eq:rieq2}
f_\ell^I(s) = \sum_{I'=0}^2\sum_{\ell'=0}^1\Pint_{s_0}^{s_2} ds' 
K_{\ell \ell'}^{I I'} (s,s')\,
\mbox{Im}\, t_{\ell'}^{I'}(s') \fs
\end{eqnarray}
The experimental input used to evaluate these integrals will be
discussed in section \ref{sec:exp_input}, together with the one for the
elasticity parameters of the $S$- and $P$-waves.

One of the main tasks we are faced with is the construction of the numerical 
 solution of the integral equations (\ref{eq:rieq1}) in the interval
$4M_\pi^2\leq s\leq s_0$, for a given 
 input $\{a_0^0,a_0^2, f_\ell^I,\eta_\ell^I,d_\ell^I\}$. Once a solution 
is known, the real part of the amplitude can be calculated 
with these equations, also in the region $s_0\leq s\leq s_1$. 

\setcounter{equation}{0}
\section{On the uniqueness of the solution}
\label{sec:unique}

The literature concerning the mathematical
structure of the Roy equations
was reviewed in the introduction. In the following, we first
discuss the situation for the single channel case -- which is simpler, but
clearly shows the salient features -- and then describe the generalization 
to the three channel problem we are actually faced with. 
For a detailed analysis, we refer the reader to two recent papers on the 
subject \cite{Gasser Wanders,wamultich} and the references quoted therein.

\subsection{Roy's integral equation in the one-channel case}

If we keep only the diagonal, singular Cauchy kernel in (\ref{eq:req1}),
the partial wave relations decouple, and the left hand cut in
  the amplitudes disappears.
Each one of the three partial wave amplitudes then obeys the following 
conditions:

\noindent
i) In the interval between
the  threshold $s=4M_\pi^2$ and the
matching point $s=s_0$, the real part is given by a 
dispersion relation
\begin{equation}\label{eq:1ch1}
\Re t(s)=a+(s-4M_\pi^2)\,{1\over \pi}\Pint_{4M_\pi^2}^\infty ds'\,
{\Im t(s')\over ( s'-4M_\pi^2)\,(s'-s)}\per
\end{equation}
ii)
Above $s_0$, the imaginary 
part $\Im t(s)$ is a given input function 
\begin{equation}\label{eq:1ch2}
\Im t(s) =A(s),\qquad s\geq s_0\per
\end{equation}
iii) For simplicity, we 
take the matching point in the elastic region, so that 
\begin{equation}\label{eq:1ch3}
t(s)=\frac{1}{\sigma(s)}e^{i\delta(s)}\sin{\delta(s)}\co\hspace{2em}
4M_\pi^2\leq s\leq s_0\co
\end{equation}
 where  $\delta(s)$ is real 
and vanishes at threshold. 
We refer the reader to \cite{Gasser Wanders} 
for a  precise formulation of the regularity
 properties required from  the amplitude and from the input absorptive 
part. As a minimal condition, we must require
\begin{equation}\label{eq:1ch4}
\lim_{s\nearrow s_0}\,\Im\,t(s)=A(s_0)\fs
\end{equation}
Otherwise, the principal value integral does not exist at the
matching point.

 Equations~(\ref{eq:1ch1})--(\ref{eq:1ch4}) constitute
the mathematical problem we are faced with in this case:
Determine the amplitudes
$t(s)$ that verify these equations  for  a given input 
of scattering length $a$ and  absorptive part $A(s)$. 
Once a solution is known, the real part of the
amplitude above $s_0$ is obtained from the dispersion relation
(\ref{eq:1ch1}), and $t(s)$ is then defined on $4M_\pi^2\leq s < \infty$.
The following points summarize the results relevant in our
context: 
\begin{enumerate}
\item
Elastic unitarity reduces the problem to the
determination of the   real function $\delta(s)$, defined in the  interval
$4M_\pi^2\leq s\leq s_0$. The amplitude $t(s)$ is then obtained 
from (\ref{eq:1ch3}).

\item
A given input $\{a,A(s)\}$ does not, in general, fix the solution
uniquely -- in addition, the value of the phase at the matching point
plays an important role.
Indeed, let $t(s)$  be a solution and suppose first that the phase at the
matching point is positive. For 
$0 < \delta(s_0) < \pi/2$, the infinitesimal neighbourhood of
$t(s)$ does not contain further solutions. For $\delta(s_0)
> \pi/2$, however, the neighbourhood contains an $m$-parameter family of
solutions. The integer $m$ is determined by the value of the phase at the
matching point ($\left[x\right]$ is the largest integer
not exceeding $x$):
\bea\label{eq:1ch5}
m=\left[2\delta(s_0)\over \pi\right]\fs\eea
 For a monotonically increasing
phase, the index $m$  counts the number of times
$\delta(s)$ goes through multiples of $\pi/2$ as $s$ varies
 from threshold to the matching point. We illustrate the situation for 
$m=0,1,2$ in figure~\ref{fig:1ch}.
\begin{figure}[t]
\begin{center}
\begin{tabular}{ccc}
\epsfig{file=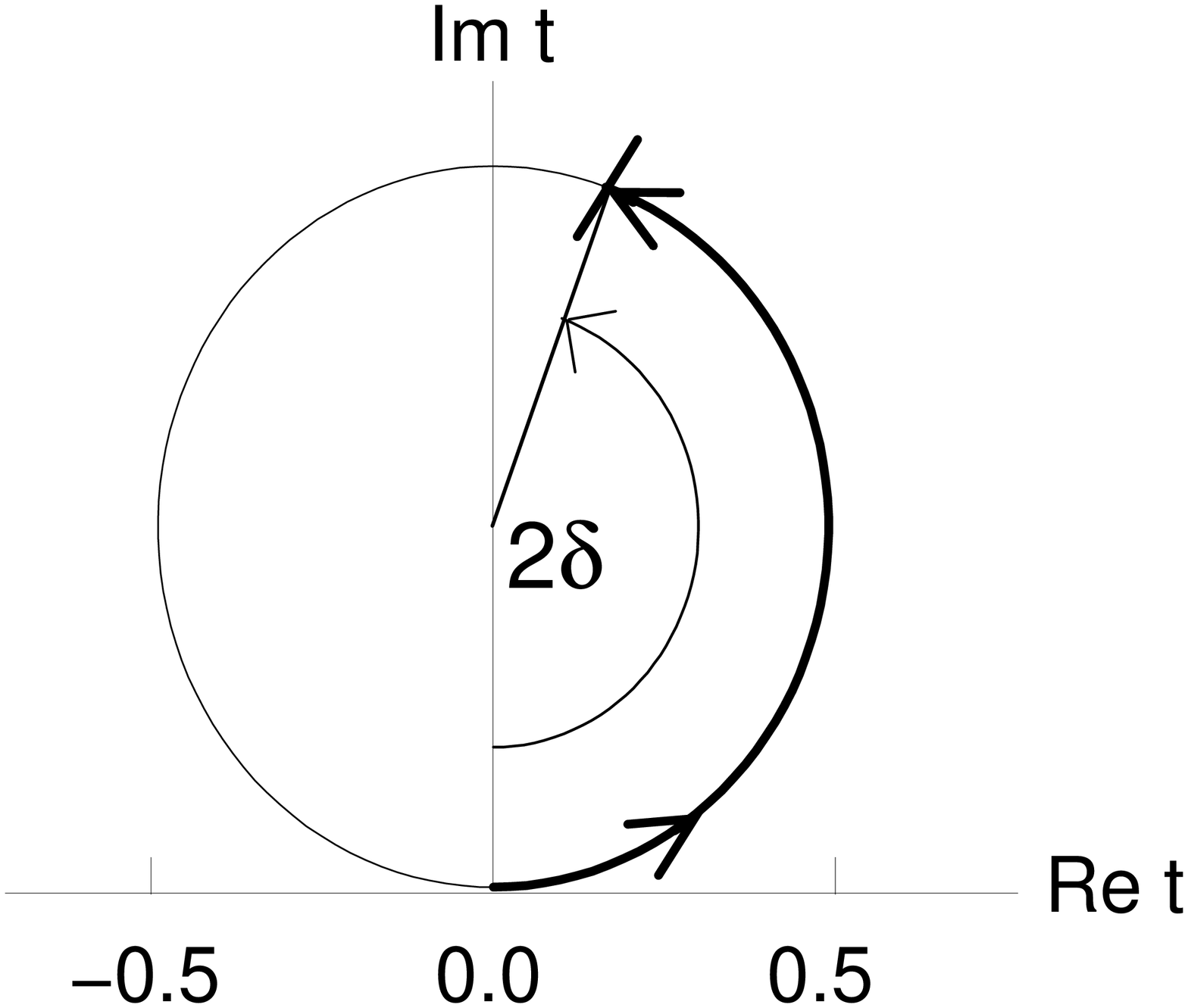,height=4.0cm,angle=-90} &
\epsfig{file=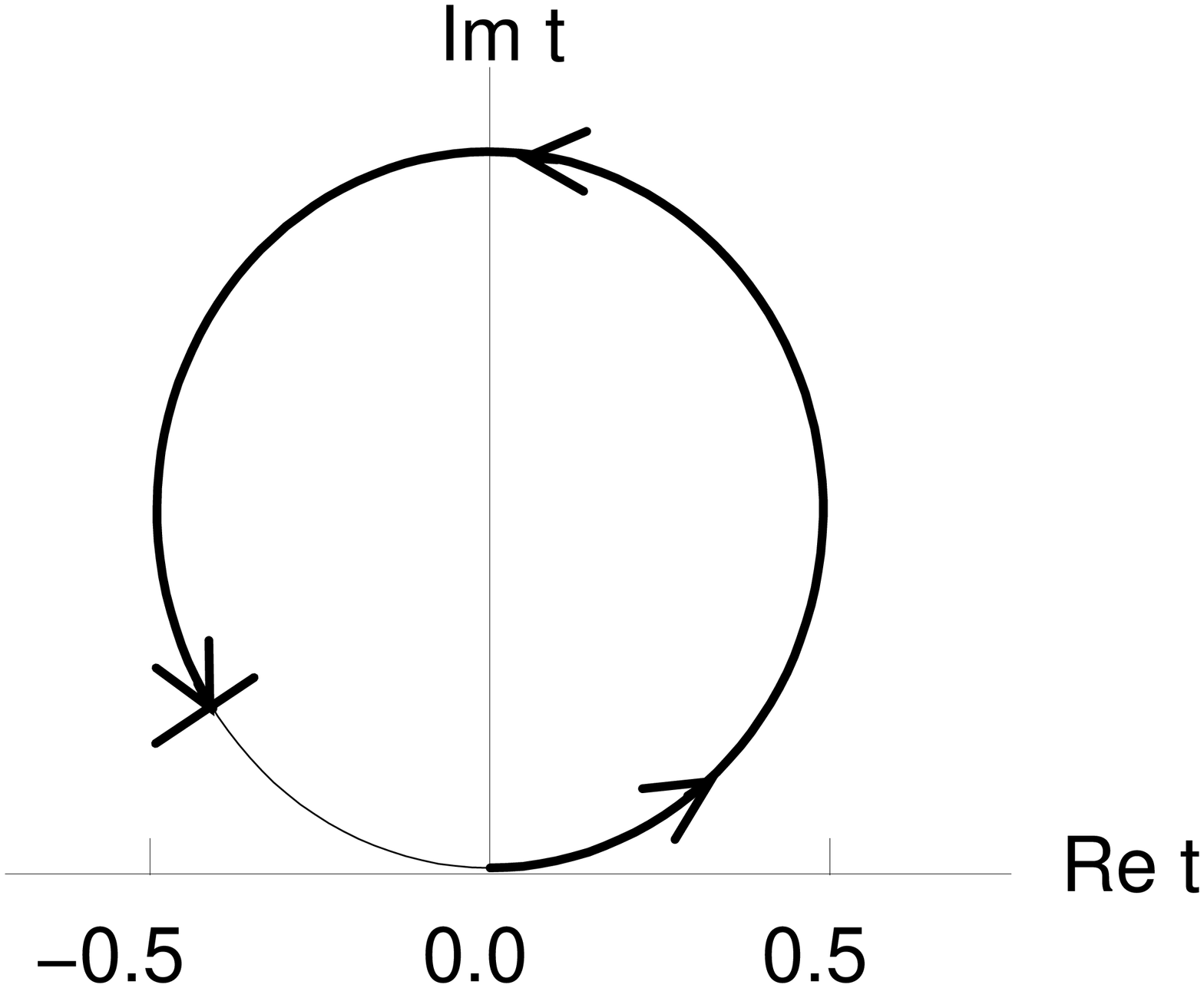,height=4.1cm,angle=-90} &
\epsfig{file=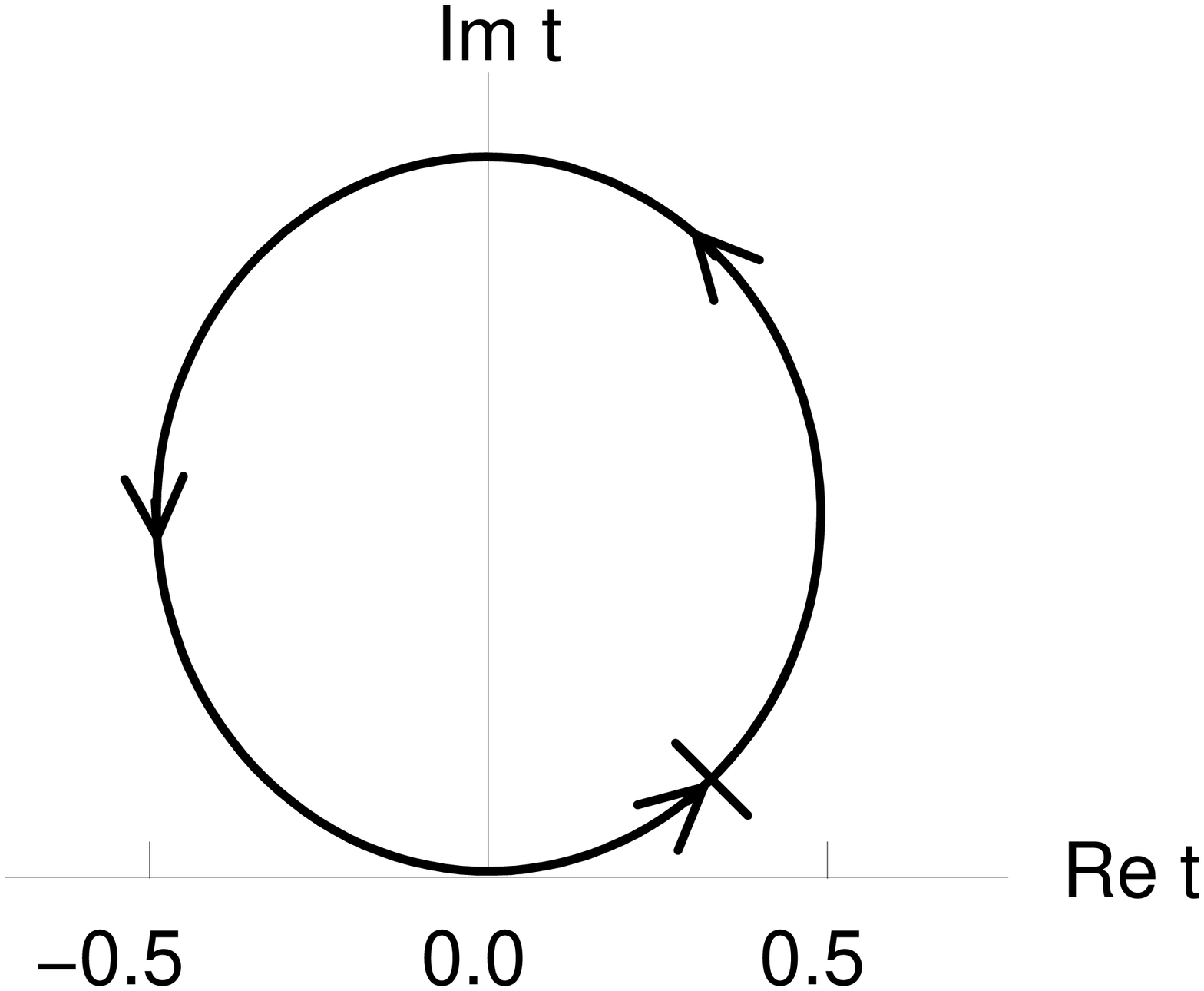,height=4.1cm,angle=-90}\\
&\\
\hspace{-1cm}a) $m=0$ & \hspace{-1cm}b) $m=1$ & \hspace{-1cm}c) $m=2$
\end{tabular}
\end{center}
\caption{Boundary conditions on the phase $\delta(s_0)$ for solving Roy's
integral equation. Figs. a,b,c represent the cases
$0<\delta(s_0)<\pi/2$, $\pi/2<\delta(s_0)<\pi$ and 
$\pi<\delta(s_0)<3\pi/2$, respectively. In fig.~c, the phase winds around the
Argand circle slightly more than once.\label{fig:1ch}}
\end{figure}
\item If the value of the phase at the matching point 
is negative, the problem does not in general have a solution.
In order for the problem to be soluble at all, the input must be tuned.
For $-\pi/2<\delta(s_0)<0$, for instance, we may keep the absorptive part
$A(s)$ as it is, but tune the scattering length $a$. This situation 
may be characterized 
by $m=-1$: Instead of having a family of solutions
containing free parameters, the input is subject to a constraint.
Once a solution does exist, it is unique in the sense 
that the infinitesimal neighbourhood does not 
contain further solutions.

\item
Consider now the case displayed in fig.~\ref{fig:1ch}a, where the phase
at the matching point is below $\pi/2$. 
This corresponds to the situation encountered 
in the coupled channel case, for our
choice of the matching point.
 According to the above statements, a given input $\{a,A(s)\}$ 
then generates a  locally
unique solution -- if a solution exists at all. 
We take it that uniqueness also holds globally, see
\cite{n/d}. 
\end{enumerate}
The solution may be constructed  in the 
following manner: Consider a
family of unitary amplitudes, parametrized through
$c_1,\ldots , c_n$. For any given amplitude, evaluate the
right and left hand sides of eq.~(\ref{eq:1ch1}) and calculate 
the square of the difference at $N$ points in the 
interval $4M_\pi^2\leq s\leq s_0$. Finally, minimize the sum of
these squares by choosing $c_1, \dots , c_n$ accordingly. Since the
 solution is unique, it suffices to find one with this method -- it is
then the only one.

\subsection{Cusps}
In general, the solutions are not regular at the matching point,
but have a cusp (branch point) there: $\delta(s)=\delta(s_0)+ 
C(s_0-s)^\gamma+\ldots\,$, with $\gamma > 0$. The phenomenon 
arises from our formulation of the problem --
the physical amplitude is regular there.
We conclude that, even if a mathematical solution can be constructed
for a given input $\{a,A(s)\}$, it will in 
general not be acceptable physically, because it contains a
fictitious singularity at the matching point. The behaviour of the phase 
is sensitive to the value of the exponent: If $\gamma$ is close to 1, the 
discontinuity in the derivative is barely visible, 
while for small values of $\gamma$, it manifests itself very 
clearly.

The strength of the singularity is determined by the constant $C$, whose value
depends on the input used. In particular,
if the scattering length $a$ is varied, while the 
absorptive part $A(s)$ is kept fixed, the size of $C$ changes. We may search 
for the value of $a$ at which 
$C$ vanishes. Although the singularity does not disappear entirely
even then, it now only manifests itself in the derivatives of the function
(for the solution to become analytic at $s_0$, we would need to also 
adapt the input for $A(s)$). In view 
of the fact that our solutions
are inherently fuzzy, because the values of the
input are subject to experimental uncertainties, we consider solutions 
with $C\simeq 0$ 
or $\gamma\simeq 1$ as
physically acceptable and refer to these as solutions without cusp.

The search for solutions without cusp can be implemented as follows. 
Instead of fixing $a$, constructing solutions 
in the class of functions with a cusp and then determining the value of $a$ at
which the cusp disappears, we may simply consider parametrizations that do not
contain a cusp, treating the scattering length $a$ as a free parameter,
on the same footing as the set $c_1,\ldots , c_n$ 
used to parametrize the phase shift and minimizing the difference between  
the left and right hand sides of 
eq.~(\ref{eq:1ch1}). We have verified that if a solution
without cusp does exist, this procedure indeed finds it: Allowing for the
presence of cusps does not lead to a better minimum.

The net result of this discussion is that the scattering length $a$ must match
the input for $A(s)$ -- it does not represent an independent parameter. When
solving the Roy equations, we can at the same time also determine the value of
$a$ that belongs to a given input for the high energy absorptive part.
The conclusion remains valid even if the matching point is above the
first inelastic threshold, provided the elasticity parameter
$\eta$ is known and sufficiently smooth at the matching point.
For a thorough analysis of the issue, we  
refer to \cite{wamultich}. 

\subsection{Uniqueness in  the multi-channel case}

In the multichannel case, we need to determine three functions
$\delta_0^0,\delta_1^1$ and $\delta_0^2$ for a given input 
 $\{a_0^0,a_0^2, f_\ell^I,\eta_\ell^I,d_\ell^I\}$. 
The multiplicity index $m$ of the infinitesimal neighbourhood of a given
solution is
displayed in table \ref{tab:3chm} \cite{wamultich}, for various values 
of the matching point $s_0$.
The table contains the following information. In the situations indicated
with the labels I and II, the infinitesimal neighbourhood
 of a given solution contains a family of solutions, characterized by   
2 and 1 free parameters, respectively. In case III, the solution is unique
in the sense that the neighbourhood does not contain further solutions,
while in case IV a solution only exists if the input is subject 
to a constraint ($m=-1$,  compare paragraph 3 in section 6.1). 
In order to uniquely characterize the solution in  case I, for instance,
we thus need to fix two more parameters -- in addition to the input --
say the position of the $\rho$ resonance and its width, or the
position of the $\rho$ resonance and the value of $s$ where the $I=0$
phase passes through $\pi/2$, and similarly for II. In the following, we stick
to  case III, where the solution is unique for
a given input. As discussed above, each of the three partial waves 
will in general develop a cusp at the matching point $s_0$, unless some
of the input parameters take special values.

\begin{table}\begin{center}
\begin{tabular}{c|l|l|l|r}
&$\rule[-0.7em]{1.7em}{0em}$ range of $s_0$ &\hspace{0.8em}range of 
$\delta_0^0$&\hspace{0.7em}range of $\delta_1^1$&$m$\\ 
\hline
I&$\rule{0em}{1.3em}\rule{1.8em}{0em}1<\Ezero<1.15$&
$\rule{0.7em}{0em}\pi <\delta_0^0 < \frac{3}{2}\,\pi$ & 
$\frac{1}{2}\,\pi < \delta_1^1 < \pi$&2\\
II&$\rule{0.5em}{0em}0.86<\Ezero<1$&$\frac{1}{2}\,\pi < \delta_0^0 < \pi$ & 
$\frac{1}{2}\,\pi < \delta_1^1 < \pi$&1\\
III&$\rule{0.5em}{0em}0.78 <\Ezero<0.86$&$\rule{0.8em}{0em}0< \delta_0^0 < 
\frac{1}{2}\,\pi$ & 
$\frac{1}{2}\,\pi < \delta_1^1 < \pi$&0\\
IV&$\rule{0.5em}{0em}0.28<\Ezero<0.78$&$\rule{0.8em}{0em}0< \delta_0^0 < 
\frac{1}{2}\,\pi$ & 
$\rule{0.9em}{0em}0<\delta_1^1 < \frac{1}{2}\,\pi$&$-1$\\
\end{tabular}\end{center}\vspace*{-1em}
\caption{Multiplicity of solutions in the coupled channel case. The
  multiplicity index $m$ is
  the number of free parameters occurring in the solutions of the Roy
  equations, if the
matching point $s_0$ is in the interval indicated (in GeV units). 
Also displayed is the
variation of the physical phases $\delta_0^0$ and $\delta_1^1$ on
that interval.\label{tab:3chm}}
\end{table}

The situation encountered in practice is the following.  
Let $0.1<a_0^0 <0.6$, and let
$f_\ell^I$, $\eta^I_\ell$ and $d_\ell^I$ be fixed as well. For an
arbitrary value of the scattering length $a_0^2$, the solution in general 
develops a strong cusp in the $P$-wave.
This cusp can be removed by tuning 
$a_0^2\rightarrow\bar{a}_0^2$, using for instance the method described in the
single channel case above. Remarkably, it turns out that the solutions so
obtained are nearly free of cusps in the two $S$-waves as 
well. The problem manifests itself almost exclusively in the
$P$-wave, because our matching point is rather close to the mass of
the $\rho$, where the imaginary part shows a pronounced 
peak. If $a_0^2$ is chosen to slightly differ from the optimal value
$\bar{a}_0^2$, a cusp in the $P$-wave is clearly seen. We thus obtain a
relation between the scattering lengths $a_0^0$ and $a_0^2$. This is how
the so-called {\em universal curve},
discovered a long time ago \cite{Morgan Shaw}, shows up in our framework. 
We will discuss the properties 
of this curve in detail below. 

In principle, we might try to also fix $a_0^0$ with this method, 
requiring that there be no cusp in one of the two $S$-waves. 
The cusps in these are very weak, however -- the procedure does not allow us 
 to accurately pin down the second scattering
length. The choice $a_0^0=-0.2$, for instance, still leads to a fully 
acceptable solution. On the
other hand, we did not find a solution in the class of smooth
functions for $a_0^0=-0.5$.
 This shows that the analyticity properties that are not encoded in
the Roy integral equations (\ref{eq:rieq1}) do constrain the range of
admissible values for $a_0^0$, but since 
that range is very large, the constraint is not of immediate
interest, and we do not consider the matter further. In our numerical work,
we consider values in the range $0.15<a_0^0<0.30$ and use the center of
this interval, $a_0^0=0.225$, as our reference point.

\setcounter{equation}{0}
\section{Experimental input}\label{sec:exp_input}
In this section, we describe
the experimental input used for the elasticity below the matching point
at $\Ezero=0.8\,\mbox{GeV}$ and for the imaginary parts of the $S$- and 
$P$-waves in the energy interval between 
$\Ezero$ and $\Etwo=2\,\mbox{GeV}$. The references are listed in
\cite{saclay}--\cite{E852} and for an overview, we refer to 
\cite{Morgan Pennington handbook,Ochs Newsletter}.
The evaluation of the contributions from the higher partial waves 
and from the asymptotic region ($s>s_2$) is discussed in detail in 
appendix \ref{background}.

\subsection{Elasticity below the matching point}
The Roy equations allow us to determine the phase shifts of the 
$S$- and $P$-waves only if -- on the interval between threshold and 
the matching point -- the corresponding elasticity parameters
$\eta_0^0(s)$, $\eta_1^1(s)$ and $\eta_0^2(s)$ are known.  
On kinematic grounds,
the transition $2\pi\rightarrow 4\pi$ is the only inelastic channel open 
below our matching point, $\Ezero=0.8\,\mbox{GeV}$. 
The threshold for this reaction is at 
$E=4\,M_\pi\simeq 0.56\,\mbox{GeV}$,
but phase space strongly suppresses the transition at low energies -- 
a significant inelasticity only sets in above the matching point. 
In particular, 
the transition $\pi\pi\rightarrow K\bar{K}$, 
which occurs for $E>2\,M_K\simeq 0.99\, \mbox{GeV}$, does generate a 
well-known, pronounced structure in the elasticity parameters of the waves 
with $I=0,1$. Below the matching point,
however, we may neglect the inelastic reactions altogether and
set
\bea 
\eta^0_0(s)=\eta^1_1(s)=\eta^2_0(s)=1\co\hspace{2em}\sqrt{s}<0.8\,
\mbox{GeV}\fs\nonumber\eea

We add a remark concerning the effects generated
by the inelastic reaction 
$2\pi\rightarrow 4\pi$, which are analyzed  in  ref.~\cite{bugg}. 
In one of the phase shift analyses
given there (solution A), the inelasticity $1-\eta_1^1(s)$ 
reaches values of order 4\%,
already in the region of the $\rho\,$-resonance.  
The effect is unphysical -- it arises because the parametrization 
used does not account for the strong phase space suppression at the
$4\pi$ threshold\footnote{We thank Wolfgang Ochs for this remark.}. 
For the purpose of the analysis performed 
in ref.~\cite{bugg}, which focuses on the region above 1 GeV,
this is immaterial, but in our context, it matters:
We have solved the Roy 
equations also with that representation for the elasticities. The result
shows significant distortions, in particular in the $P$-wave.

\subsection{Input for the $I=0,1$ channels}

The experimental information on the $\pi \pi$ phase shifts in
the intermediate energy region comes mainly from the reaction $\pi N
\rightarrow \pi \pi N$.  A rather involved analysis is necessary
to extract the $\pi \pi$ phase shifts from the raw data, and
several different representations  for the phases and
elasticities are available in the literature.  The main
source of experimental information is still the old measurement of the
reaction $\pi^- p \rightarrow \pi^- \pi^+ n $ by the CERN--Munich
(CM) collaboration \cite{Grayer}, but there are also older, statistically less
\begin{figure}[htb]
\begin{center}
\leavevmode
\includegraphics[angle=-90,width=13cm]{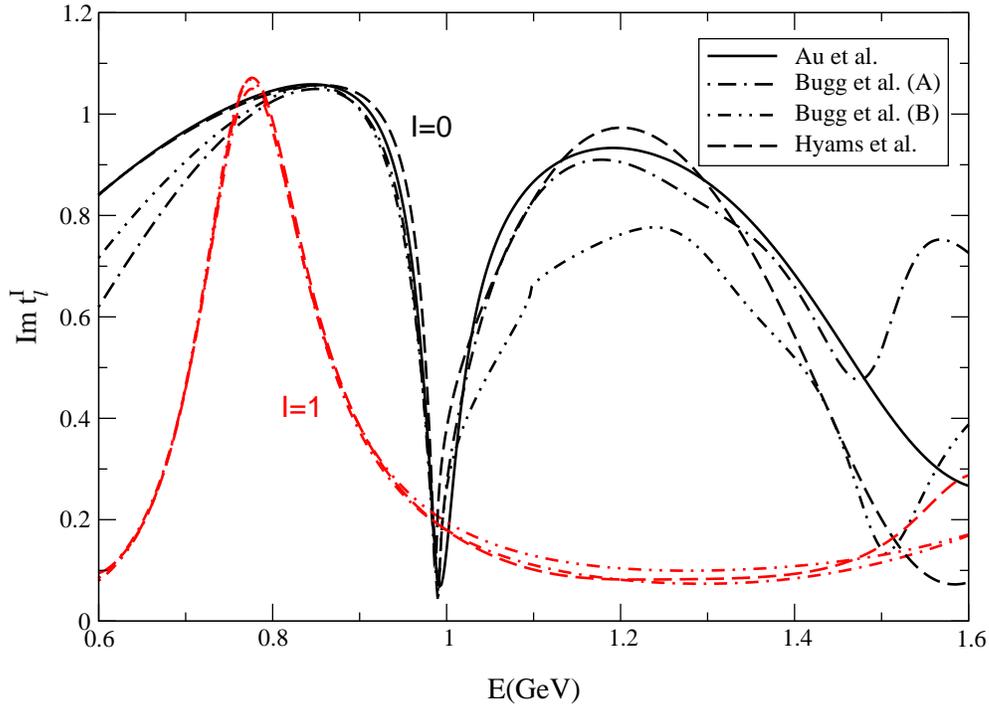}
\end{center}
\caption{\label{fig:ImAHSZ} Comparison of the different input we used for the
  imaginary parts of the $I=0$ and $I=1$ lowest partial waves above the
  matching point at 0.8 GeV.} 
\end{figure}
precise data, for instance from Saclay \cite{saclay} and Berkeley 
\cite{Protopopescu}, as well as 
newer ones, such as the data of the CERN-Cracow-Munich collaboration
concerning pion production
on polarized protons \cite{Becker} and those on the reaction 
$\pi^- p\rightarrow
\pi^0\pi^0 n$, obtained recently by the 
E852 collaboration at Brookhaven \cite{E852}.
For a detailed discussion of the available experimental information,
we refer to \cite{Morgan Pennington handbook,bugg,Ochs Newsletter}.

\begin{figure}[thb]
\begin{center}
\leavevmode
\includegraphics[angle=-90,width=12cm]{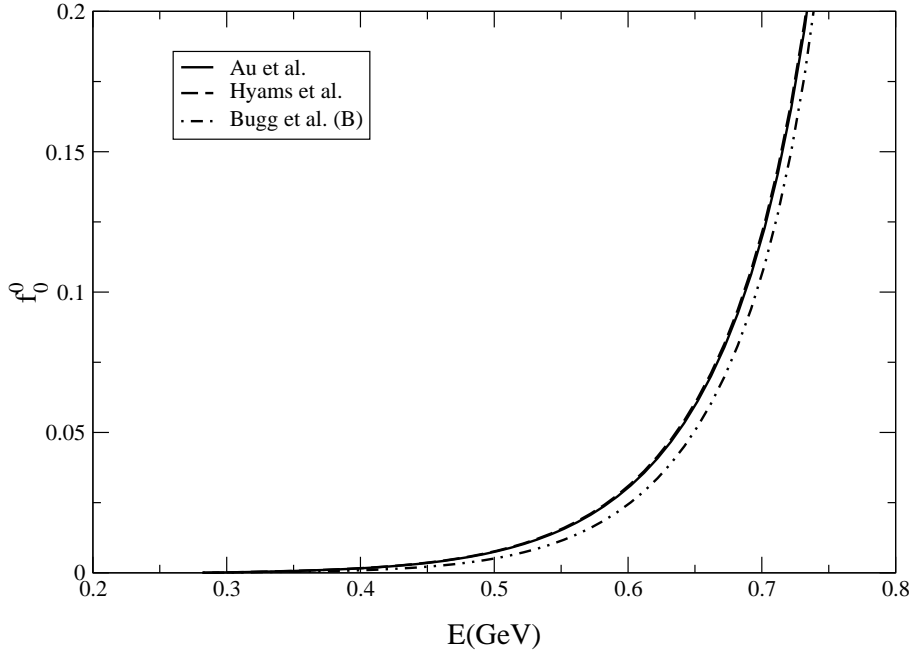}
\end{center}
\vspace*{-0.5cm}
\caption{\label{fig:RoyAHSZ} Comparison of the results obtained for the
  dispersion integral $f^0_0$ with the various imaginary parts
  shown in fig.~\protect{\ref{fig:ImAHSZ}}.}
\end{figure}

For our purposes, energy-dependent analyses are most convenient, 
because these yield analytic expressions for the imaginary parts, so that
the relevant integrals can readily be worked out.
To illustrate the differences between these analyses, we plot the
corresponding imaginary parts in
fig.~\ref{fig:ImAHSZ}, both for the $I=0$ $S$-wave and for the
$P$-wave. 
The representations of refs.~\cite{hyams,au,bugg} do not 
extend to 2 GeV, but they do cover the range between 0.8 and 1.7 GeV.   
Unitarity ensures that the contributions generated by the
imaginary parts of the $S$- and $P$-waves in the region between 
1.7 and 2 GeV are very small, so that we may use these representations
also there without introducing a significant error. For the $P$-wave, 
the differences between the
various parametrizations are not dramatic, but 
for the $I=0$ $S$-wave, they are quite substantial. Despite these
differences, the result obtained for the dispersive integrals are 
similar, at least in the range where we are solving the Roy equations.
This can be seen in fig.~\ref{fig:RoyAHSZ}, where we plot the value
of the dispersion integral $f^0_0$, defined in eq.~(\ref{eq:rieq2}). 
The only visible difference is between parametrization B of 
ref.~\cite{bugg} and the others. In order of magnitude, the effect is
comparable to the one occurring if the scattering length $a^0_0$ 
is shifted by $0.01$.
It arises from the difference in the behaviour of the $S$-wave imaginary part 
in the region between 1 and 1.5 GeV.  
The phase shift analysis of Protopopescu et al.~\cite{Protopopescu} 
does not cover that region, as it only extends to 1.15 GeV, but
those of Au, Morgan and Pennington \cite{au} as well as Bugg,
Sarantsev and Zou \cite{bugg} do. Both of these include, aside from the CM
data, additional experimental information, not included in the analysis of
Hyams et al.~\cite{hyams}. 

In the following, we rely on the representation of Au et al.~\cite{au} for
the $S$-wave and the one of Hyams et al.~\cite{hyams} for the $P$-wave (the
analysis of Au et al.~does not include the $P$-wave). We have verified
that, using~\cite{hyams} also for the $S$-wave would not change our results
below the matching point, beyond the uncertainties to be attached to the
solutions, anyway.  On the other hand, Au et al.~\cite{au} yield a more
consistent picture above the matching point -- for this reason
we stick to that analysis.  More precisely, we use the solution denoted by
$K_1$(Etkin) in ref.~\cite{au}, table I. That solution contains a narrow
resonance in the 1 GeV region, which does not occur in the other phase
shift analyses.  In our opinion, the extra state is an artefact of the
representation used: A close look reveals that the occurrence of this state
hinges on small details of the $K$-matrix representation. In fact, 
the resonance disappears if two of the $K$-matrix coefficients are slightly
modified, for instance with
$(-c_{12}^0,-c_{22}^0)=(3.1401,2.8447)\rightarrow (3.2019,2.6023)$.

\subsection{Phase of the $P$-wave from $e^+e^-\rightarrow \pi^+\pi^-$
and $\tau\rightarrow \pi^-\pi^0\,\nu_\tau$}
\label{sec:Pwave}
For the $P$-wave, the data on the processes $e^+e^-\rightarrow \pi^+\pi^-$
and $\tau\rightarrow \pi^-\pi^0\,\nu_\tau$ yield very useful, independent
information. The corresponding transition amplitude is proportional to the
pion form factor $F_{\ind e.m.}(s)$ of the electromagnetic current and to
the form factor $F_{\ind V}(s)$ of the charged vector current,
respectively. The data provide a measurement of the quantities $| F_{\ind
e.m.}(s) |$ and $| F_{\ind V}(s) |$ in the time-like region, $s>4M_\pi^2$.

In the isospin limit, the two
form factors coincide: The currents only differ by an isoscalar operator 
that carries odd $G$-parity, so that the pion matrix elements thereof vanish.
While the isospin breaking effects 
in $| F_{\ind V}(s) |$ are very small, $\rho-\omega$ interference 
does produce a pronounced structure in the electromagnetic form 
factor. The $\omega$-resonance
generates a second sheet pole in the isoscalar matrix elements, 
at $s=(M_\omega-i\,\frac{1}{2}\,\Gamma_\omega)^2$. 
The residue of the pole is small, of order $O(m_d-m_u,e^2)$, but 
in view of the small width of the $\omega$, the denominator also
nearly vanishes for $s=M_\omega^2$. Moreover, the pole 
associated with the exchange of a $\rho$ occurs in
the immediate vicinity of this point, so that the transition amplitude
involves a sum of two contributions that rapidly change with $s$, both in
magnitude and phase. Since the
interference phenomenon is well understood, it can be corrected for. When this
is done, the data on the two processes $e^+e^-\rightarrow \pi^+\pi^-$ and
$\tau\rightarrow \pi^-\pi^0\nu$ are in remarkably good agreement
(for a review, see \cite{Eidelman Jegerlehner,aleph}).

We denote the phase of the vector form factor by $\phi(s)$,
\bea F_{\ind V}(s) =| F_{\ind V}(s)|\,e^{\,i\,\phi(s)} \fs\nonumber \eea 
In the elastic region $4M_\pi^2<s<16 M_\pi^2$, the final state interaction
exclusively involves $\pi\pi$ scattering, so that the Watson
theorem implies that the phase $\phi(s)$
coincides with the $P$-wave phase shift,
\bea \phi(s)=\delta_1^1(s)\co\hspace{2em}4M_\pi^2 < s < 16 M_\pi^2\fs\nonumber
\eea
In fact, phase space suppresses the inelastic channels also in this case
-- the available data on the decay channel $\tau\rightarrow 4\,\pi\,\nu_\tau$
show that, for $E< 0.9\,\mbox{GeV}$, the inelasticity is below 1\%, so that
the phase of the form factor must agree with the $P$-wave phase shift, to high
accuracy \cite{Lukaszuk}. 

In the region where the singularity generated by $\rho\,$-exchange dominates,
in particular also in the vicinity of our matching point, 
the form factor is well represented by a resonance term and a slowly
varying background. Quite a few such representations may be found in the
recent literature. Since the uncertainties in the data (statistical as
well as systematic) are small, these parametrizations agree quite well.
In the following, we use the Gounaris-Sakurai representation of
ref.~\cite{cleo} as a reference point. That representation involves a
linear superposition of three resonance terms, associated with
$\rho(770)$, $\rho(1450)$ and $\rho(1700)$. We have investigated the 
uncertainties to be attached to this representation by (a) comparing the 
magnitude of the form factor with the available data\footnote{We are indebted
  to Simon Eidelman and Fred Jegerlehner for providing us with these.}, (b)
comparing it with other parametrizations, (c) varying the
resonance parameters in the range quoted in ref.~\cite{cleo} 
and (d) using the fact that analyticity imposes a strong
correlation between the phase of the form factor and its magnitude.
On the basis of this analysis, we conclude that the $e^+e^-$ and $\tau$ data
determine the phase of the $P$-wave at $0.8 \,\mbox{GeV}$ to 
within an uncertainty of $\pm 2^\circ$. A detailed comparison between the
phase of the form factor and the solution of the Roy equations for the 
$P$-wave will be given in section \ref{sec:rho}.

\subsection{Phases at the matching point}\label{phases at the matching point}

In the framework of our analysis, the input used for $s\geq s_0$ 
enters in two ways: (i) it specifies the value of the three phases
at the matching point and (ii) it determines the contributions to the 
Roy equation integrals from the region above that point. Qualitatively, we are
dealing with a boundary value problem: At threshold, the phases vanish,
while at the matching point, they are specified by the input.
The solution of the Roy equations then yields the proper interpolation 
between these boundary values. The behaviour of the
imaginary parts above the matching point is less important than 
the boundary values, because it only affects 
the slope and the curvature of the solution.
\begin{table}
\begin{center}
\begin{tabular}{llll}
$\rule{1.7em}{0em}\delta_0^0$ & $\rule{1.7em}{0em}\delta_1^1$ & 
$\rule{0.5em}{0em}\delta_1^1-\delta_0^0\rule{2em}{0em} $& 
reference$\rule[-0.5em]{0em}{0em}$ \\  \hline 
       81.7 $\pm$ 3.9  &  105.2 $\pm$ 1.0&   23.4 $\pm$ 4.0&
\cite{ochs_phd,hyams}$\rule{0em}{0.95em}$\\
       90.4 $\pm$ 3.6 &  115.2 $\pm$ 1.2 &   24.8 $\pm$ 3.8&
\cite{EM} s-channel moments\\
       85.7 $\pm$  2.9 &   116.0  $\pm$ 1.8 &   30.3 $\pm$ 3.4 &
\cite{EM} t-channel moments $\rule[-0.45em]{0em}{0em}$ \\
\hline
       81.6  $\pm$ 4.0  &  108.1         $\pm$ 1.4 &   26.5 $\pm$ 4.2
&\cite{Protopopescu} table VI$\rule{0em}{0.95em}$\\ 
       80.9  &      105.9 &   25.0 & 
 \cite{ochs_phd,hyams}$\rule{0em}{1em}$\\
       79.5  &      106.1 &   26.5 & 
 \cite{bugg} solution A\\
       79.9  &      106.8 &   26.9 & 
 \cite{bugg} solution B\\
       80.7   &    $\rule{1em}{0em} -$ &   $\rule{0.5em}{0em}-$&
\cite{au} solution K$_1$\\
       82.0  &      $\rule{1em}{0em}-$ &   $\rule{0.5em}{0em}-$&
\cite{au} solution K$_1$(Etkin) $\rule[0.7em]{0em}{0em}$
\end{tabular}
\end{center}
\vspace{-1em}
\caption{Value of the phases $\delta_0^0$ and $\delta_1^1$ at 0.8 GeV. 
The first three rows stem from analyses of the data
at a fixed value of the energy (``energy independent''), 
while the remaining entries
are obtained from a fit to the data that relies on an explicit 
parametrization of the energy dependence (``energy dependent analysis'').} 
 \label{tab:phase_matching}
\end{table}

We now discuss the available information for the phases $\delta_0^0$
and $\delta_1^1$ at the matching
point. The values obtained from the high energy, high statistics 
$\pi N\rightarrow\pi\pi N$ experiments
are collected in table \ref{tab:phase_matching}.  
In those cases where the published numbers do not directly apply at 
$0.8 \,\mbox{GeV}$, we have used a quadratic interpolation between the three 
values of the energy closest to this one. 
The errors given in the third column are obtained by adding
 those from the first two columns in quadrature.
For the energy dependent entries, the error analysis is more involved -- 
only ref.~\cite{Protopopescu} explicitly quotes an error.
The scatter seen in the table partly arises from the fact that different
methods of analysis are used. The corresponding systematic uncertainties
are not covered by the error bars quoted in the individual phase shift
analyses: Taken at face value, the numbers listed in the table
are contradictory, particularly in the case of the $P$-wave. For a thorough
discussion of the experimental discrepancies, we refer to 
\cite{Ochs Newsletter}. 

As discussed above,  both the statistical and the systematic uncertainties of
the $e^+e^-$ and $\tau$ data are considerably smaller.
They constrain the phase of the $P$-wave at 0.8 GeV to a narrow range, 
centered around the
value $\delta_1^1(s_0)=108.9^\circ$ obtained with the Gounaris-Sakurai
representation of the form factor in ref.~\cite{cleo}:
\bea\label{d1} \delta_1^1(s_0)=108.9^\circ\pm 2^\circ\fs\eea
The comparison with the numbers listed in the second column of the table
shows that this value is within the range of the results obtained from 
$\pi N\rightarrow\pi\pi N$.

Unfortunately, the $e^+e^-$ and $\tau$ data only concern the $P$-wave.
To pin down the $I=0$ $S$-wave, we observe that 
the overall phase of the scattering amplitude drops
out when considering the difference $\delta_1^1-\delta_0^0$, so that 
one of the sources of systematic error is absent.
Indeed, the third column in the table shows that the outcome of the
various analyses is consistent with the assumption
that the fluctuations seen are of statistical origin. 
The statistical average of the energy independent analyses yields
$\delta_1^1(s_0)-\delta_0^0(s_0)=26.6^\circ \pm 3.7^\circ$, with
$\chi^2=2$ for 2 degrees of freedom (as the numbers are based on the 
same data, we have inflated the error bar -- the number given is 
the mean error of the three data points). 
The remaining entries in the table neatly confirm this result. Combining
it with the one in the fourth row, which is based on
independent data, we finally arrive at 
\bea\label{d01} \delta_1^1(s_0)-\delta_0^0(s_0)=26.6^\circ \pm 2.8^\circ\fs\eea

Since the value for $\delta_1^1$ comes from the data on the form factor,
while the one for the difference $\delta_1^1-\delta^0_0$ is based on the
reaction $\pi N\rightarrow \pi\pi N$, these numbers are independent, so
that it is legitimate to combine them. Adding errors quadratically, we
obtain \bea \label{d0}\delta^0_0(s_0)=82.3^\circ\pm 3.4^\circ\fs\eea

In the following, we rely on the two values for the phases at the matching
point given in eqs.~(\ref{d1}) and (\ref{d0}). We emphasize that the $\pi
N\rightarrow \pi\pi N$ data are consistent with these -- in fact, the
result of the energy-dependent analysis quoted in the fourth row of the
table is in nearly perfect agreement with the above numbers. We are
exploiting the fact that the $e^+e^-$ and $\tau$ data strongly constrain
the behaviour of the $P$-wave in the region of the $\rho$, thus reducing
the uncertainties in the value of $\delta_1^1$ at the matching point.

For the principal value integrals to exist, we need to continuously connect
the values of the imaginary parts calculated from the phases at the
matching point with those of the phase shift representation we wish to
use. This can be done, either by slightly modifying the parameters
occurring in the representation in question or with a suitable
interpolation of the phases between the matching point and $K\bar{K}$
threshold. We have checked that our results do not depend on how that is
done, as long as the interpolation is smooth. Note that, for the
representation $K_1$(Etkin) \cite{au} -- our reference input for the
imaginary part of the $I=0$ $S$-wave -- an interpolation is not needed: The
last row of table \ref{tab:phase_matching} shows that, at the matching
point, this representation nearly coincides with the central value in
eq.~(\ref{d0}).

\subsection{Input for the $I=2$ channel}
\begin{figure}[htb]
\leavevmode\centering
\includegraphics[width=12cm]{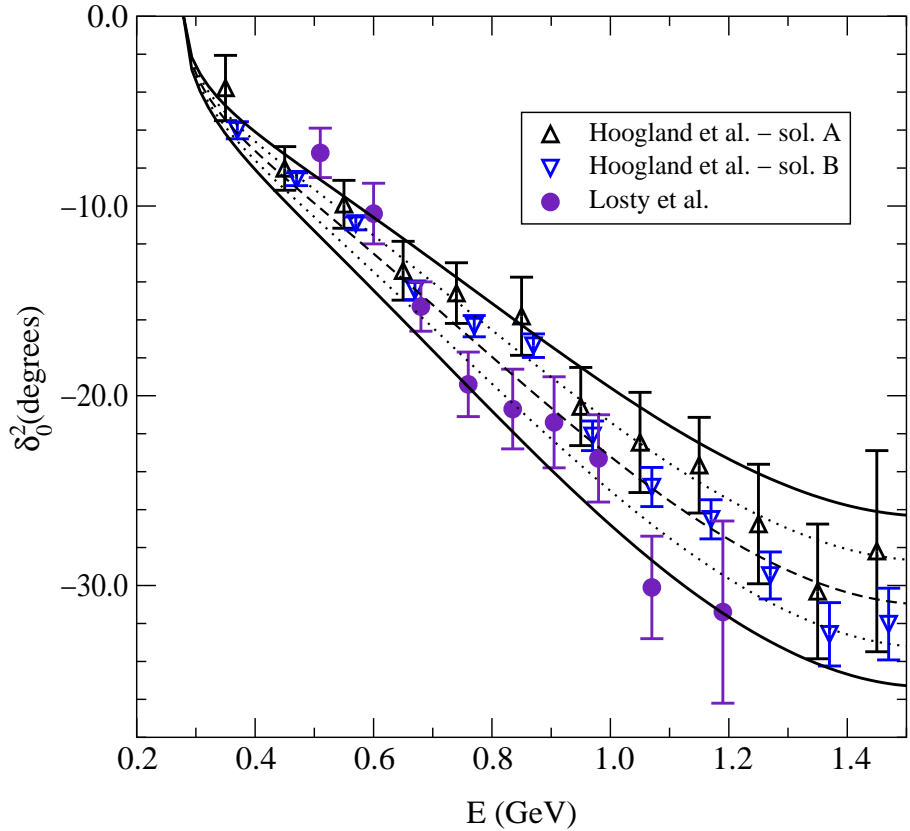}
\caption{\label{fig:I2} Different data sets for the $S$-wave in the $I=2$
  channel and curves that we have used as input in the Roy equation
  analysis.}
\end{figure}
The uncertainties in this channel are rather large.  The current
experimental situation is summarized in fig.~\ref{fig:I2}, where we show
the data points from the two main experiments \cite{losty,hoogland}, and
five different parametrizations that we will use as input. The central one
is our best fit to the data of the Amsterdam--CERN--Munich collaboration
(ACM) \cite{hoogland} solution B (which we call from now on ACM(B)) with a
parametrization {\em \`a la} Schenk \cite{schenk}. To cover the rather wide
scatter of the data, we have varied the input in this channel, using the
five curves shown in the figure, together with $\eta^2_0=1$ (note that for
the Roy equation analysis, only the value of the scattering length $a^2_0$
and the behaviour of the imaginary part above 0.8 GeV matter).

\setcounter{equation}{0}
\section{Numerical solutions}
\label{numerical solutions}
In the preceding section, the input required to evaluate 
the r.h.s.~of our system of equations was discussed in 
detail. In the present section, we describe the numerical method used to 
solve this system and illustrate the outcome with an example.

\subsection{Method used to find solutions}
We search for solutions of the Roy equations by numerically minimizing the
square of the difference between the left and right hand sides
of eq.~(\ref{eq:rieq1}) in the region between threshold and 0.8 GeV. 
As we are neglecting the inelasticity in this region,
the real and imaginary parts of $t^I_\ell(s)$ are determined by a
single real function, the phase $\delta_\ell^I(s)$. 
In principle, the minimization
should be performed over the whole space of physically acceptable functions
$\{\delta_0^0(s),\; \delta_1^1(s),\; \delta_0^2(s)\}$, but for obvious
practical reasons we restrict ourselves to functions described by a simple
parametrization. We will use the one proposed by Schenk some time ago
\cite{schenk}, allowing for an additional parameter in the polynomial part:
\begin{equation}
\tan \delta_\ell^I = \sqrt{1-{4 M_\pi^2 \over s}}\; q^{2 \ell} \left\{A^I_\ell
+ B^I_\ell q^2 + C^I_\ell q^4 + D^I_\ell q^6 \right\} \left({4
  M_\pi^2 - s^I_\ell \over s-s^I_\ell} \right) \; \; ,
\end{equation}
The first term represents the scattering length, while the second is
related to the effective range: 
\bea \label{effrange}
a^I_\ell = A^I_\ell\co\hspace{2em}
b^I_\ell =B_\ell^I + {4  \over s^I_\ell -4 M^2_\pi}\, A_\ell^I -
\frac{1}{ M_\pi^2}\,\left(A^I_\ell \right)^3 \delta_{\ell\, 0} \fs
\eea 
In each channel, one of the five parameters is fixed in order to ensure the
proper value of the phase at $s_0$.  Moreover the $S$-wave scattering
lengths $a^0_0$ and $a^2_0$ are identified with the two constants that
specify the subtraction polynomials in the Roy equations. As discussed
in sect.~\ref{sec:unique}, we need to tune the value of $a_0^2$ in order to
avoid cusps. Treating this parameter on the same footing as the others, we
are dealing altogether with $15-3-1=11$ free variables, to be determined by
a minimization procedure. Our choice of $s_0$ ensures that the solution is
unique, and therefore the method is safe: The choice of a bad
parametrization would manifest itself in a failure of the minimization
method -- the minimum would not yield a decent solution.

The square of the difference between the left and right hand sides of
the Roy equations is calculated at 22 points between threshold and $s_0$
for each of the three waves, so that the sum of squares ($\sos$) contains
66 terms. The minimization of the function ($\sos$) over $11$ parameters
can be handled by standard numerical routines \cite{NAG}. Our procedure
does generate decent solutions: The differences between the left and right hand
sides of the Roy equations are not visible on our plots -- they are
typically of order $10^{-3}$. The equations could be solved even more
accurately by allowing for more degrees of freedom in the parametrization
of the phases, but, in view of the uncertainties in the input, the accuracy
reached is perfectly sufficient. Note also that the exact solution
corresponding to a given input contains cusps. We have checked that these
are too small to matter: Enlarging the space of functions on which the
minimum is searched by explicitly allowing for such cusps in the
parametrization of the phases, we find that the solutions remain
practically the same.

\subsection{Illustration of the solutions}

To illustrate various features of our numerical solutions, we freeze for a
moment all the inputs and analyze the properties of the
specific solution we then get. The input for the imaginary parts
above $s_0$ is the following:
For the $I=0$ wave, we use the parametrization labelled 
$K_1$ (Etkin) of Au et al.~\cite{au}. In the case of 
the $I=1$ wave, we rely on the
energy--dependent analysis of Hyams et al.~\cite{hyams}, smoothly modified
between $s_0$ and $4 M_K^2$ to match the value 
$\delta_1^1(s_0)=108.9^\circ$. For the
$I=2$ wave, we take the central curve in fig.~\ref{fig:I2}. The driving
terms are specified in eq.~(\ref{numerical driving terms}). Moreover we
fix $a_0^0=0.225$.
\begin{figure} 
\leavevmode \centering
\includegraphics[angle=-90,width=12cm]{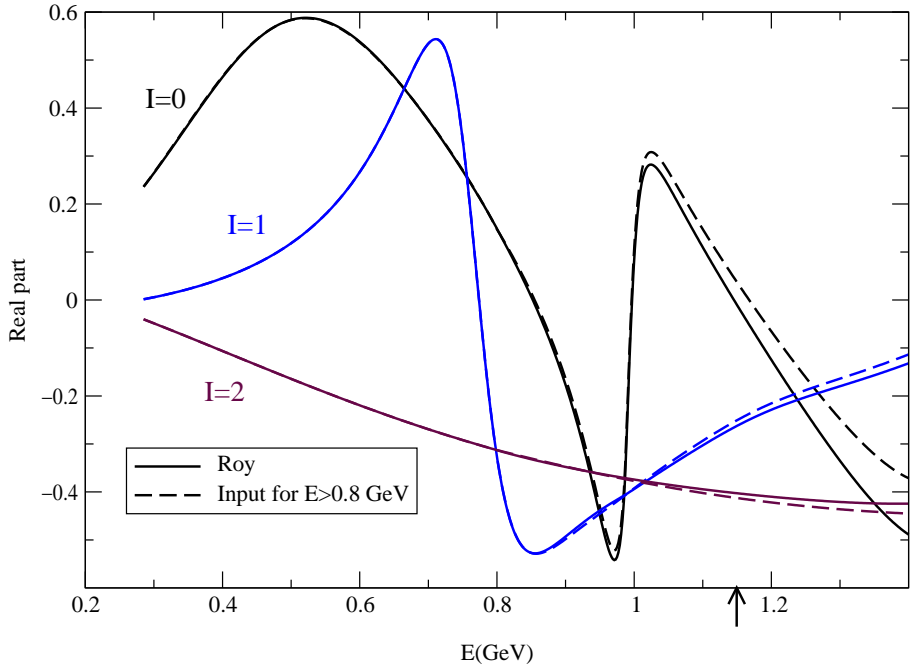}
\caption{\label{fig:roy_a225} Numerical solution of the Roy equations for
$a_0^0=0.225$, $a_0^2=-0.0371$ (the value of $a_0^0$ corresponds to the center
of the range considered while the one of $a_0^2$ results if 
the input used for $\mbox{Im}\,t^2_0$ is taken from the  central curve in 
fig.~\protect{\ref{fig:I2}}). The arrow indicates the limit of validity of
the Roy equations.}
\end{figure}
With this input,
the minimization leads to $a^2_0=-0.0371$ and the Schenk 
parameters take the values listed in table \ref{tab:standardsolution}, 
in units of $M_\pi$.
\begin{table}
\centering
\begin{tabular}[th]{|l|r|r|r|}
\hline
 &  \qquad   $I=0\rule{1.5em}{0em}$ &  \qquad  
$I=1\rule{1.5em}{0em}$
 &\qquad   $I=2\rule{1.5em}{0em}$\\
\hline
$A^I_\ell$& $\rule{0em}{1.2em} 0.225\rule{1.6em}{0em}$ & $ 3.63 \cdot 10^{-2}$
& $-3.71
\cdot 10^{-2}$\\ 
$B^I_\ell$& $ 0.246\rule{1.6em}{0em}$& $ 1.34 \cdot 10^{-4}$ & $-8.55
\cdot 10^{-2}$\\ 
$C_\ell^I$&$-1.67 \cdot 10^{-2}$& $-6.98 \cdot 10^{-5}$ & $-7.54
\cdot 10^{-3}$\\
$D_\ell^I$&$-6.40 \cdot 10^{-4}$& $ 1.41 \cdot 10^{-6}$ &
$ 1.99 \cdot 10^{-4}$\\
$\rule{0.2em}{0em}s_\ell^I$&$ 36.7\rule{1.8em}{0em}$ & $ 
30.7\rule{1.8em}{0em}$ & 
$-11.9\rule{1.8em}{0em}$\\
\hline
\end{tabular}
\caption{\label{tab:standardsolution}Schenk parameters of the solution shown 
in fig.~\ref{fig:roy_a225}.}
\end{table}

\noindent
The plot in fig.~\ref{fig:roy_a225} shows that the numerical solution is
indeed very good: Below $s_0$, it is not possible to distinguish the two
curves representing the right and left hand sides of 
eq.~(\ref{eq:rieq1}). For this solution we found as a minimum $\sos = 2.1
\cdot 10^{-5}$, which corresponds to an average difference between the right 
and left hand sides of about $6 \cdot 10^{-4}$.

Having solved the Roy equations in the low--energy region, we now have a
representation for the imaginary parts of the three lowest partial waves
from threshold up to $s_2$. Since the driving terms account for all 
remaining contributions, we can then calculate the Roy representation for
the real parts from threshold up to 1.15 GeV (full lines in
fig.\ref{fig:roy_a225}). On the same plot, above $s_0$,
we also show the real part of the partial wave representation
that we used as an input for the imaginary parts (dashed lines).  
The comparison shows that the input we are
using is well compatible with the Roy equations (we should stress at this
point that in none of the phase--shift analyses which we are using as input
the Roy equations have been used).

\setcounter{equation}{0}
\section{Universal band}
\label{sec:UB}
As we have discussed in the preceding sections, for a given value of
$a_0^0$ and fixed input, the Roy equations admit a solution without cusp
only for a single value of $a_0^2$. By varying the input value of $a_0^0$,
the Roy equations define a function $a_0^2=F(a_0^0)$ that is known in the
literature as the ``universal curve'' \cite{Morgan Shaw}.  The experimental
uncertainties in the input above 0.8 GeV convert this curve into a band. The
universal band is the area in the $(a_0^0,a_0^2)$ plane that is allowed by
the constraints given by the $\pi \pi$--scattering data above 0.8 GeV and
the Roy equations. In this section we give a more precise definition of our
universal band, and calculate it accordingly.

\begin{figure} 
\leavevmode \centering
\includegraphics[angle=-90,width=13cm]{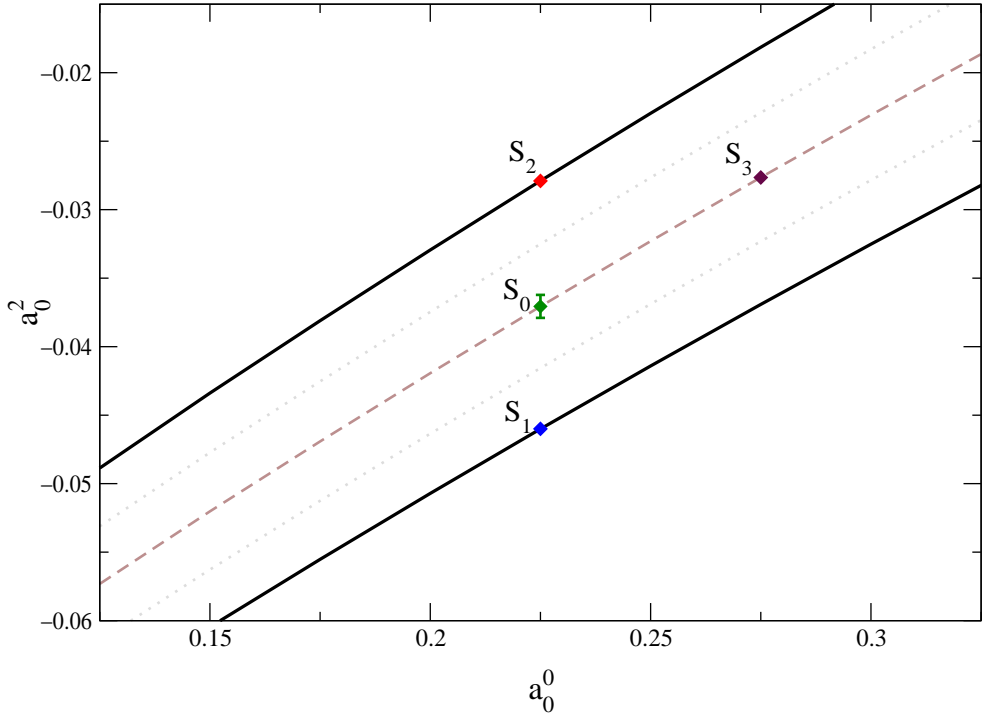}
\caption{\label{fig:UB} Universal band. The five lines
  correspond to the five different curves shown in
  fig.~\protect{\ref{fig:I2}} (the top line, for instance, 
results if  the input for  $\Im t_0^2$ in the region above 0.8 GeV
  is taken from the top curve in that figure). 
$S_0$ marks our reference point: $a_0^2=0.225$, $a_0^2=-0.0371$. The bar
attached to it indicates the uncertainty in $a_0^2$ due to the one
in the phase $\delta_0^0$ at the matching point -- 
the most important remaining source of error if the input for 
$\Im t_0^2$ is held fixed.}
\end{figure}

We first point out that the universal
curve $a_0^2=F(a_0^0)$ depends rather mildly on the input in the $I=0$ and
$I=1$ channel (a more quantitative statement concerning this
dependence is given below). For this reason, we only consider 
the uncertainties in the input for the $I=2$ channel. The available data in 
this channel are shown in fig.~\ref{fig:I2}, together with five different 
curves that we have used as input. For each one of these, 
we obtain a universal curve, which nearly represents a straight line
in the $(a_0^0,a_0^2)$ plane. The resulting five lines are shown in
fig.~\ref{fig:UB}.  The central one is well represented by the
following second degree polynomial: 
\be 
\label{eq:UB}
a_0^2 =-0.0849+0.232\, a_0^0-0.0865\, (a_0^0)^2  \fs 
\ee 
The analogous representations for the top and bottom lines read:
\bea a_0^2 \al =\al -0.0774+0.240\, a_0^0-0.0881\, (a_0^0)^2\co
\nonumber \\
 a_0^2 \al=\al-0.0922+0.225 \, a_0^0-0.0847\, (a_0^0)^2\fs
\eea
The region between these two solid lines is our universal band.
It is difficult to make a precise statement in probabilistic terms of how
unlikely it is that the physical values of the two scattering lengths are
outside this band. With our rather generous
choice of the two extreme curves, we consider it fair to say that
the experimental information above the matching point essentially excludes 
such values. In fact, we will argue below that the theoretical
constraints arising from the consistency of the Roy equations above the
matching point restrict the admissible region even further.

We now turn to the
dependence of the universal curve $a_0^2=F(a_0^0)$ on the input in the $I=0$
and $I=1$ channels, keeping the one for $I=2$ fixed. Changes in the input 
above $2 M_K$ are practically invisible at threshold: If we keep the phase 
shifts at the
matching point fixed, the three different available inputs for the $I=0$
and $I=1$ channels yield values of $a_0^2$ that differ by less than one
permille. The phase shifts at $s_0$ are the only relevant
factor here. Moreover, for the value of $a^2_0$,
$\delta_0^0(s_0)$ is much more important than 
$\delta^1_1(s_0)$: Shifts of $\delta_1^1(s_0)$ by
$\pm2^\circ$ change the value of $a_0^2$ roughly by a permille, but a change
by $\pm3.4^\circ$ in $\delta_0^0(s_0)$ induces a shift of $\Delta a_0^2
=\pm 8.4 \cdot 10^{-4}$, which amounts to two percent. Even so, this is much
smaller than the width of the band, as can be seen in fig.~\ref{fig:UB}.

We have also varied $\Ezero$ within the bounds 0.78 and 0.86 GeV and found that
the dependence of the relation $a_0^2=F(a_0^0)$ on $s_0$ is rather
weak. To exemplify, we mention that for the solution with $a_0^0=0.25$ at the
center of the universal band, a shift from $\Ezero=0.8$ GeV to 0.85 GeV 
changes $a_0^2$ by $10^{-3}$. 

\setcounter{equation}{0}
\section{Consistency}
\label{sec:math}
It takes a good balancing of the various terms occurring in the Roy 
equations for the partial waves not to violate the unitarity
limit. In the case of the $S$-wave with $I=0$, for instance, the contribution
to $\mbox{Re}\,t_0^0$ that arises from the subtraction term $k_0^0(s)$
is very large already at 1 GeV: 
The solution shown in fig.~\ref{fig:roy_a225} corresponds to
$a_0^0=0.225$ and $a_0^2=-0.0371$, so that $k_0^0(s)=2.7$ for
$s=1\,\mbox{GeV}^2$. 
As the energy grows, the term increases and reaches $k_0^0(s_1)=3.6$ 
at the upper end
of the region where our equations are valid, $s_1=68\,M_\pi^2$. 
Unless the contributions from the dispersion integrals nearly
compensate the subtraction term, the unitarity limit, 
$|\,\mbox{Re}\,t_0^0\,|\leq (2\sigma)^{-1}\simeq\frac{1}{2}$ is violated.
The example in fig.~\ref{fig:roy_a225} demonstrates that we do find
solutions for which such a cancellation takes place, with values of 
$a_0^0$, $a_0^2$ that are within the universal band.

It is striking that, above the matching point, this solution very closely
follows the real part of the input. In a restricted sense, this is necessary
for the solution to be acceptable physically: The solution is obtained by
identifying the imaginary part above the matching point with 
the one obtained from a particular representation of the partial waves. 
The Roy equations then determine the real part of the amplitude in the
region below $\Eone=1.15\,\mbox{GeV}$. If the result were very different from
the real part of the particular representation used, we would have to 
conclude that this representation cannot properly describe the physics. 
This amounts to a consistency condition: Above the matching point, the Roy
solution should not strongly deviate from the real part of the input.
The condition can be met only if the cancellation discussed above takes place,
but it is stronger. The example in fig.~\ref{fig:roy_a225} demonstrates that
there are solutions that obey the consistency condition remarkably well, 
indicating that our apparatus is indeed working properly.

We will discuss the consistency condition on a quantitative level below.
Before entering this discussion, we briefly comment on a different aspect of
our framework: the stability of the solutions. The behaviour 
below 0.8 GeV is not sensitive to 
the uncertainties in the input used for the imaginary parts above 1 GeV. 
We can modify that part of the input quite substantially, and without
changing anything else (not even below $s_0$) still get a decent solution
from threshold up to the limit of validity of our equations. 
Naturally, if we do not 
modify the Schenk parameters that define
the phase below $s_0$, the Roy equations are not strictly obeyed,
but the deviation from the true solution is quite small. The reason is that,
if $s$ is small, the kernels 
$K^{II'}_{\ell \ell'}(s,s')$ strongly suppress the contributions from the
region where $s'$ is large. The term $K^{00}_{00}(s,s')$, for instance, 
has the following expansion for $s'\gg s$: 
\bea 
K^{00}_{00}(s,s') = {1\over 9} \left\{ 11
s^2 - 10 s (4M_\pi^2) - (4M_\pi^2)^2 \right\} {1\over s^{\prime\,3}} + 
O\left(\frac{1}{s^{\prime\,4}}\right) \;
\; .  
\nonumber\eea
The interval above 1 GeV only 
generates very small contributions to the integrals on the r.h.s.~of the Roy 
equations, if these are evaluated in the region below the matching point.

We now take up the consistency condition and first observe that, once
a solution has a consistent behaviour above the matching point,
reasonable changes in the input above 1 GeV lead to solutions that
also obey the consistency condition: It looks as if the Roy equations were 
almost trivially satisfied, behaving like an identity for $E> 1$ GeV.
Is this consistent behaviour automatic, or does it depend crucially on part of
the input ?

\begin{figure}
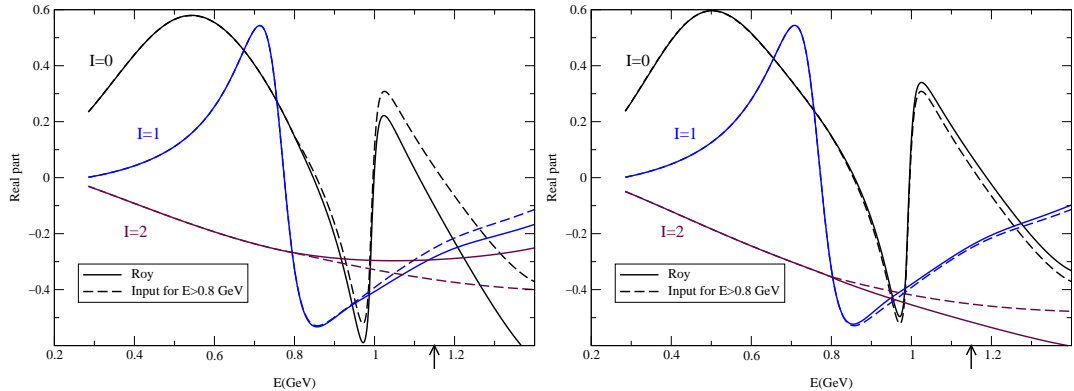
 
\leavevmode \centering
\includegraphics[angle=-90,width=7cm]{f8_a225_example_hi}
\includegraphics[angle=-90,width=7cm]{f8_a225_example_lo}
\caption{\label{fig:mathC} Solutions of the Roy equations for $a_0^0=0.225$
and two extreme values for $a_0^2$. The left figure corresponds to 
the point $S_2$ in fig.\ref{fig:UB},  while the one on the right
shows the solution for $S_1$. The arrows indicate the limit of validity of
the Roy equations.}
\end{figure}
The answer to this question can be found in fig.~\ref{fig:mathC}, where we
show two solutions obtained with the same value of $a_0^0$ as
in fig.~\ref{fig:roy_a225}, but different inputs for $\Im t_0^2$: The solution
on the left is obtained by using the top curve in fig.~\ref{fig:I2}
instead of the central one ($a_0^2=-0.0279$
instead of $a_0^2=-0.0371$). The solution on the right corresponds to the 
bottom curve
in fig.~\ref{fig:I2}, where $a_0^2=-0.0460$. 
The figure clearly shows that the consistent picture which
we have at the center of the universal band 
is almost completely lost if we go to the upper border of this band: 
It is by no means trivial that we at all find solutions for 
which the output is consistent with the input.

The fact that the peaks and valleys seen in the solutions 
mimic those in the input can be understood on the basis of analyticity alone:
The curvature above the matching point arises from the behaviour of the
imaginary parts there. The relevant term 
is the one from the principal value integral,
\bea \mbox{Re}\,t(s)=\frac{1}{\pi}\Pint_{4M_\pi^2}^{s_2}
ds'\,\frac{\mbox{Im}\,t(s')}{s'-s}+r(s)\fs\nonumber\eea 
The remainder, $r(s)$ contains the contributions associated with
the subtraction polynomial,
the left hand cut, the higher partial waves, as well as 
the asymptotic region. On the interval
$s_0<s<s_1$,  it varies only slowly and is well approximated
by a first order polynomial in $s$. 

The representations of the partial 
wave amplitudes that we are using as an input are specified in
terms of simple functions. In the vicinity of the region where we are 
comparing their real parts with the Roy solutions, these
are analytic in $s$, except for the cut along the positive real axis.
Hence they also admit an approximate
representation of the above form -- the contributions from distant
singularities are well approximated by
a first order polynomial. Disregarding the interpolation needed
to match the
representation with the prescribed value of the phase at $s_0$,
their imaginary parts coincide with the one of the corresponding Roy solution
above the matching point. 
The small differences occurring in the interpolation region
and below the matching point do not generate an important difference
in the curvature. We conclude that
the difference between the Roy solution and the real part of the
input must be linear in $s$, to a good approximation. Moreover, within the 
accuracy to which our solutions obey the Roy equations, the
two expressions agree at the matching  
point, by construction. Accordingly, the relation can be written in the form
\bea\label{beta} \mbox{Re}\,t(s)\rule[-0.3em]{0.1em}{0em}_{\mbox{\tiny
    Roy}}=\mbox{Re}\,t(s)\rule[-0.3em]{0.1em}{0em}_{\mbox{\tiny input}}
+(s-s_0)\,\beta\fs\eea
We have checked that this relation indeed holds to sufficient accuracy, 
for all three partial waves.
This does not yet explain why the solution follows the real part 
of the input, but shows that it must do so up to a term linear in $s$ that
vanishes at the matching point. In particular, if the difference between
input and output is small at the upper end of validity of our equations, 
then analyticity ensures that the same is true
in the entire region between the matching point and that energy (in this
interval, $s$ varies by about a factor of two). 

\begin{figure}[t] 
\leavevmode \centering

\includegraphics[angle=-90,width=14cm]{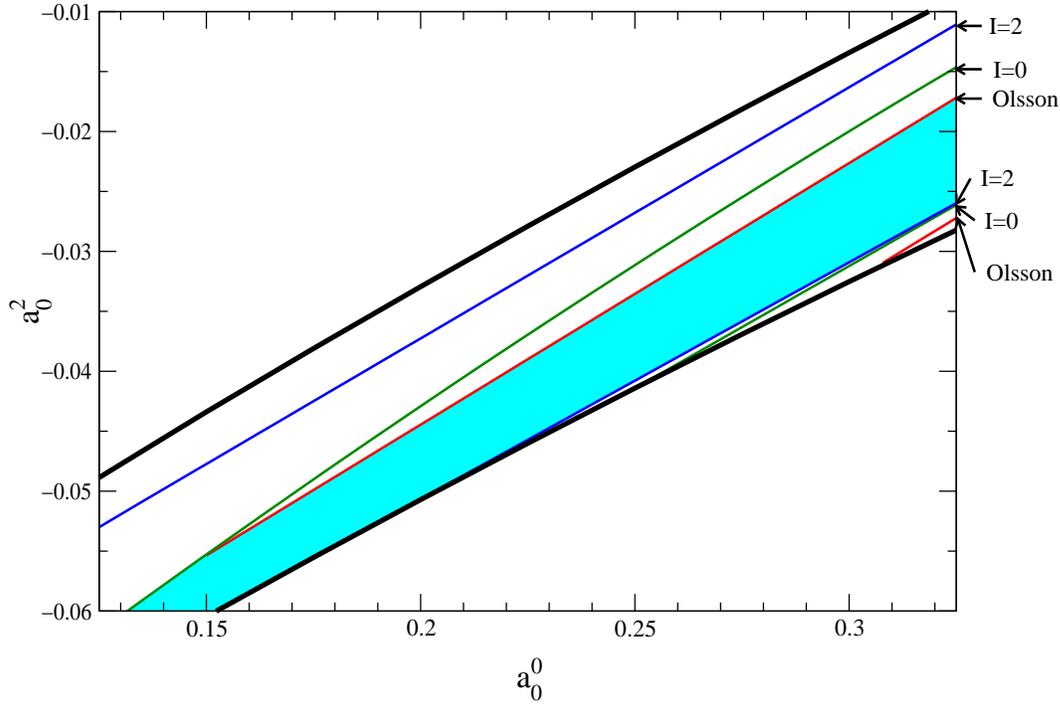}
\caption{\label{fig:Olsson}  Regions inside which the consistency
  condition is met. The band between the two blue lines is for the
  condition in the $I=2$ channel, whereas the one between the two green
  lines is for the $I=0$ channel. The two red lines delimit the band
  inside which the Olsson sum rule is satisfied. The shaded area gives the
  intersection of the three bands.}
\end{figure}

In view of the uncertainties attached to our input, we cannot require
the Roy equations to be strictly satisfied also above the matching point.
The band spanned by the two green lines  
in fig.~\ref{fig:Olsson} shows the
region in the $(a_0^0,a_0^2$) plane, where the solution for
$\mbox{Re}\,t_0^0(s)$ differs from the real part of the input by less than 
0.05 (expressed in terms of the parameter $\beta$ in eq.~(\ref{beta}), 
this amounts to 
$|\beta^0_0|<0.07\,\mbox{GeV}^{-2}$).
Likewise, the band spanned by the two blue lines represents the
region where $|\mbox{Re}\,t_0^2(s)_{\mbox{\tiny Roy}}-
\mbox{Re}\,t_0^2(s)_{\mbox{\tiny input}}|<0.05$, so that 
$|\beta^2_0|<0.07\,\mbox{GeV}^{-2}$. The corresponding band
for the $P$-wave is much broader -- in this channel, the consistency 
condition is rather weak and is met everywhere inside the universal band. 
We conclude that, in the lower half of the universal band, 
all three waves show a consistent behaviour, while for the 
upper quarter of the band, this is not the case 
(the situation at the upper border 
is shown on the left in fig.~\ref{fig:mathC}).

It is not difficult to understand why the consistency condition is strongest
for the $I=0$ $S$-wave. In this connection, the most important term in the 
Roy equations is the
one from the subtraction polynomial -- 
the solution can satisfy the
consistency condition only if the term proportional to $s$ 
is nearly cancelled by a linear
growth of the remaining contributions. The term generates the contribution 
$(\beta_0^0,\beta_1^1,\beta^2_0)=
(6,1,-3)\times (2a_0^2-5a_0^2)/(72\,M_\pi^2)$ 
to the coefficients that describe the difference between output and
input for the three lowest partial waves. 
The subtraction
polynomial thus contributes twice as much to $\beta_0^0$ as to 
$\beta^2_0$, so that the consistency band for the $I=2$ wave
must be about twice as broad as the one for the $I=0$ wave, while the one 
for the $P$-wave must roughly be six times broader. At the qualitative level, 
these features are indeed born out in the figure, but  
we stress that the term from the
subtraction polynomial is not the only one that matters -- those arising
from the integrals also depend on the values of $a_0^0$ and $a^2_0$. 
The two green lines correspond to a variation in $a^2_0$ by about $\pm 0.004$.
Increasing $a^2_0$ by 0.004, the value of the subtraction term
$k_0^2(s_1)$ decreases by 0.10. The fact that the
lines correspond to a change in $\mbox{Re}\,t^0_0(s_1)$ of 
only $\pm 0.05$ implies that the contributions from the integrals reduce the
shift by a factor of 2. Also, if only the subtraction
term were relevant, the consistency bands would be determined by 
the combination $2a_0^2-5a_0^2$ and thus have a slope of $\frac{2}{5}$.
Actually, these bands are roughly parallel to the
universal band, whose slope is positive, but smaller by about a factor of 2.  

\setcounter{section}{10}
\setcounter{equation}{0}
\section{Olsson sum rule}\label{Olsson sum rule}
In the Roy equations, the imaginary parts above the matching point and
the two subtraction constants $a_0^0$, $a_0^2$ appear as independent 
quantities. The consistency condition interrelates the two in such a manner
that the 
contributions from the integrals over the imaginary parts nearly cancel
the one from the subtraction term. In fact, a relation of this type can be
derived on general grounds.

The fixed-$t$ dispersion relation (\ref{fixedt}) contains two subtractions.
In principle, one subtraction suffices,
for the following reason.
The $t$-channel $I=1$ amplitude
\bea T^{(1)}(s,t)\equiv \frac{1}{6}\,\{2\, T^0(s,t)+3\,T^1(s,t)-5\,T^2(s,t)\}
\nonumber\eea
does not receive a Pomeron 
contribution and thus only grows in proportion to $s^{\alpha_\rho(t)}$ for
$s\rightarrow
\infty$. The dispersion relation
(2.4), however, does contain terms that grow linearly with $s$.
For the relation to be consistent with Regge asymptotics, 
the contribution from the subtraction term must
cancel the one from the dispersion integral\footnote{In the case of the
  $t$-channel amplitudes with $I=0$ and $I=2$,  
the fixed-$t$ dispersion relation (2.4) does ensure the proper asymptotic
behaviour.}. 
At $t=0$, this condition reduces to the Olsson sum
rule, which relates the subtraction constants to an integral over the imaginary
parts \cite{olsson sum rule}:  
\bea \hspace*{-2.5em}2\, a_0^0-5\,a_0^2\al=\al
\frac{M_\pi^2}{8\pi^2}\int_{4M_\pi^2}^\infty\!ds\;
\frac{2\,\mbox{Im}\,T^0(s,0)+3\,\mbox{Im}\,T^1(s,0)-5\,
\mbox{Im}\,T^2(s,0)}{s\,(s-4M_\pi^2)}\,.\eea
It is well known that this sum rule converges only slowly -- 
the contributions from the asymptotic region cannot be neglected. 
We split the integral into four pieces,
\bea 2\,a_0^0-5\,a_0^2=O_{\ind SP}  +O_{\ind D}+O_{\ind F}+
O_{as}\fs\nonumber\eea
The first term represents the contributions from the
imaginary parts of the $S$- and $P$-waves in the region below 2 GeV,
which are readily worked out, using our Roy solutions on the interval from
threshold to 0.8 GeV and the input phase shifts on the remainder. The result is
not very sensitive to the input used and is
well approximated by a linear dependence on the scattering
lengths,
\bea O_{\ind SP}= 0.483 \pm 0.011\,  +1.13\,
(a_0^0-0.225)-1.01\,(a^2_0+0.0371)\fs\nonumber\eea 
The remainder is closely related to the moments $I^I_n$ introduced in appendix
B.1: here, we are concerned with the case $n=-1$.
The term $O_{\ind D}$ describes the contribution from the 
imaginary part of the $D$-waves, in the interval from threshold to
2 GeV. The relevant experimental information is discussed in
appendix B.3, where we also explain how we estimate the uncertainties. 
The numerical result reads $ O_{\ind D} =0.061 \pm 0.004$, including
the small, negative contribution from the $I=2$ $D$-wave. The bulk stems from
the tensor meson $f_2(1275)$: In the narrow width approximation, this
contribution amounts to 0.063.  For the analogous contribution due to the
$F$-wave, we obtain  $ O_{\ind F} = 0.017  \pm 0.002$ (in narrow width
approximation, the term generated by the $\rho_3(1690)$ yields 0.013).  
Those from the asymptotic region are dominated by the
leading Regge trajectory -- as noted above, the Pomeron does not
contribute. Evaluating the asymptotic contributions 
with the formulae given in appendix B.4, 
we obtain $O_{as}=0.102\pm 0.017$. Collecting terms, this
yields
\bea 2\,a_0^0-5\,a_0^2=0.663 \pm 0.021 + 1.13\,(a_0^0-0.225)-1.01\,
(a^2_0+0.0371)\fs\eea
The result corresponds to a band in the $(a_0^0,a_0^2)$ plane:
\bea  a^2_0=-0.044 \pm 0.005 +0.218\,(a_0^0-0.225)\fs\eea 
The band is spanned by the two red lines shown in fig.~\ref{fig:Olsson}.
One of these nearly coincides with the lower border of the universal 
band, while the other runs near the center. The Olsson sum rule thus
imposes roughly
the same relation between $a^0_0$ and $a^2_0$ as the consistency condition.
Note that the asymptotic contributions
are numerically quite important here: The term $O_{as}$ amounts to
a shift in $a^2_0$ of $-0.026\pm 0.004$. The fact that -- in the region where 
our solutions are internally consistent --  the sum rule
is indeed obeyed, represents a good check on our asymptotics. 

The Olsson sum rule ensures the proper asymptotic behaviour of the
amplitude only for $t=0$. In order for the terms that grow linearly with $s$
to cancel also for $t\neq 0$, the imaginary part of the $P$-wave must 
obey an entire family of sum rules. The matter is discussed in detail in
appendix C.1, where we demonstrate that one of these offers a further, 
rather sensitive test of our framework. The relationship between the 
Roy equations and those proposed by Chew and Mandelstam 
\cite{Chew Mandelstam} is described in appendix C.2, where we also comment 
on the asymptotic behaviour of the dispersion integrals that occur on the 
r.h.s.~of the Roy equations for the $S$- and $P$-waves.  

\section{Comparison with experimental data}

In our framework, the only free parameter is $a_0^0$. Comparing our
Roy equation solutions to data, we can determine the range of $a_0^0$
consistent with these, as well as a corresponding
range for $a_0^2$. This experimental determination of the two $S$-wave
scattering lengths is the final scope of the present analysis and the main
subject of the present section. Data on the $\pi \pi$ amplitude are
available in a rather wide range of energies (we do not indicate the upper
limit in energy when this exceeds 1.15 GeV, the limit of validity of our
equations):
\begin{itemize}
\item
$K_{e4}$ data for the combination $\delta_0^0-\delta_1^1$ ($2 M_\pi \leq E
\leq 0.37$ GeV);
\item
ACM and Losty et al.~data for $\delta_0^2$ (0.35 GeV $\leq E$); 
\item
Data on the vector form factor -- according to the discussion in
section \ref{sec:Pwave}, these can safely be converted into values for
$\delta_1^1$ in the region of the $\rho$ ($0.5 \leq E \leq 0.9$ GeV);
\item
CERN--Munich, and Berkeley data in the channels with $I=0$ and $I=1$ (0.5 GeV 
$\leq E$);
\end{itemize}
In the Roy equations, $a_0^0$ and $a^2_0$ exclusively enter through the 
subtraction polynomials, specified in eq.~(\ref{eq:subconst1}). 
Those relevant for the $S$-waves contain a 
constant contribution given by the scattering length and a term proportional
to $(s-4M_\pi^2)\times (2a^0_0-5a_0^2)$. In the $I=0$ wave, that term
is larger
than $a_0^0$ from $E \sim 0.5$ GeV on. For the $I=2$ wave, the linear
term starts dominating over $a_0^2$ even earlier. Since $t^1_1(s)$ vanishes 
at threshold, the corresponding subtraction polynomial exclusively
involves the linear term. This implies that, except in the vicinity of
threshold, the behaviour of the solutions is sensitive only
to the combination $2 a_0^0-5 a_0^2$ of scattering lengths -- roughly the 
combination that characterizes the universal band. Accordingly,
only data that reach down close to threshold give a direct handle to
separately determine $a_0^0$ and $a_0^2$. In fact, only those  
coming from $K_{e4}$ decays meet this condition. 
 
There is another threshold in energy that is obviously relevant for our
approach: the matching point $s_0$. We will make a clear distinction
between data points below $s_0$ and those at higher energies. 
The comparison to data
above $s_0$ can hardly yield any information on the scattering lengths, because
the behaviour of our solutions at those energies very strongly depends on the
input used for the imaginary parts: The uncertainties in 
the experimental input completely cover the dependence of the solutions on 
the scattering lengths -- we will discuss this in detail below.
Instead, we analyze the requirement that the solution is consistent 
with the input for $s>s_0$, in the sense discussed in section \ref{sec:math}.
This condition turns out to be practically independent of the input used for
the imaginary parts above $s_0$ and does therefore yield a meaningful
constraint on $2 a_0^0-5 a_0^2$.
\begin{figure}[t] 
\leavevmode \centering
\includegraphics[angle=-90,width=13cm]{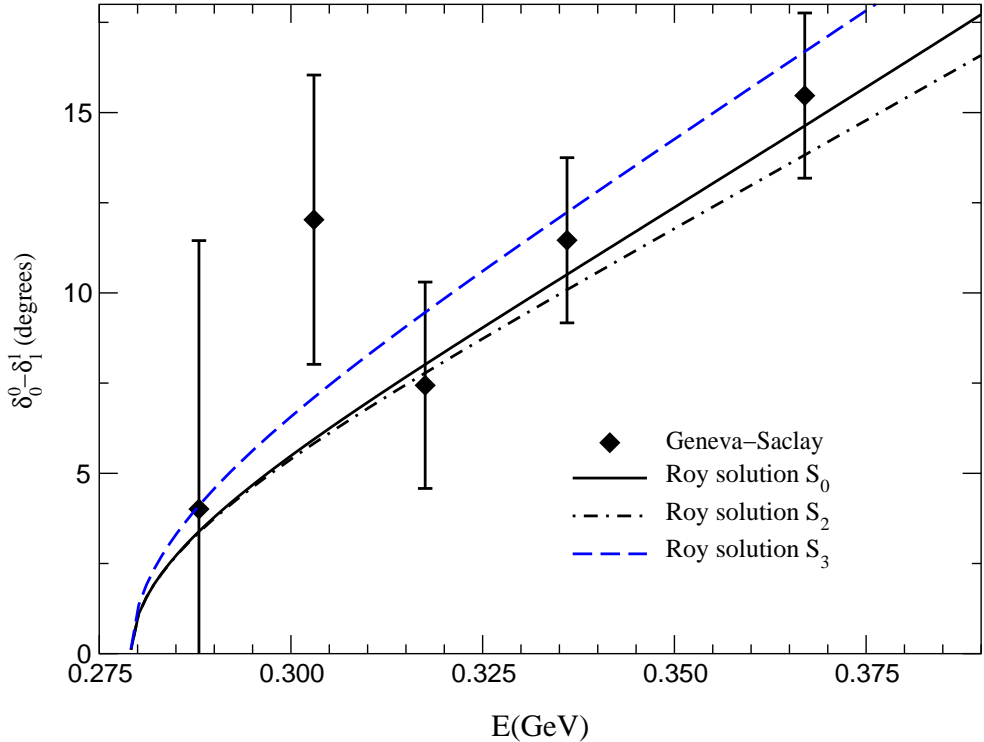}
\caption{\label{fig:kl4} Comparison of our Roy solutions 
  for different values of the scattering lengths with the data of the
  Geneva--Saclay collaboration, Rosselet et
  al.~\protect{\cite{Rosselet}}. The full, dash-dotted and dashed lines
  correspond to the points $S_0$, $S_2$ and $S_3$ in
  fig.~\protect{\ref{fig:UB}}.}
\end{figure}

\begin{figure}[thb]
\includegraphics[angle=-90,width=13cm]{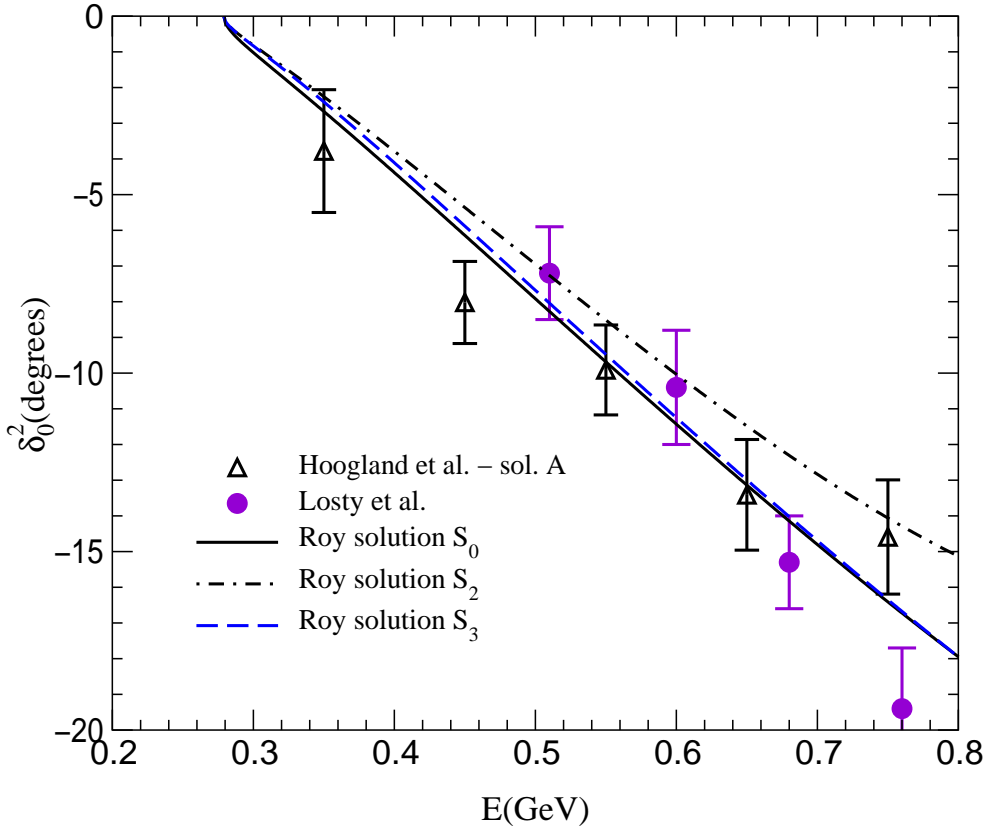}
\caption{\label{fig:acmlo} Comparison of our Roy solutions with the
  data on $\delta^2_0$ obtained by the ACM collaboration
  \protect{\cite{hoogland}} and by Losty et 
  al.~\protect{\cite{losty}}. The full, dash-dotted and dashed lines 
correspond to the points $S_0$, $S_2$ and $S_3$ in
  fig.~\protect{\ref{fig:UB}}.}
\end{figure}

\subsection{Data on $\delta_0^0-\delta_1^1$ from $K_{e4}$, and on
  $\delta_0^2$ below 0.8 GeV}

Let us first consider the $K_{e4}$ data. The comparison between our
solutions and the high-statistic data of the Geneva--Saclay collaboration
\cite{Rosselet} is shown in fig.~\ref{fig:kl4}, for various values of the
scattering lengths. The figure confirms the simple intuition that these
data are mainly sensitive to $a_0^0$. In accordance with previous analyses
\cite{Nagels}, we find that they roughly constrain $a_0^0$ to the
range between $0.18$ and $0.3$.

\begin{figure} 
\leavevmode \centering
\includegraphics[angle=-90,width=12cm]{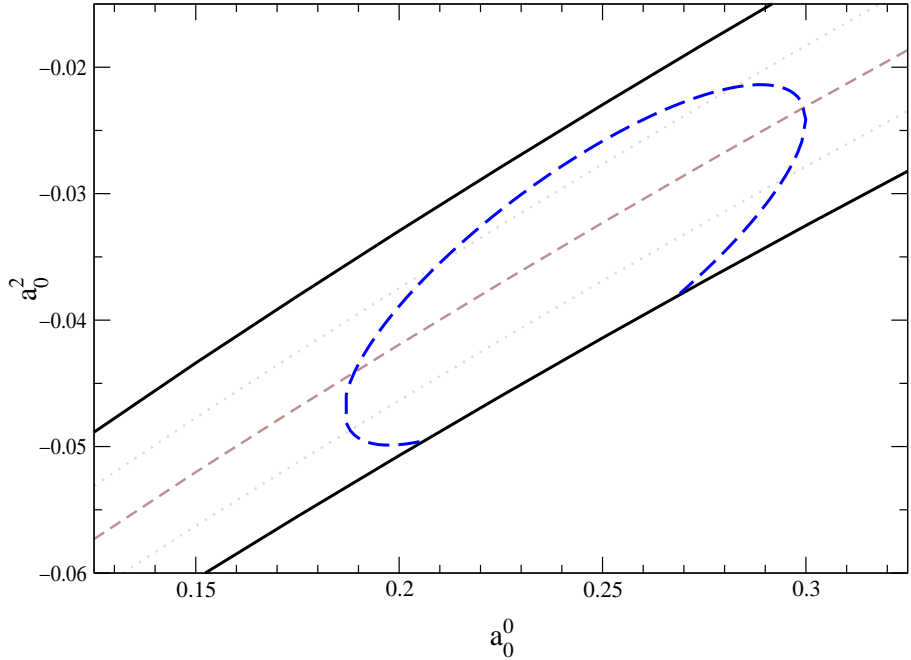}
\caption{\label{fig:kl4_acm_lo} Range selected by the data below 0.8 GeV.
The dashed line represents the 68\% C.L. contour obtained by combining the
Geneva--Saclay data on $K_{e4}$ decay with those from ACM(A) on
$\delta_0^2$.} 
\end{figure}

As for the low--energy data in the $I=2$ channel, we should stress that
this wave is quite strongly constrained once $\delta_0^2(s_0)$ is
fixed. Because of the absence of any structures between threshold and 0.8
GeV, once we fix $\delta_0^2(s_0)$, the only freedom is in the way the phase
approaches zero at threshold, i.e. in the value of $a_0^2$ -- which depends
on $a_0^0$. Fig.~\ref{fig:acmlo} shows that, at fixed $\delta_0^2(s_0)$,
even a sizeable change in $a_0^0$ is barely visible in the $I=2$ phase. The
only important factor here is the value of the phase at the matching point:
The comparison with the experimental data basically tells us which value of
$\delta_0^2(s_0)$ is preferred.

A quantitative statement can be made in terms of $\chi^2$, and in principle
we could calculate three different $\chi^2$-values, based on the three sets
of data shown in fig.~\ref{fig:I2}. Two of these, however, represent two
different analyses of the same set of $\pi N \to \pi \pi N$ data. Their
difference is a clear sign of the presence of sizeable systematic
errors. We have estimated the latter using the difference, point by point,
between the two analyses A and B of ref.~\cite{hoogland}, and added this in
quadrature to the statistical errors. As reference we have used the ACM(A)
set of data, but have checked that interchanging it with the one of Losty
et al.~does not give significantly different results.  The corresponding
$\chi^2$, combined with the one obtained from the $K_{e4}$ data, has a
minimum $\chi^2_{min}=5.1$ (with 8 d.o.f.) at $a_0^0=0.242$,
$a_0^2=-0.0357$. The contour corresponding to 68\% confidence level
($\chi^2=\chi^2_{min}+2.3$) is shown in fig.~\ref{fig:kl4_acm_lo}: The
range $0.18<a_0^0<0.3$ is dictated by the $K_{e4}$ data, whereas the $I=2$
data exclude the upper border of the band.
 
\subsection{The $\rho$ resonance.}
\label{sec:rho}
\begin{figure}[thb] 
\leavevmode 
\centering
\includegraphics[angle=-90,width=14.2cm]{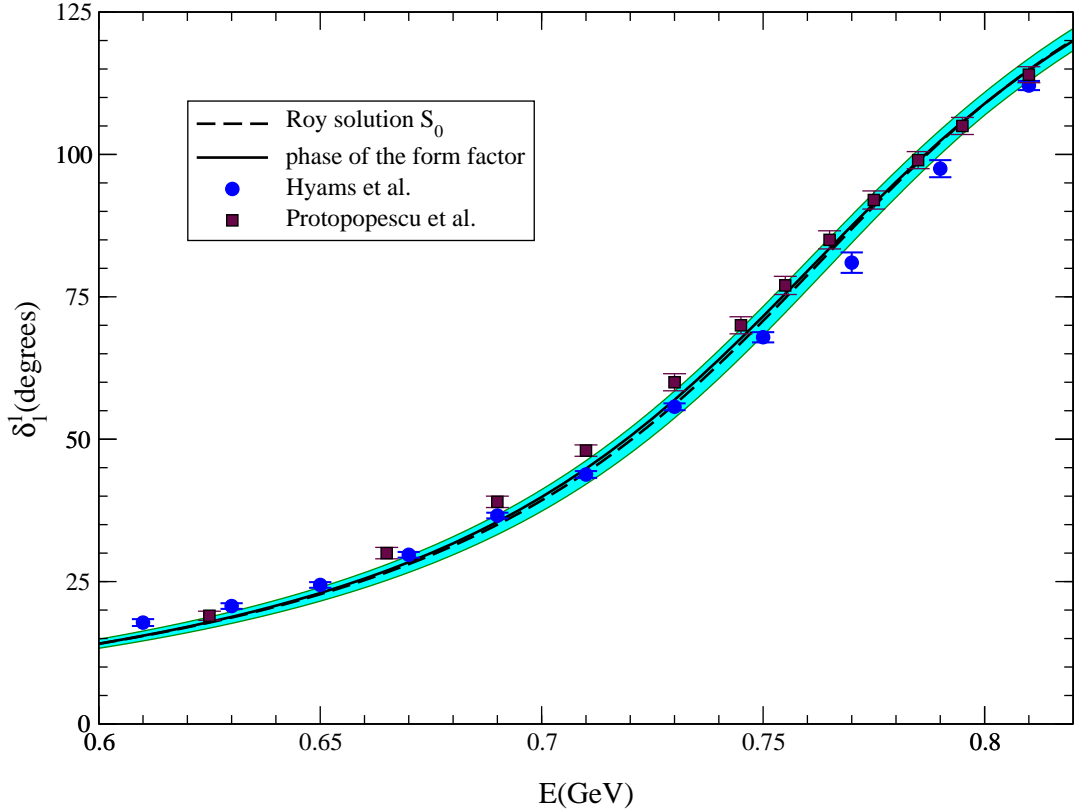}
\caption{\label{fig:P-wave phase shift} P-wave phase shift. The band 
shows the result of our analysis, obtained by varying the input 
within its uncertainties, while the data points indicate the phase
shift measured in the process $\pi N\rightarrow\pi\pi N$ by the CERN-Munich
collaboration. The full line represents the phase of the vector form factor
(Gounaris-Sakurai fit of ref.~\cite{cleo}).}
\end{figure}

The input used at the matching point implies that the $P$-wave phase shift
must pass through $90^\circ$ somewhere between threshold and 0.8 GeV -- the
Roy equations determine the place where this happens and how rapidly the
phase must grow with the energy there.  The solutions turn out to be very
stiff: Varying the values of $a_0^0$ and $a_0^2$ within the universal band,
and also varying the input for the imaginary parts above 0.8 GeV within the
experimental uncertainties, we obtain the narrow band of solutions shown in
fig.~\ref{fig:P-wave phase shift}.

In this figure, the energy range
only extends to $0.82\,\mbox{GeV}$, for the following reason: Our solutions 
move along the Argand circle only below the matching point.
At higher energies, the real part of the partial wave calculated from the 
Roy equations does not exactly match the imaginary part used as an input:
  unless we correct the latter,
the elasticity $\eta_1^1$ differs from unity, already before
the inelastic channels start making a significant contribution. 
If the consistency condition is met well, the departure from unity
is small, but it can become as large as 5\% if we go to the extreme of the 
consistency region shown in fig.\ref{fig:Olsson}. 
This means that it does not make much
sense to extract the value of the phase without adjusting the imaginary
part. The proper way to do this is to extend the interval on which 
the Roy equations are solved, but we did not carry this out.

In the region $0.7\,\mbox{GeV}<0.82\,\mbox{GeV}$, the result
closely follows the data of the CERN-Munich collaboration. 
Below 0.7 GeV, however, the data are in conflict with the
outcome of our analysis: The five lowest data points 
are outside the range allowed by the Roy equations, a problem 
noted already in ref.~\cite{BFP2}. In our opinion, we are
using a generous estimate of the uncertainties to be attached to our
input. Note, in particular, that at those energies, the driving terms
barely contribute. We conclude that the discrepancy between our result and
the CERN-Munich phase shift analysis occurring on the left wing of the
$\rho$ is likely to be attributed to an underestimate of the experimental
errors. As discussed below, the comparison with the $e^+e^-$ and $\tau$
decay data corroborates this conclusion.

Concerning the resonance parameters, we first give the ranges of mass and
width that follow if, in the vicinity of the resonance, the phase shift is
approximated by a Breit-Wigner formula\footnote{The difference between
$M_\rho^2\pm i\,M_\rho\Gamma_\rho$ and $(M_\rho\pm \frac{i}{2}\,
\Gamma_\rho)^2$ is beyond the accuracy of that approximation. The second is
obtained from the first with the substitution $M_\rho^2\rightarrow
M_\rho^2-\frac{1}{4}\,\Gamma_\rho^2$, $M_\rho\Gamma_\rho\rightarrow
M_\rho\Gamma_\rho$, which increases the value of $M_\rho$ by about 4 MeV. } 
\bea 
e^{2\,i\,\delta_1^1(s)}=\frac{M^2_\rho+i\,\Gamma_\rho M_\rho-s}
{M_\rho^2-i\,\Gamma_\rho M_\rho-s}\co\hspace{2em} \mbox{tg}\,
\delta_1^1(s)=\frac{\Gamma_\rho M_\rho}{M_\rho^2-s}\fs\nonumber
\eea 
In this approximation,
the mass of the resonance is the real value of the
energy where the phase passes through $90^\circ$ and the width may be
determined from the value of the slope $d \delta^1_1/ds$ at resonance. The
solutions contained in the band shown in the figure
correspond to the range  
$M_{\rho}=774\pm 3\,\mbox{MeV}$ and $ 
\Gamma_{\rho}=145\pm 7\,\mbox{MeV}$,  
to be compared with the average values obtained by the Particle
Data Group, 
$M_\rho=770.0\pm 0.8 \,\mbox{MeV}$, $\Gamma_\rho= 150.7\pm 1.1\,\mbox{MeV}$
\cite{PDG}.
  
The only process independent property of the resonance is the position of 
the corresponding pole -- the above numbers specify this position only
approximately. To determine it more accurately, we first observe 
that the Roy equations yield a representation of 
the partial wave $t_1^1(s)$ on the first sheet, in terms of the imaginary
parts along the real axis. The first sheet
contains both a right and a left hand cut. 
We need to analytically continue the function from the upper
rim of the right hand cut into the lower half plane (second sheet). 
The difference between the values obtained in this manner and those
found by evaluating the Roy representation 
in the lower half plane is given by the analytic continuation
of the imaginary part,
\bea \mbox{Im}\,t_1^1(s)=\frac{1}{\sigma(s)}\,\sin^2\!\delta_1^1(s)
\fs\nonumber \eea 
On the first sheet, $t_1^1(s)$ does not have singularities. 
Hence a pole can only arise from the continuation of the imaginary part.
Indeed, the function $\sin^2\!\delta_1^1(s)$ contains the term
$\exp 2\,i\,\delta_1^1(s)$, which has a pole below 
the real axis. The position is readily worked out
with the explicit, algebraic parametrization of the phase that we are
using. The result illustrates an observation made long ago \cite{Pisut Roos,
Lang, Bohacik}:
The pole mass is lower than the energy at which the phase goes
through $90^\circ$, by about 10 MeV: 
For the band shown in the figure, the pole position varies in the 
range
\bea M_\rho = 762.5 \pm 2\,\mbox{MeV}\co\hspace{2em}\Gamma_\rho =142 \pm
7\,\mbox{MeV} \fs 
\nonumber\eea

The $e^+e^-$ and $\tau$ data neatly confirm the conclusion reached above: 
The phase of the form factor is in perfect agreement with the 
behaviour of the $P$-wave that follows from the Roy equations, 
but differs from the data of the CERN-Munich phase shift analysis,
particularly below 0.7 GeV. In our opinion,
the information obtained about the behaviour on the left wing of the resonance
on the basis of the reactions $e^+e^-\rightarrow\pi^+\pi^-$ and
$\tau\rightarrow \pi^-\pi^0\nu$ is more reliable than the one obtained from
$\pi N\rightarrow \pi\pi N$. The fact that the Roy equations
are in good agreement with the $e^+e^-$ and $\tau$ data is very encouraging. 
\begin{figure} 
\leavevmode \centering
\includegraphics[angle=-90,width=12cm]{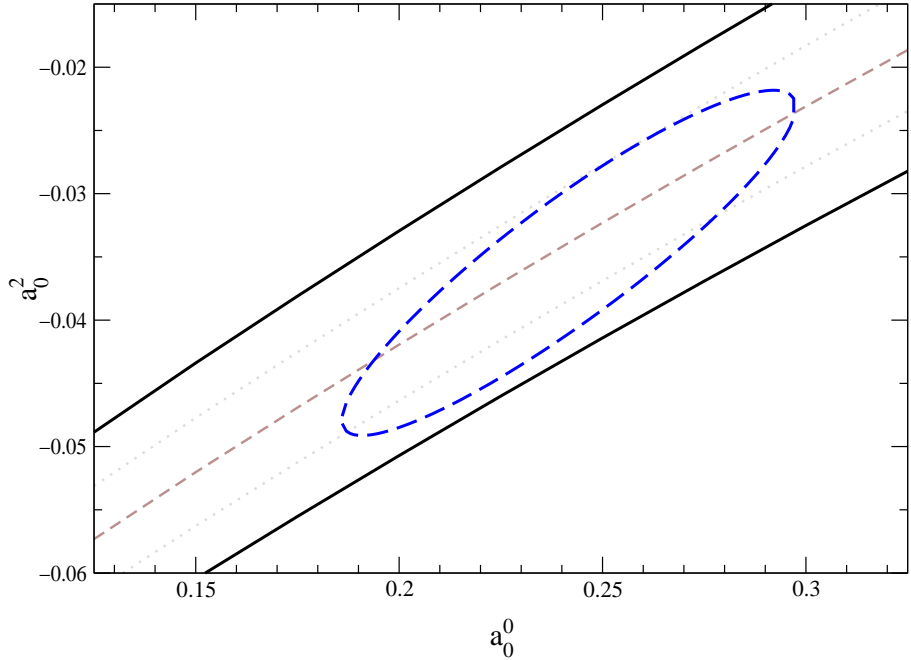}
\caption{\label{fig:lo_and_CLEO} 68\% C.L. contour obtained by combining 
all relevant low energy data: $K_{e4}$ decay,  ACM(A) data on
$\delta_0^2$ below 0.8 GeV and results for $\delta_1^1$
 extracted from the $e^+e^-$ and $\tau$ data on the pion form factor.}
\end{figure}

In view of the clean determination of the $P$-wave phase
shift through $e^+e^-$ and $\tau$ experiments, we find it instructive to 
draw fixed
$\chi^2$-contours in the $(a_0^0, a_0^2)$ plane. To do so, we first need to
attach an error bar to the curve representing the phase shift. In section
\ref{phases at the matching point}, we estimated the
uncertainty in $\delta_1^1(s_0)$ at $\pm 2^\circ$ or $\pm 2\%$. As we go 
down in energy, the relative precision of the
determination of the phase decreases: A generous estimate of the
uncertainty at $\sqrt{s}=0.5$ GeV is 10\% or $\pm 0.6^\circ$. A smooth
interpolation between these two values is our estimate of the experimental
error bar (below that energy, the $e^+ e^-$ and $\tau$ data become scarce 
and have sizeable uncertainties). To construct the $\chi^2$ we 
have compared our solutions to the
experimentally determined phase shift at five points between 0.5 and 0.75
GeV. Combining this $\chi^2$ with those from the data on $K_{e4}$ decays and
on $\delta_0^2$ below 0.8 GeV, we obtain the 68\% C.L. area drawn in
fig.~\ref{fig:lo_and_CLEO}. The minimum of the $\chi^2$ is now 5.4 (with 13
d.o.f.). The position of the minimum is barely shifted: It now occurs at 
$a_0^0=0.240$, $a_0^2=-0.0356$. In other words, at the place where the 
$\chi^2$ of the $K_{e4}$ data on
$\delta_0^0-\delta_1^1$ and those on $\delta_0^2$ had a minimum, the
$\chi^2$ relative to the data on the form factor
is practically zero and also has a minimum. In view of the fact that the
uncertainties in $\delta_1^1$ are very small, this is quite remarkable. The
data on the $P$-wave do not change the position of the minimum, but shrink
the ellipse along the width of the universal band. As expected, they do not 
reduce the range of allowed values of $a_0^0$.

\subsection{Data on the $I=0$ $S$-wave below 0.8 GeV}
\label{I0low}

\begin{figure} 
\leavevmode \centering
\includegraphics[angle=-90,width=12cm]{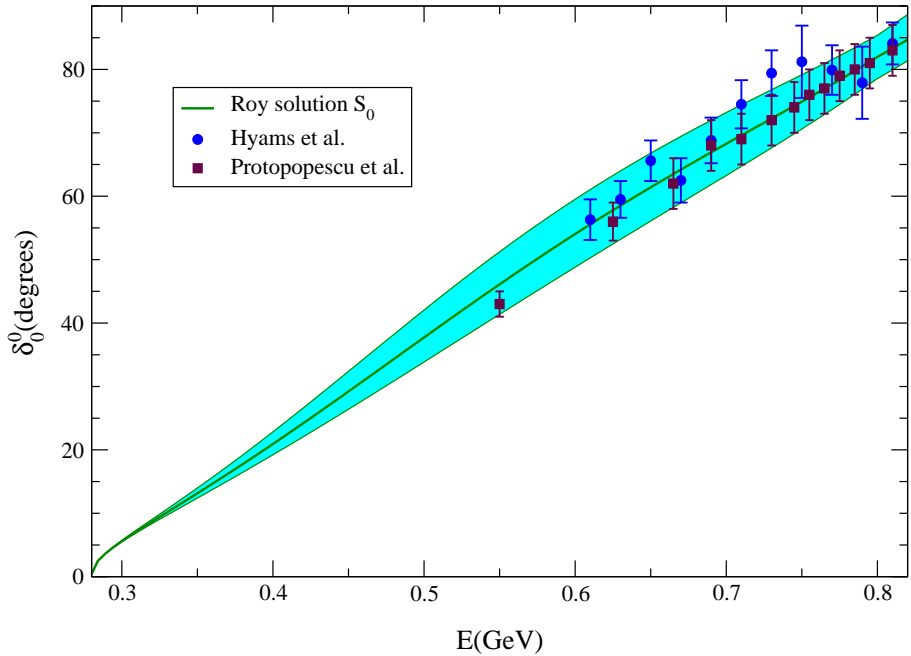}
\caption{\label{fig:Swave} Comparison between the Roy solution for the
  $S$-wave and the phase shift analyses of the CERN-Munich~(circles) and 
  Berkeley~(squares) collaborations. The band shows the uncertainties 
  in the Roy solution, which are dominated by those in $a_0^0$ and $a_0^2$. } 
\end{figure}

In fig.~\ref{fig:Swave} we compare the $S$-wave obtained from our
Roy equation solutions with the available data: CERN-Munich \cite{hyams}
and Berkeley \cite{Protopopescu}. The band shown is a representation
of the uncertainties in the solution, which have two main
sources: the uncertainty in 
$\delta_0^0(s_0)$ and the one in $\delta^2_0(s_0)$ 
(width of the universal band). The central curve shows our reference 
solution $a_0^0=0.225$, $a_0^2=-0.0371$. The uncertainties
indicated do not account for the changes occurring if the value 
$a_0^0=0.225$ is modified. Changing this value within reasonable bounds, 
however, brings the solution out of the band only below 0.4 GeV, already far
below the first data point.
The figure shows good agreement with the data, especially so for the
Berkeley data set. The CERN-Munich data set shows a certain structure, 
which does not occur in our solutions -- in view of the uncertainties in the
data, this difference does not represent a problem.

Despite the positive picture which emerges from the comparison, we refrain
from using these data to draw confidence--level contours in the $(a_0^0,
a_0^2)$ plane. The $S$-wave phase shifts have been extracted simultaneously 
with the $P$-wave. As discussed in the preceding section, these
are affected by systematic errors which are at least as large as the
statistical ones. The same must be true for the data in the $I=0$ channel,
so that a quantitative comparison with the Roy solutions is barely
significant.

\subsection{Data above 0.8 GeV}
\label{I0high}
\begin{figure}[t]
\leavevmode \centering
\includegraphics[angle=-90,width=12cm]{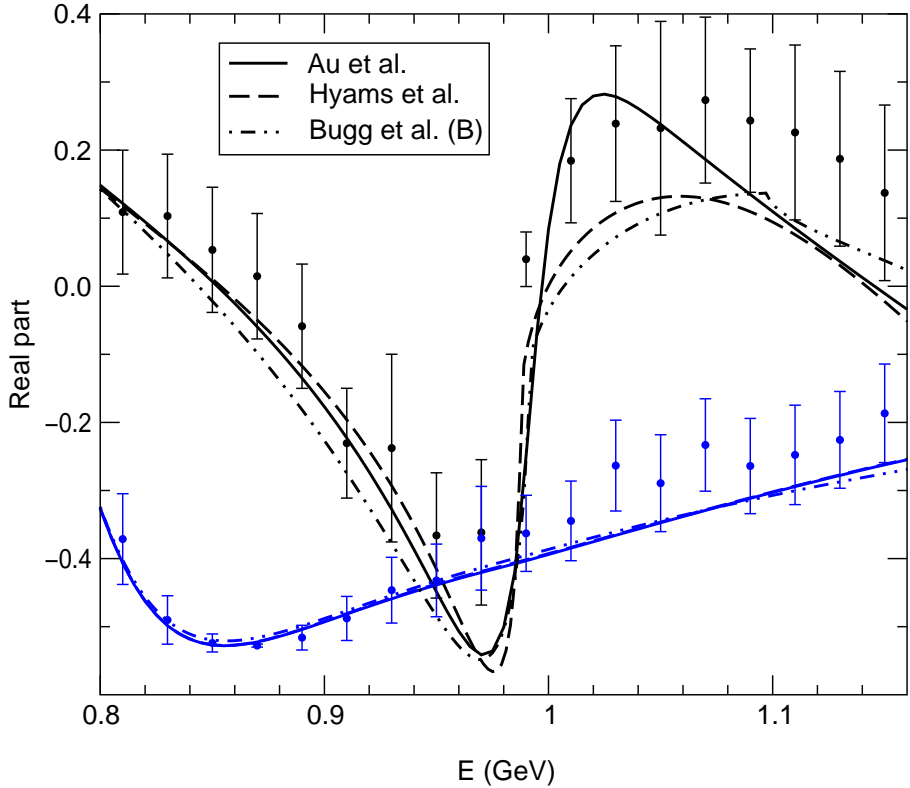}
\caption{\label{fig:ochs_cmp} Behaviour of the solutions above the
matching point. The curves show the solutions obtained with
three different inputs for the imaginary parts.
The data points are taken from the energy independent
analysis of the CERN-Munich data \cite{hyams}.
The $I=0$  $S$-wave is shown in black, the $I=1$ $P$-wave in blue.}
\end{figure}

The Roy equations are valid up to $\Eone=1.15$ GeV. In
fig.~\ref{fig:ochs_cmp}, we show three different solutions for
the $I=0$ and $I=1$ partial waves, in the region above the matching
point.
They are obtained by using three different inputs for the imaginary
parts
(note that the curves represent our solutions, not the real parts
of the input). The
figure shows that the differences are substantial,
especially in the $S$-wave, despite the fact that, below
$\Ezero=0.8\,\mbox{GeV}$, the three solutions are
practically identical, for all three waves. Evidently, above the
matching
point, the Roy solutions are
very sensitive to the input used for the imaginary parts.

It is not difficult to understand why that is so. As discussed in detail
in
section 10, the solutions follow the real parts of
the representation that is used as input (see
fig.~\ref{fig:roy_a225} for the case of Au et al. -- in the other two
cases,
the picture is similar). The real parts of the three representations
differ
considerably.
Moreover, all of these are systematically lower
than the ``data points'' in fig.~\ref{fig:ochs_cmp},
which show the result of an energy independent analysis of the
CERN-Munich data
\cite{hyams}.  In view of this, it is not surprising that the three Roy
solutions are quite different and that they are also systematically
lower than
the data points.
 
We conclude that a comparison of the Roy solutions with the data
in the region above the matching point does not yield reliable
information
about the values of the two $S$-wave scattering lengths and we
do therefore not show confidence--level
contours relative to data above 0.8 GeV.

\setcounter{equation}{0}
\section{Allowed range for $a_0^0$ and $a_0^2$}

\begin{figure}[thb] 
\leavevmode \centering
\includegraphics[angle=-90,width=12cm]{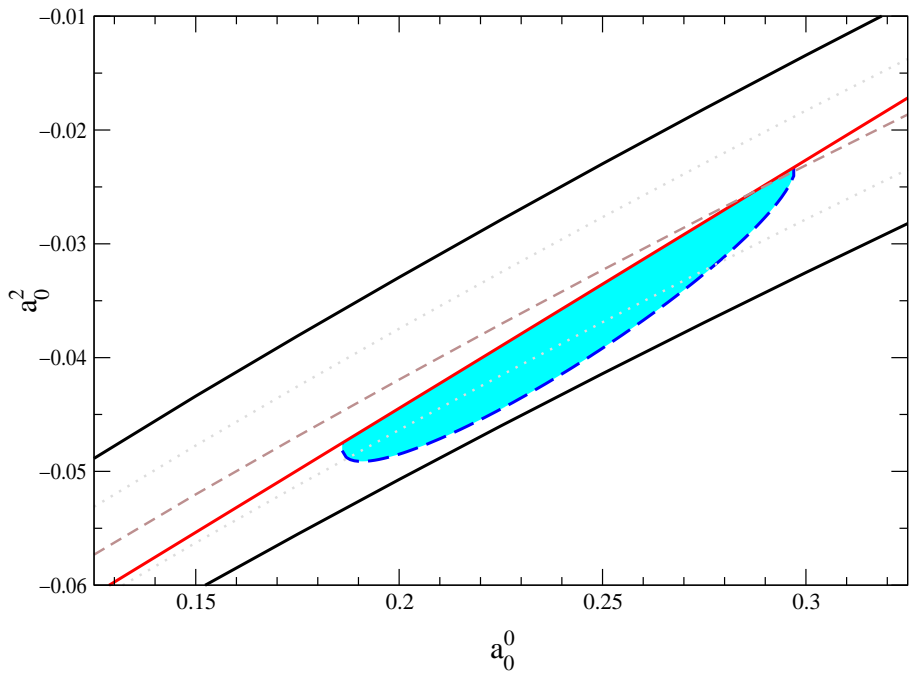}
\caption{\label{fig:summary} Intersection of the ellipse in 
fig.~\protect{\ref{fig:lo_and_CLEO}} (68\% C.L. relative to 
the data on $K_{e4}$ decay, on $\delta_0^2$ 
  and on the form factor) with the bands allowed by
  the consistency condition in all the three channels and by the 
Olsson sum rule.}
\end{figure}

The above discussion has made clear that we can rely only on two rather
solid sources of experimental information to determine the two $S$-wave
scattering lengths: the data on $K_{e4}$ and those on the $P$-wave in the
$\rho$ region. The former determine a range of allowed values for $a_0^0$
while the latter yield a range for the combination $2 a_0^0-5 a_0^2$. 
The consistency condition and the Olsson sum rule impose
further constraints. 
Figure \ref{fig:summary} summarizes
our findings: We have superimposed the ellipse of 
fig.~\ref{fig:lo_and_CLEO} with the lines that delimit the
consistency bands for the two $S$-waves, as well as those relevant for the 
Olsson sum rule.
The allowed range for $a_0^0$ and $a_0^2$ is the intersection
of the ellipse with the band where the Olsson sum rule is obeyed within the
estimated errors. In that region, the solutions also satisfy the
consistency condition.

We find it quite remarkable that the data on the shape
of the $\rho$ resonance, the consistency condition and the Olsson sum
rule all show a preference for the lower part of the
universal band. This gives us confidence that our conclusion on
which region in the $(a_0^0, a_0^2)$ plane is allowed by the present
experimental information is rather solid.
Once the new data on $K_{e4}$ decays will become available, the allowed
range in $a_0^0$ will become much narrower, and we will have a very small
ellipse. The prospects of making a real precision test of the predictions
for the two $S$-wave scattering lengths in the near future, appear to
be very good, in particular also in view of the pionium experiment under way
at CERN \cite{dirac}.

The $\pi N\rightarrow\pi\pi N$ data do provide essential information 
concerning the input of our calculations, but, as discussed in sections
\ref{I0low} and \ref{I0high}, they do not impose a firm constraint on the 
scattering lengths (incidentally, these data also prefer the lower half of the
universal band). This is unfortunate, because the power of the Roy equations
(unitarity, crossing symmetry and analyticity) 
is that of connecting regions of very
different energy scales. The behaviour of the two $S$-waves in the
immediate vicinity of threshold is determined by the scattering
lengths. In the combination $2 a_0^0 - 5 a_0^2$, these also determine the
linear growth of the subtraction polynomial: As we discussed in detail in
section 10, the large contribution from the
polynomial must be compensated to a high degree of accuracy by the
dispersive integrals. We therefore expect that a reanalysis of the 
$\pi N\rightarrow\pi\pi N$ data based on the Roy equations 
would lead to a rather stringent constraint on the allowed region, 
as it would make full use of the information contained in these data -- 
in our opinion, the
existing phase shift analyses are a comparatively poor substitute.

\setcounter{equation}{0}
\section{Threshold parameters}
\subsection{$S$- and $P$-waves}
As shown in ref.~\cite{Wanders sum rules}, the effective ranges of the 
$S$- and $P$-waves and the $P$-wave scattering length 
can be expressed in the form of sum rules, involving integrals over 
the imaginary parts of the scattering amplitude and the combination
$2a_0^0-5a_0^2$ of $S$-wave scattering lengths. 
The sum rules may be derived from the 
Roy representation by expanding the r.h.s.~of
eq.~(\ref{eq:rieq1}) in $q^2$ and reading off the coefficients according to
eq.~(\ref{eq:threxp}). In the case of the $S$-wave effective ranges,
the expansion can be interchanged with the integration over the imaginary
parts only after removing the threshold singularity.
This can be done by supplementing the integrand with a term proportional to
the derivative 
\bea \frac{d}{ds}\,\frac{1}{\sqrt{ s(s-4\,M_\pi^2)}}=-
\frac{h(s)}{\{ s(s-4\,M_\pi^2)\}^{2}}\co\hspace{1em}
h(s)= (s-2M_\pi^2)
\sqrt{s(s-4M_\pi^2)}\fs\nonumber\eea
In this notation, the sum rules 
may be written in the form: 
\bea 
 b_0^0  \al =\al  \frac{1}{3 M_\pi^2}\, (2 a_0^0-5 a_0^2) 
+\,\frac{16 }{3\pi}\int_{4 M_\pi^2}^{s_2} \frac{ds}{\{s(s-4
  M_\pi^2)\}^2}\left\{ 4M_\pi^2 (s-M_\pi^2)
\,\mbox{Im}\,t_0^0(s)
\right.\no   \al\al\left. 
-\,9 M_\pi^2 (s-4M_\pi^2)\,\mbox{Im}\,t_1^1(s) 
 +\, 5 M_\pi^2(s-4 M_\pi^2)\, \mbox{Im}\,t_0^2(s)
-\mbox{$\frac{3}{2}$}(a_0^0)^2\,h(s)\right\}\no
\al\al-\frac{8}{\pi}\int_{s_2}^\infty\frac{ds}{\{s(s-4
  M_\pi^2)\}^2}\,(a_0^0)^2\,h(s) +b^0_{0\, d}\co\label{sum rules}\\  
b^2_0\al=\al-\frac{1}{6 M_\pi^2}\, (2a_0^0-5a_0^2)  
+\,\frac{8 }{3\pi}\int_{4 M_\pi^2}^{s_2}
\frac{ds}{\{s(s-4M_\pi^2)\}^2}\left\{ 2 M_\pi^2(s-4 M_\pi^2) 
\,\mbox{Im}\,t_0^0(s) \right.\no\al\al\left.
+\,9 M_\pi^2(s-4 M_\pi^2)\,\mbox{Im}\,t_1^1(s) 
 +\, M_\pi^2 (7s -4 M_\pi^2)\, \mbox{Im}\,t_0^2(s) -3
(a_0^2)^2\,h(s)\right\}\no
\al\al-\frac{8}{\pi}\int_{s_2}^\infty\frac{ds}{\{s(s-4
  M_\pi^2)\}^2}\,(a_0^2)^2\,h(s)
+b^2_{0\,d} \co \no  
a^1_1\al=\al \frac{1}{18 M_\pi^2}\,(2a^0_0-5a^2_0)  
+\,\frac{8 M_\pi^2}{9\pi}\int_{4M_\pi^2}^{s_2} \frac{ds}{\{s(s-4 
  M_\pi^2)\}^2}\left\{ -2(s-4 M_\pi^2)\,   
\mbox{Im}\,t_0^0(s) \right.\no \al\al\left.\hspace{2cm} +\,9(3 s-4
M_\pi^2)\,\mbox{Im}\,t_1^1(s) + 5(s -4 M_\pi^2) \, 
\mbox{Im}\,t_0^2(s)\right\}+a^1_{1\,d}\co\no 
b^1_1\al= \al \frac{8 }{9\pi}\int_{4 M_\pi^2}^{s_2}
\frac{ds}{\{s(s-4 M_\pi^2)\}^3}\left\{-2(s-4 M_\pi^2)^3\, \mbox{Im}\,t_0^0(s)
 +9\left(\rule[-0.4em]{0em}{1em}3\,s^3-12\, s^2 M_\pi^2 \right.\right.\no 
\al\al\left.\left.\rule[-0.4em]{0em}{1em}+\,48\, s M_\pi^4-64M_\pi^6\right)
 \, \mbox{Im}\,t_1^1(s) + 5(s-4 M_\pi^2)^3\,\mbox{Im}\,t_0^2(s)\right\}
+ b^1_{1\,d}  \;. \nonumber
\eea 
The integrals only involve the imaginary
parts of the $S$- and $P$-waves and are cut off at 
$s=s_2$. The contributions from higher energies, as well as those 
from the imaginary parts of the
partial waves with $\ell=2,3,\ldots$ are contained in the
constants $b^0_{0\,d}$, $b^2_{0\,d}$, $a^1_{1\,d}$, $b^1_{1\,d}$.
By construction, these represent derivatives of the driving terms at 
threshold,
\vskip .2cm
\noindent
$ d_0^0(s)=q^2 b^0_{0\,d}+O(q^4)\,,\hspace{0.5em}
d^1_1(s)=q^2 a^1_{1\,d} +q^4 b^1_{1\,d}+O(q^6)\,,\hspace{0.5em}
d_0^2(s)=q^2 b^2_{0\,d}+O(q^4)\,.
$
\vskip.2cm
The numerical
\begin{table}
\hspace{-1em}
\begin{tabular}{|l|r|r|r|r|r|r|l|l|r|l|}
\hline
$\rule{0em}{1.1em}$&A$\rule{0.8em}{0em}$&B$\rule{0.8em}{0em}$&
C$\rule{0.5em}{0em}$
&D$\rule{0.5em}{0em}$&E$\rule{0.2em}{0em}$&total$\rule{0.2em}{0em}$&
$\rule{0.8em}{0em}\Delta_1$&
$\rule{0.3em}{0em}\Delta_2$
&ref.\cite{Nagels}$\rule{0.8em}{0em}$&$\rule{0.5em}{0em}$ units\\
\hline
$b^0_0$&2.12&$.45$&$-.03$& .02&.00 &
$2.56$&$ \pm .02\rule{0.45em}{0em}$&$ \rule[-0.3em]{0em}{1.4em}
{}^{+.28}_{-.12}$ &$2.5 \pm 0.3$&$10^{-1}M_\pi^{-2}$\\
$b_0^2$&$-1.06$&$.26$ &$.02$ &.01&.00&
$-.77$&$\pm.003$&$\rule[-0.3em]{0em}{1.5em} 
{}_{-.07}^{+.03}$&$-.82\pm .08$ &$10^{-1}M_\pi^{-2}$\\
$a_1^1$&3.53&$-.03$&$.13$ &$-.01$&.01&
$3.63$&$\pm.02\rule{0.45em}{0em}$&$\rule[-0.3em]{0em}{1.5em} 
{}_{-.11}^{+.29}$&$3.8\pm0.2$&$10^{-2}M_\pi^{-2}$\\   
$b_1^1$&$\rule[0.3em]{1em}{0.03em}\rule{0.7em}{0em}$&$4.05$&$1.39$&$-.07$&.08
&$5.45$&$\pm .13\rule{0.45em}{0em}$&
$\rule[-0.5em]{0em}{1.6em}{}_{-.44}^{+.35}$&&
$10^{-3}M_\pi^{-4}$\\ 
\hline
$a^0_2$&$\rule[0.3em]{1em}{0.03em}\rule{0.7em}{0em}$&$1.29$&$.28$ &.07&.03&
$1.67$&$\pm.01\rule{0.4em}{0em}$&$\rule[-0.4em]{0em}{1.5em}{}_{-.06}^{+.15}$&
$1.7 \pm .3\rule{0.5em}{0em}$&$10^{-3} M_\pi^{-4}$\\ 
$b_2^0$&$\rule[0.3em]{1em}{0.03em}\rule{0.7em}{0em}$&$-3.48$&$-.04$&.25&.02&
$-3.25$&$\pm
.07\rule{0.4em}{0em}$&$\rule[-0.4em]{0em}{1.5em}{}_{-.87}^{+.34}$&& 
$10^{-4} 
M_\pi^{-6}$\\
$a^2_2$&$\rule[0.3em]{1em}{0.03em}\rule{0.7em}{0em}$&$1.67$&$-.51$&.35&.02&
$1.53$&$\pm
.07\rule{0.4em}{0em}$&$\rule[-0.4em]{0em}{1.5em}{}_{-.45}^{+1.1}$& 
$1.3 \pm 3\rule{0.75em}{0em} $&$10^{-4}M_\pi^{-4}$\\ 
$b^2_2$&$\rule[0.3em]{1em}{0.03em}\rule{0.7em}{0em}$&$-3.10$&$-.09$&.06&.02&
$-3.11$&$\pm .07\rule{0.4em}{0em}$&
$\rule[-0.4em]{0em}{1.5em}{}_{-.95}^{+.41}$&&$10^{-4}M_\pi^{-6}$\\
$a^1_3$&$\rule[0.3em]{1em}{0.03em}\rule{0.7em}{0em}$&$5.11$&$.26$&.05&.01&
$5.43$&$\pm .1\rule{0.8em}{0em}$&$\rule[-0.4em]{0em}{1.5em}{}_{-.72}^{+1.6} $&
$6\pm2\rule{0.75em}{0em} $&
$10^{-5}M_\pi^{-6}$\\ 
$b^1_3$&$\rule[0.3em]{1em}{0.03em}\rule{0.7em}{0em}$&$-3.96$&$-.01$&.01&.01&
$-3.95$&$\pm .08\rule{0.4em}{0em}$&$\rule[-0.5em]{0em}{1.5em}
{}_{-1.9}^{+.89}$&&$10^{-5}M_\pi^{-8}$\\
\hline
\end{tabular}
\caption{\label{tab:SPDF}Threshold parameters of the $S$-, $P$-, $D$- and 
$F$-waves. The significance of the entries in columns A--E is specified in the
text. The column $\Delta_1$ indicates the uncertainty due to
  the error bars in the experimental input at and above 0.8 GeV, whereas
  $\Delta_2$ shows the shifts occurring if $a_0^0$ and $a_0^2$ are varied
  within the ellipse of fig.~\ref{fig:lo_and_CLEO}, according to
  eqs.~(\protect{\ref{SPT_pol}}) and (\protect{\ref{DFT_pol}}).} 
\end{table}
values obtained within our framework are given in the upper half of 
table \ref{tab:SPDF}, where we also show the numbers quoted in the compilation
of Nagels et al.~\cite{Nagels}, which are based on
the analysis of Basdevant, Froggatt and Petersen \cite{BFP2}. 
In accordance with the 
literature, we use pion mass units. Since the relevant physical scale 
is of the order of 1 GeV, the numerical values rapidly decrease
with the dimension of the quantity listed. The columns A -- E 
indicate the following contributions to the total\footnote{The numbers given
for the total include the tiny additional contributions 
to $b_0^0$ and $b_0^2$ that arise from the integrals 
over $h(s)(a_0^0)^2$ and $h(s)(a_0^2)^2$ in the interval
$s_2<s<\infty$. Numerically, these amount to 
$\delta b_0^0=-6.3\!\cdot\! 10^{-4} \,M_\pi^{-2}$ and $\delta b_0^2=-1.7
\!\cdot\! 10^{-5}\,M_\pi^{-2}$.}:  
\begin{description}
\item{A.} Contribution from the subtraction term 
$\propto 2a_0^2-5a_0^2$.
\item{B.} Imaginary parts of the $S$- and $P$-waves on the interval
$4\,M_\pi^2<s<s_0$. This contribution 
is evaluated with the Roy solutions described in the text. 
\item{C.} Imaginary parts of the $S$- and $P$-waves in the range
$s_0<s<s_2$. Here, we are relying on the experimental information,
discussed in section \ref{sec:exp_input}.
\item{D.} Imaginary parts of the higher partial waves in the range
$4\,M_\pi^2<s<s_2$. These are calculated 
in the same manner as for the driving terms of the $S$- and
$P$-waves (see appendix \ref{B3}).
\item{E.} Asymptotic contributions, $s>s_2$. These are evaluated with the
representation given in appendix \ref{B4}. 
\end{description}
For the reasons discussed earlier, we use $\Ezero=0.8\,\mbox{GeV}$,
$\Etwo=2\,\mbox{GeV}$. The values quoted in columns A and B are obtained with
our reference solution,  $a_0^0=0.225$, $a_0^2=-0.0371$, which corresponds to
the point $S_0$ in fig.~\ref{fig:UB}. 

The table shows that the result for $b_0^0$, $b_0^2$, $a_1^1$, $b_1^1$
is dominated by the contributions from the subtraction term and from the 
imaginary parts of the
$S$- and $P$-waves. The higher partial waves
and the asymptotic region only yield tiny corrections.
The sum D+E represents the contribution from the driving terms.
In the evaluation of these terms, which is discussed in detail in appendix
\ref{B5}, we have constrained the polynomial fit with the relevant
derivatives at threshold, so that
the numerical values of the four constants  $b^0_{0\,d}$, $b^2_{0\,d}$,
$a^1_{1\,d}$, $b^1_{1\,d}$ are correctly reproduced by
the corresponding terms in the 
representation (\ref{numerical driving terms}). 

The uncertainty given in column $\Delta_1$ of table \ref{tab:SPDF}
only accounts for the noise seen in our evaluation for the specific values
$a_0^0=0.225$, $a_0^2=-0.0371$ 
(errors in columns B--E added up quadratically). The sensitivity to 
these two parameters is well represented 
by linear relations of
the form\footnote{For $0.15\leq a_0^0 \leq 0.30$
the representation holds inside the universal band to better than 4\%.
Similar relations also follow directly from the
representation of the $S$- and $P$-waves given in appendix \ref{explicit 
numerical solutions}, but since the threshold region does not carry particular
weight when solving the Roy equations, these do not have the same 
accuracy.}
\bea
\label{SPT_pol}
\begin{array}{rrlrr}
 b_0^0=\al 2.56\times10^{-1}M_\pi^{-2} 
&\!\!\!\left\{\right.\!  1\,
+\al 3.2\, \Delta a^0_0\! \al\, -12.7 \Delta a^2_0\left.\right\}
\,,\\
b_2^0=\al -0.77\times10^{-1}M_\pi^{-2} 
&\!\!\left\{\right.\! 1\, 
+\al 2.5\, \Delta a^0_0\al \!-\,7.6\,\Delta a^2_0\left.\right\}\,,\\  
a_1^1=\al 
 3.63\times10^{-2}M_\pi^{-2} 
&\!\!\left\{\right. \! 1 \,
+\al 2.3\,\Delta a^0_0\al\! -\,7.8\,\Delta a^2_0\left.\right\}\,,\\  
b_1^1 =\al 
5.45\times10^{-3}M_\pi^{-4} 
&\!\!\left\{\right. \! 1 \,
+\al 0.1\,\Delta a^0_0\al \!-\, 5.7\,\Delta a^2_0\left.\right\}\,,\\  
\end{array}
\eea 
with $\Delta a^0_0=a_0^0-0.225$, $\Delta a^2_0=a_0^2+0.0371$.
Using this representation, the 1$\sigma$ ellipse of 
fig.~\ref{fig:lo_and_CLEO} can be translated into 1$\sigma$ ranges for the
various quantities listed in the table -- these are shown in
column $\Delta_2$ (since our reference point is not at the center of the 
ellipse, the ranges are asymmetric). 

The table neatly demonstrates that the two $S$-wave scattering lengths
are the essential low energy parameters -- the uncertainty in the result
is due almost exclusively to the one in 
$a_0^0$, $a_0^2$. This is to be expected on general grounds 
\cite{Leutwyler Frascati}: The integrals occurring in the above sum rules 
are rapidly convergent, so
that only the behaviour of the partial waves in the threshold region matters.
The uncertainties in the input used for the imaginary parts above the matching
point only enter indirectly, through their effect on the $S$- and
$P$-waves in the threshold region. We did not expect, however, that the
effect would be as small as indicated in the table and add a few comments
concerning this remarkable finding.

In order to document the statement  
that the uncertainties which we are attaching to the phenomenological input
of our calculation (behaviour of the imaginary parts above the matching point,
elasticity, driving terms) only have a minute effect on the result for the 
threshold parameters, we find it best to give the numerical size of this
effect (column $\Delta_1$ of the table). We repeat that the numbers
quoted there merely indicate the noise seen in our 
evaluation -- we do not claim to describe the
scattering amplitude to that accuracy. Isospin breaking, for instance, cannot
be neglected at that level of precision. 

The reason why the threshold parameters are insensitive to the uncertainties
of our input is the following.
As discussed in detail in sections \ref{sec:unique}--\ref{sec:UB}, the
solutions of the Roy equations in general exhibit a cusp at the matching
point. If the imaginary parts above 0.8 GeV and the value of $a_0^0$ are
specified, there is a solution with physically acceptable behaviour in the
vicinity of the matching point only if the parameter $a^2_0$ is chosen
properly. In other words, there is a strong correlation between the
behaviour of the imaginary parts and the parameters $a_0^0$, $a_0^2$.
As we are selecting a specific value for these parameters, 
we are in effect subjecting the imaginary parts to a constraint. 
For this reason, the uncertainties in the input can barely be
seen in the output for the threshold parameters -- the main effect is 
hidden in $a_0^0$, $a_0^2$. 
The correlation just described originates in the fact that one of 
the two subtraction constants is superfluous: The combination
$2\,a_0^0-5\,a_0^2$  may be represented as a convergent dispersion
integral over the imaginary part of the amplitude. 

The correlation is illustrated by the lines in fig.~\ref{fig:UB},
which correspond to the specific parametrization of the 
input used for the imaginary part of the $I=2$ $S$-wave shown in
fig.~\ref{fig:I2}. As there is very little experimental information about the
energy dependence of this partial wave, we have worked out the
change in the Roy solutions that occurs if 
this energy dependence is modified above the matching point. 
The result for the threshold parameters turns out to be 
practically unaffected. Also, we have varied the driving terms within
the uncertainties given in section \ref{section driving terms}. Again, the
response in the threshold parameters can barely be seen. 

\subsection{$D$- and $F$-waves}
\label{B6}
Similar sum rules also hold for the threshold parameters of the higher partial
waves. The contributions from the
imaginary parts of the $S$- and $P$-waves 
are obtained by expanding the kernels occurring in the
Roy equations for the $D$- and $F$-waves around threshold. We write the result
in the form
\bea
a^0_2\al=\al\frac{16}{45\pi} \int_{4M_\pi^2}^{s_2} 
\frac{ds}{s^3\,(s-4M_\pi^2)} \,
\left\{ (s-4M_\pi^2)\,\mbox{Im}\,t_0^0(s)+
9(s+4M_\pi^2)\,\mbox{Im}\,t_1^1(s)\right.\no\al\al \hspace{4em}+\left.
5 (s-4M_\pi^2)\,\mbox{Im}\,t_0^2(s)
\right\}+a^0_{2\,d}\co\no
b^0_2\al=\al-\frac{32}{15\,\pi}\int_{4M_\pi^2}^{s_2} 
\frac{ds}{s^4\,(s-4M_\pi^2)} \,\left\{(s-4M_\pi^2)\,\mbox{Im}\,t_0^0(s)
-3(s-12M_\pi^2)\,\mbox{Im}\,t_1^1(s)\right.\no\al\al \hspace{4em}+\left.
5 (s-4M_\pi^2)\,\mbox{Im}\,t_0^2(s)
\right\}+b^0_{2\,d}\co\no
a^2_2\al=\al\frac{8}{45\pi} \int_{4M_\pi^2}^{s_2} 
\frac{ds}{s^3\,(s-4M_\pi^2)} \,
\left\{2 (s-4M_\pi^2)\,\mbox{Im}\,t_0^0(s)
-9(s+4M_\pi^2)\,\mbox{Im}\,t_1^1(s)\right.\no\al\al \hspace{4em}+\left.
 (s-4M_\pi^2)\,\mbox{Im}\,t_0^2(s)
\right\}+a^2_{2\,d}\co\\
b^2_2\al=\al-\frac{16}{15\,\pi}\int_{4M_\pi^2}^{s_2} 
\frac{ds}{s^4\,(s-4M_\pi^2)} \,\left\{2(s-4M_\pi^2)\,\mbox{Im}\,t_0^0(s)
+3(s-12M_\pi^2)\,\mbox{Im}\,t_1^1(s)\right.\no\al\al \hspace{4em}+\left.
 (s-4M_\pi^2)\,\mbox{Im}\,t_0^2(s)
\right\}+b^2_{2\,d}\co\no
a^1_3\al=\al\frac{16}{105\pi} \int_{4M_\pi^2}^{s_2} 
\frac{ds}{s^4\,(s-4M_\pi^2)} \,
\left\{2 (s-4M_\pi^2)\,\mbox{Im}\,t_0^0(s)+
9(s+4M_\pi^2)\,\mbox{Im}\,t_1^1(s)\right.\no\al\al \hspace{4em}-\left.
5 (s-4M_\pi^2)\,\mbox{Im}\,t_0^2(s)
\right\}+a^1_{3\,d}\co\no
b^1_3\al=\al-\frac{128}{105\,\pi}\int_{4M_\pi^2}^{s_2} 
\frac{ds}{s^5\,(s-4M_\pi^2)} \,\left\{2(s-4M_\pi^2)\,\mbox{Im}\,t_0^0(s)
+36M_\pi^2\,\mbox{Im}\,t_1^1(s)\right.\no\al\al \hspace{4em}-\left.
5 (s-4M_\pi^2)\,\mbox{Im}\,t_0^2(s)
\right\}+b^1_{3\,d}\co\nonumber
\eea
where $a^0_{2\,d},\,b^0_{2\,d},\ldots$ contain the contributions
from $s>s_2$ as well as those from the higher partial waves.
The evaluation of these contributions, however, meets with problems that
we need to discuss in some detail.

First, we note that the definition of the driving terms in eq.~(\ref{dt}) is
suitable only for the analysis of the $S$- and $P$-waves. For $\ell\geq 2$,
the functions $d^I_\ell(s)$ contain a branch cut at threshold,
so that these quantities are complex. In order to solve the Roy 
equations for the $D$-waves, for instance, the contributions generated by 
their imaginary parts need to be isolated, using a 
different decomposition of the right hand side of these equations.
As far as the scattering lengths and effective ranges are concerned,
however, only the values of the functions $d^I_\ell(s)$ 
and their first derivatives at threshold are needed, which are real.

A more subtle problem arises from the fact that the explicit form of the 
kernels occurring in the Roy equations for the higher partial waves
depends on the choice of the partial wave projection.
As discussed in detail in
ref.~\cite{Atkinson}, the definition (\ref{intrep}) -- which we used
in our analysis of the $S$- and $P$-waves -- does 
not automatically ensure that the threshold behaviour of the
partial waves with $\ell\geq 3$ starts with the power $q^{2\ell}$.
The problem arises from the fact that the solution of the 
Roy equations leads to a crossing symmetric scattering amplitude
only if the imaginary parts of the higher partial waves satisfy sum rules
such as the one in eq.~(\ref{SRpw}). In particular, the expansion of
the $F$--wave in powers of $q$ in general starts with 
\bea\mbox{Re}\,t^1_3(s)=x_3^1\,q^4+a_3^1\,q^6+b^1_3\,q^8+\ldots\nonumber\eea 
For the fictitious term $x_3^1$ to be absent, the imaginary parts of the
higher partial waves
must obey a sum rule. In fact, we have written down the relevant sum rule
already: equation (\ref{SRpw}). The derivation given in section \ref{B2}
shows that this constraint ensures crossing symmetry of the terms
occurring in the expansion of the scattering amplitude around threshold, up to
and including contributions of $O(q^4)$. The threshold expansion of the
partial waves with $\ell\geq 3$ thus only starts at $O(q^6)$ if this
condition holds  -- in particular $x_3^1$ then vanishes. 
The sum rule that allows us 
to pin down the asymptotic contributions to the driving
terms for the $S$- and $P$-waves thus at the same time also 
ensures the proper threshold behaviour of the $F$--waves.
The absence of a term of $O(q^6)$ in the $G$-waves 
leads to a new constraint, which could be derived in the same manner, etc.
Note that the contributions from the imaginary parts of the $S$- and
$P$-waves are manifestly crossing symmetric -- the constraints imposed
by crossing symmetry exclusively concern the higher waves\footnote{The
family of sum rules discussed in appendix \ref{C1} does not follow
from crossing symmetry, but from an asymptotic condition that
goes beyond the Roy equations. As shown there, those sum rules do
tie the imaginary part of the $P$-wave to the higher partial waves.}.

The $F$-wave scattering length occurs in the expansion of the amplitude 
around threshold among the contributions of $O(q^6)$, two powers 
of $q$ beyond the term just discussed. In the numerical analysis, 
we thus need to make sure that the sum rule holds to high precision 
if we are to get a reliable value in this manner. 
For the effective range, the situation is even worse. This indicates that
for the numerical analysis of the higher partial waves, the extension of the
range of validity 
of the Roy equations achieved if the standard partial wave projection
(\ref{pwp1}) is replaced by (\ref{pwp}) generates considerable complications.

For the evaluation of the threshold parameters, this extension is not
needed -- we may use the 
partial wave projection (\ref{pwp1}), for which the problem discussed above
does not occur. In particular, $x_3^1$ then automatically vanishes, so that
the evaluation of the scattering lengths and effective ranges does not
pose special numerical problems. 
To evaluate those
from the asymptotic region, we expand the fixed-$t$ dispersion relation
(\ref{fixedt}) in powers of $t$.  
The results obtained for $a_0^0=0.225$,
$a^2_0=-0.0371$ are listed in the lower half of table \ref{tab:SPDF}.

The dependence on the $S$-wave scattering lengths may again be 
represented (to better than 6\% inside the
universal band for $0.15\leq a_0^0 \leq 0.30$) 
with a set of linear relations:
\bea
\label{DFT_pol}
\begin{array}{rrlrr}
 a_2^0=\al  1.67\times10^{-3}M_\pi^{-4} 
&\!\!\!\left\{\right. \! 1\, 
+\!\!\!\!\!\al 2.6\, \Delta a_0^0\! \al -8.6\, 
\Delta a^2_0\left.\right\}\,,\\
b_2^0=\al-3.25\times10^{-4}M_\pi^{-6} 
&\!\!\!\left\{\right.\! 1\, 
+\al 6.6\, \Delta a_0^0\!\al -17\,\Delta a^2_0\left.\right\}\,,\\  
a_2^2=\al 
 1.53\times10^{-4}M_\pi^{-4} 
&\!\!\!\left\{\right.\! 1 \,
+\al 14\,\Delta a_0^0\!\al-25\,\Delta a^2_0\left.\right\}\,,\\  
b_2^2 =\al  
-3.11\times10^{-4}M_\pi^{-6} 
&\!\!\!\left\{\right.\! 1 \,
+\al 6.2\,\Delta a_0^0\!\al -11\,\Delta a^2_0\left.\right\}\,,\\  
a_3^1 =\al  
5.43\times10^{-5}M_\pi^{-6}
&\!\!\! \left\{\right.\! 1 \,
+\al 5.5\,\Delta a_0^0\!\al\!-\,8\,\Delta a^2_0\left.\right\}\,,\\  
b_3^1 =\al  
-3.95\times10^{-5}M_\pi^{-8} 
&\!\!\!\left\{\right.\! 1 \,
+\al 8\,\Delta a_0^0\!\al\! -\,8\,\Delta a^2_0\left.\right\}\,. 
\end{array}
\eea 
The sensitivity is more pronounced here than in the case of the
threshold parameters for the $S$- and $P$-waves. In particular,
the linear representation for the $D$-wave scattering length $a_2^2$
only holds to a good approximation if $a_0^0$ and $a_0^2$ do not deviate
too much from the central values.

\setcounter{equation}{0}
\section{Values of the phase shifts at $s=M_K^2$}

A class of important physical processes where the $\pi \pi$ phase shifts
play a relevant role is that of kaon decays. Let us recall, for instance, 
that the phase of $\varepsilon'$ is given by 
the value of $\delta_0^2-\delta_0^0+\frac{1}{2}\,\pi$ at $s=M_K^2$.
In this section, we give numerical values for the three phase shifts at
the kaon mass as they come out from our Roy equation analysis, and show
the explicit dependence on the two $S$-wave scattering lengths. In this
manner, an improved determination of the latter will immediately translate
into a better knowledge of the phases at $s=M_K^2$.
 
The decays
$K^0\rightarrow\pi \pi$ and $K^+\rightarrow \pi \pi$ concern
slightly different values of the energy. In view of the fact that
the CP-violating parameter $\varepsilon'$ manifests itself in the decays of
the neutral kaons, we evaluate the phases at $s=M_{K^0}^2$. 
Note that, in addition to this difference in the masses, there are
also isospin breaking effects in the relevant transition matrix elements.
As far as the $\pi\pi$ phases are concerned, however,
the isospin breaking effects due to $m_d-m_u$ are tiny,
because $G$-parity implies that these only occur at order $(m_d-m_u)^2$.

As in the preceding
section, we give values at the reference point $a_0^0=0.225$ and
$a_0^2=-0.0371$, and break down the errors into those due to the noise in
our calculations and those due to the poorly known values of the two
scattering lengths.  The results are shown in table
\ref{tab:mk_phase}. 
Like for the threshold parameters, the two $S$-wave
scattering lengths are the main source of uncertainty. In the present case,
the errors due to the uncertainties in our experimental
input at 0.8 GeV are not negligible, but they amount to at most
4\%.

The dependence of the central values on the two scattering lengths is well
described by the following polynomials:
\bea 
\label{dmk_pol}
\delta_0^0(M_{K^0}^2)\al=\al 37.3^\circ\left\{1+3.0 \Delta a^0_0-8.5
  \Delta a^2_0 \right\}\,, \no
\delta_1^1(M_{K^0}^2)\al=\al 5.5^\circ\left\{1+1.7 \Delta a^0_0-6.7
  \Delta a^2_0 \right\}\,, \\
\delta_0^2(M_{K^0}^2) \al=\al -7.8^\circ\left\{1+1.9 \Delta a^0_0-13
  \Delta a^2_0 \right\}\,,  \no
\delta_0^0(M_{K^0}^2) - \delta_0^2(M_{K^0}^2) \al=\al 45.2^\circ\left\{1+2.8
  \Delta a^0_0-9.4 \Delta a^2_0 \right\}\,. \nonumber
\eea

Our results are in agreement with 
refs.~\cite{Ochs Newsletter,donoghue,gassermeissner}, but are more accurate. 
In the foreseeable future, the two
$S$-wave scattering lengths will be pinned down to good precision, so that
the above formulae will fix the phases to within remarkably
small uncertainties.
\begin{table}
\begin{center}
\begin{tabular}{|c|r|l|l|}
\hline
      &\parbox{4em}{Value at\\$s=M_{K^0}^2$} &
\hspace{0.5em}$\rule[-1em]{0em}{2.5em}
\Delta_1$ &\hspace{0.5em}$\Delta_2$ \\
\hline
$\delta_0^0$ & $37.3$\hspace{1em}    & $\pm 1.4$    & 
$\rule[-0.3em]{0em}{1.4em}{}_{-1.6}^{+4.3}$  \\
$\delta_1^1$ &  $5.5$\hspace{1em}   & $\pm 0.1$    & 
$\rule[-0.3em]{0em}{1.5em}{}_{-.13}^{+.3}$  \\
$\delta_0^2$ & $-7.8$\hspace{1em}  & $\pm 0.04$    & 
$\rule[-0.3em]{0em}{1.5em}{}_{-.8}^{+.7}$    \\
$\delta_0^0-\delta_0^2$ & $45.2$\hspace{1em}   & $\pm 1.3 $   &
$\rule[-0.5em]{0em}{1.6em}{}_{-1.6}^{+4.5}$ \\
\hline
\end{tabular}
\end{center}
\caption{\label{tab:mk_phase}Values of the phase shifts at $s=M_{K^0}^2$ in
  degrees. The central value is obtained with our reference solution of the 
Roy equations, where $a_0^0=0.225$, $a_0^2=-0.0371$. 
The column $\Delta_1$ indicates the uncertainty due to
  the error bars in the experimental input at and above 0.8 GeV, whereas
  $\Delta_2$ shows the shifts occurring if $a_0^0$ and $a_0^2$ are varied
  within the ellipse of fig.~\ref{fig:lo_and_CLEO}, according to
  eq.~(\protect{\ref{dmk_pol}}).}
\end{table}

\setcounter{equation}{0}
\section{Comparison with earlier work}
The Roy equations were used to obtain information on the
$\pi\pi$ scattering amplitudes, already in the early seventies.
Most of the work done since then either follows the method of Pennington and
Protopopescu \cite{PP1,PP2} 
or the one of Basdevant, 
Froggatt and Petersen \cite{BFP1,BFP2}. In the present section, we briefly 
compare these two approaches with ours.
A review of the results obtained by means of the Roy equations is given in 
ref.~\cite{Morgan Pennington handbook}. 

To our knowledge, Pennington and Protopopescu \cite{PP1} were the
first to analyze $\pi\pi$ scattering data using Roy's equations.  In
principle, the approach of these authors is similar to ours. In our
language, they fixed the matching point at $\Ezero=0.48$ GeV.  As input data,
they relied on the $\pi\pi$ production experiment of the
Berkeley group \cite{Protopopescu}, using the data of Baton et
al.~\cite{saclay} for the $I=2$ channel (at the time they performed the
analysis, the high-energy, high-statistics CERN-Munich data \cite{hyams}
were not yet available). The Roy equations then allowed them to extrapolate
the $S$- and $P$-wave phases of Protopopescu et al.~\cite{Protopopescu} to
the region below $0.48$ GeV. Comparing the Roy-predicted real parts with the
data (this corresponds to what we call consistency), they found that these
constrain the two $S$-wave scattering lengths to the range $
a_0^0=0.15 \pm 0.07$, $a_0^2=-0.053\pm 0.028$.
In their subsequent work \cite{PP2}, they then used the Roy equations to
solve the famous Up-Down ambiguity that occurs in the analysis of the
$S$-wave.

The fact that, in their analysis, the matching point is taken below the
mass of the $\rho$ has an interesting mathematical consequence: As
discussed in section 6.3, the Roy equations do then not admit a solution
for arbitrary values of $a_0^0$, $a_0^2$, even if cusps at the matching
point are allowed for (the situation corresponds to row IV of table
\ref{tab:3chm}). To enforce a solution, one may for instance keep the input
for the imaginary parts as it is, but tune the scattering length
$a_0^2$. The result, however, in general contains strong cusps in the
partial waves with $I=0,1$. These can only be removed if the input used for
the imaginary parts above the matching point is also tuned -- the situation
is very different from the one encountered for our choice of the matching
point.

Basdevant, Froggatt and Petersen \cite{BFP1,BFP2} constructed solutions of
the Roy equations by considering several phase shift analyses and a broad
range of $S$-wave scattering lengths. The method used by these authors is
different from ours in that they relied on an analytic parametrization of
the $S$- and $P$-waves from threshold up to
$\Etwo=\sqrt{110}\,M_\pi=1.47\,\mbox{GeV}$, the onset of the asymptotic
region in their case. Some of the parameters occurring therein are
determined from a fit to the data, some by minimizing the difference
between the right and left hand sides of the Roy equations in the region
below $\Ezero=\sqrt{60}\,M_\pi=1.08$ GeV. In this manner, they construct
universal bands corresponding to the Berkeley \cite{Protopopescu}, Saclay
\cite{saclay} and CERN-Munich phases as determined by Estabrooks et
al.~\cite{EM}. The individual bands are not very much broader than the
shaded region in fig.~\ref{fig:summary}, but they are quite different from
one another: Crudely speaking, the Berkeley band is centered at the upper
border of our universal band, while the one constructed with the CERN-Munich
phases is centered at the lower border.  The Saclay band runs outside the
region where we can find acceptable solutions at all.

In order to compare their results with ours, we first note that, for the
six explicit solutions given in table 5 of \cite{BFP2}, the value of
$a_0^0$ varies between $-0.06$ and $0.59$. Only two of these correspond to
values of the $S$-wave scattering  
lengths in the region considered in the present paper: 
BKLY$_2$ and SAC$_2$. For these two,
the value of the $P$-wave phase shift at 0.8 GeV is $108.3^\circ$ and
$108.0^\circ$, respectively, remarkably close to the central value of the
range allowed by the data on the form factor, eq.~(7.2). Concerning the
value of $\delta_0^0$ at 0.8 GeV, however, the two solutions differ
significantly: While BKLY$_2$ yields $79.7^\circ$ and is thus within our
range in eq.~(7.4), the value $70.2^\circ$ that corresponds to SAC$_2$ is
significantly lower. In our opinion, that solution is not consistent with
the experimental information available today. In the interval from threshold 
to 0.8 GeV, our
solution differs very little from BKLY$_2$.  Above this energy, the
imaginary part of the $I=0$ $S$-wave in BKLY$_2$ is substantially
smaller than the one we are using as an
input. Nevertheless, the solutions are very similar at low energies,
because the behaviour below the matching point is not sensitive to the
input above 1 GeV.

\setcounter{equation}{0}
\section{Summary and conclusions}
The Roy equations follow from general properties 
of the $\pi \pi$ scattering amplitude. We have set up a
framework to solve these equations numerically. In the following, we
summarize the main features of our approach and the results 
obtained with it, omitting details -- even if these would be necessary to
make the various statements watertight.

1.  In our analysis, three energies $s_0<s_1<s_2$ play a special
role:

\vspace{0.5em}\hspace{3em}
\begin{tabular}{rcrrcr} $\Ezero$&$\!\!=\!\!$&$ 0.8\,\mbox{GeV}
\co\hspace{1em}$&$s_0$&$\!\!=\!\!$&$ 32.9\,M_\pi^2\co\hspace{1em}$\\
$\Eone $&$\!\!=\!\!$&$ 1.15\,\mbox{GeV}\co\hspace{1em}$&
$s_1$&$\!\!=\!\!$&$ 68\, M_\pi^2\co\hspace{1em}$\\
$\Etwo $&$\!\!=\!\!$&$ 2\,\mbox{GeV}\co\hspace{1em}$&
$s_2$&$\!\!=\!\! $&$ 205.3\,  M_\pi^2\fs\hspace{1em}$\end{tabular}

\vspace{0.3em}\noindent
We refer to the point $s_0$
as the matching point: At this energy, the region where we calculate
the partial waves meets the one where we are relying
on phenomenology. The point $s_1$ indicates the upper end of the
interval on which the Roy equations are valid, while $s_2$ is the 
onset of the asymptotic region. 

2. Given the strong dominance of 
the $S$- and $P$-waves, we solve the Roy equations only for these, and only
on the interval $4M_\pi^2<s<s_0$, that is on the lower half of their range of 
validity. In that region, the contributions generated by 
inelastic channels are negligibly small. There, 
we set $\eta_0^0(s)=\eta_1^1(s)=\eta_0^2(s)=1$. 
In the interval from $s_0$ to $s_2$, we evaluate
the imaginary parts with the available experimental information, whereas
above $s_2$, we invoke a theoretical 
representation, based on Regge asymptotics. We demonstrate that crossing 
symmetry imposes
a strong constraint on the asymptotic contributions, which 
reduces the corresponding uncertainties quite substantially -- in most of our
results, these are barely visible. 

3. The Roy equations involve two subtraction constants, which may be identified
with the two $S$-wave scattering lengths $a_0^0$, $a_0^2$. In principle, 
one subtraction would suffice: The Olsson sum rule
relates the combination $2\,a_0^0-5\,a_0^2$ to an integral over 
the imaginary parts in the forward direction 
(or, in view of the optical theorem, over 
the total cross section). This imposes a correlation between 
the input used for the imaginary parts
and the values of the $S$-wave scattering lengths, but using this constraint
ab initio would lead to an unnecessary complication of our
scheme. We instead treat the two subtraction constants as independent
parameters. The consequences of the Olsson sum rule are discussed below.

4. Unitarity converts the Roy equations for the $S$- and $P$-waves
into a set of three coupled integral equations for the corresponding 
phase shifts: The real part of the partial
wave amplitudes is given by a sum of known contributions 
(subtraction polynomial,
integrals over the region $s_0<s<s_2$ and driving terms) and certain 
integrals over their imaginary parts, extending from threshold to $s_0$. 
Since unitarity relates the real and imaginary parts in a nonlinear manner, 
these equations are inherently nonlinear and cannot be solved explicitly. 

5. Several mathematical properties of such integral equations are known, and
are used as a test and a guide for our numerical work. In
particular, the existence and uniqueness of the
solution is guaranteed only if the matching point $s_0$ is taken 
in the region between the place where
the $P$-wave phase shift goes through $90^\circ$ and the energy where
the $I=0$ $S$-wave does the same. As this range is quite narrow
($0.78\,\mbox{GeV}<\Ezero< 0.86\,\mbox{GeV}$), there is little freedom in the 
choice of the matching point -- we use $\Ezero=0.8$ GeV. According to table 
\ref{tab:3chm}, the multiplicity index
of the interval  $0.86 < \Ezero < 1 \,\mbox{GeV}$ is equal to 1. By way of 
example
($\Ezero=0.88$ GeV), we have verified that our framework indeed admits a
one--parameter family of numerical solutions if the matching point is
taken in that energy range.

6. A second consequence of the mathematical structure of the Roy
equations is that, for a given input and for a random choice of the
two subtraction constants, the solution has a cusp at $s_0$:
In the vicinity of the matching point, the solution in general exhibits
unphysical behaviour. The strength of the cusp is very sensitive to the
value of $a^2_0$. In fact, we find that the cusp disappears in the noise of
our calculation if that value is tuned properly. 
Treating the imaginary parts 
as known, the requirement that the solution is free of cusps at the matching
point determines the value of $a_0^2$ as a function of $a_0^0$. 
This is how the universal curve of Martin, Morgan and Shaw manifests 
itself in our approach.

7. The input used for the imaginary parts above the matching point is subject
to considerable uncertainties. In our framework, the values of the
$S$- and $P$-wave phase shifts at the matching point represent the 
essential parameters in this regard. In order to pin these down, we 
first make use of the fact that
the data on the pion form factor, obtained from the processes
$e^+e^-\rightarrow \pi^+\pi^-$ and $\tau\rightarrow \pi^-\pi^0\nu_\tau$,
very accurately determine the behaviour of the $P$-wave phase shift in the
region of the $\rho\,$-resonance, thus constraining the value of
$\delta_1^1(s_0)$ to a remarkably narrow range. Next, we observe that the
absolute phase of the $\pi\pi$ scattering amplitude drops out in the
difference $\delta^1_1-\delta^0_0$, so that one of the sources
of systematic uncertainty is eliminated. Indeed, the phase shifts extracted
from the reaction $\pi N\rightarrow\pi\pi N$ yield remarkably coherent 
values for this difference. Since the $P$-wave is
known very accurately, this implies that $\delta_0^0(s_0)$ 
is also known rather well. The experimental information concerning
$\delta_0^2$, on the other hand, is comparatively meagre. 
We vary it in the broad range shown in fig.~\ref{fig:I2}. 

8. The uncertainties in the experimental input for the imaginary parts and 
those in the driving terms turn the universal curve into a band in 
the $(a_0^0,a_0^2)$ plane, part of which is shown in fig.~\ref{fig:UB}. 
Outside this ``universal band'', 
the Roy equations do not admit physically
acceptable solutions that are consistent with what is known about the behaviour
of the imaginary parts above the matching point. 

9. One of the striking features of the solutions is that, 
above the matching point, they very closely
follow the real part of the partial wave used as input 
for the imaginary part, once the value of $a^2_0$ is in the proper range. 
The phenomenon is discussed
in detail in section \ref{sec:math}, where we show that, in a certain 
sense, this property represents a necessary condition 
for the solution to be acceptable physically. The region where this
consistency condition holds is shown in fig.~\ref{fig:Olsson}:
It roughly constrains the admissible values of $a_0^2$ to the lower
half of the universal band.

10. As mentioned above, the Olsson sum rule relates
the combination $2a_0^0-5a_0^2$ of scattering lengths to an integral over
the imaginary parts of the amplitude. Evaluating the integral, we find that
the sum rule is satisfied in the band spanned by the two red curves
shown in fig.~\ref{fig:Olsson}.
The Olsson sum rule thus amounts to essentially the same
constraint as the consistency condition. Presumably, the universal band
is of the same origin:
Physically acceptable solutions only exist if the subtraction constants
are properly correlated with the imaginary parts. The shaded region 
in fig.~\ref{fig:Olsson} shows the domain where all
of these conditions are satisfied. It is by no means built in from the start
that the various requirements can simultaneously be met -- in our opinion,
the fact that this is the case represents a rather thorough check
of our analysis.

11. The admissible region can be constrained further 
if use is made of experimental data below the matching point.
At the moment there are two main sources of
information on $\pi \pi$ scattering below 0.8 GeV: A few data points for
the $I=2$ $S$-wave phase shift -- which to our knowledge 
will, unfortunately, not be improved 
in the foreseeable future -- and a few data points on $\delta_0^0-\delta_1^1$
very close to threshold, as measured in $K_{e4}$ decays. 
These data do provide an important constraint. 
We compare our solutions inside the universal
band to both sets of data. As shown in fig.~\ref{fig:kl4_acm_lo}, 
the corresponding $\chi^2$ contours nicely fit inside the
universal band. The net result for the allowed range of the parameters 
is shown in fig.~\ref{fig:summary}, 
which summarizes our findings. 

12. To our knowledge, the Roy equation analysis is the only method that 
allows one to reliably translate low energy data on the scattering amplitude
into values for the scattering lengths. As discussed above, the available data 
do correlate the value of $a^2_0$ with the one of $a^0_0$.
Unfortunately, however, the value of $a_0^0$ as such
is not strongly constrained: In agreement with
earlier analyses, we find that these data are consistent with any value of 
$a_0^0$ in the range from 
0.18 to 0.3. 

13. The new experiments at Brookhaven \cite{e865} and at
DA$\Phi$NE \cite{kloe} will yield more precise information in the
very near future. We expect that the analysis of the forthcoming results
along the lines described in the present paper 
will reduce the error in $a_0^0$ by about a factor of three. 
Moreover, the pionic atom experiment under way at CERN
\cite{dirac} will allow a direct measurement of $|a_0^0-a_0^2|$ and thus
confine the region to the intersection of the corresponding, 
approximately vertical strip with the region shown in fig.~\ref{fig:summary}.

14. The two subtraction constants $a_0^0$, $a_0^2$ are the essential
parameters at low energies: If these were known, our method would allow us
to calculate the $S$- and $P$-wave phase shifts below 0.8 GeV to an amazing
degree of accuracy. The parameters $a_0^0$, $a_0^2$ act like a
filter: If the solutions are sorted out according to the values of these
parameters, the noise due to the uncertainties in our input practically
disappears, because variations of that input require a corresponding 
variation, either in $a_0^0$ or in $a_0^2$ -- otherwise, the behaviour of the 
solution near the matching point is unacceptable. A simple explicit
representation for the $S$- and $P$-wave phase shifts
as functions of the energy is given in appendix
\ref{explicit numerical solutions}. The representation explicitly displays the
dependence on $a_0^0$, $a_0^2$.

15. We have also analyzed the implications for
the scattering lengths of the $P$-, $D$- and $F$-waves, as well as
for the various effective ranges. The fact that $a_0^0$ and $a_0^2$ are the
essential low energy parameters manifests itself also here: If we change the 
input in the Roy equations within the uncertainties,
but keep $a_0^0$ and $a_0^2$ constant, the values of the various threshold 
parameters vary only by tiny amounts, typically around one percent or
less. The main source of uncertainty in the determination of the threshold
parameters is by far the one attached to the $S$-wave scattering lengths.

16. If the energy approaches the matching point,
the uncertainties in the experimental input, naturally, come more directly
into play. Also, the uncertainties in the driving terms grow rather rapidly
with the energy. At the kaon mass, however, these are still very small.
We have analyzed the phase shifts at $E=M_K$ in detail, because these
represent an important ingredient in the calculation for
various decay modes of the $K$ mesons. The result shows that the uncertainties
are dominated by those in $a_0^0$, $a_0^2$, also at that energy. We conclude
that the future precision data on $K_{\ell_4}$-decay and on 
pionic atoms  will translate, via the Roy equations, 
into a rather precise knowledge
of the $\pi \pi$ scattering amplitude (not only the lowest three partial
waves) in the entire low--energy region, extending quite far above
threshold.

17. In the present paper, we followed the phenomenological path and 
avoided making use of chiral symmetry, in order not to bias the
data analysis with theoretical prejudice. 
A famous low energy theorem \cite{Weinberg 1966} 
predicts the values of the two basic
low energy parameters in terms of the pion decay constant. 
The prediction holds to leading order in an expansion in
powers of the quark masses. 
The corrections arising from the higher order terms in the chiral expansion
are now known to order $p^6$ (two loops) \cite{pipi6}.
We plan to match the chiral perturbation theory 
representation of the scattering amplitude 
with the phenomenological one obtained in the present paper \cite{Roy_chiral}.
This should lead to a very sharp prediction for $a_0^0$ and $a_0^2$. 
The confrontation of the prediction with the
forthcoming results of the precision measurements will subject chiral
perturbation theory to a crucial test.

\subsection*{Acknowledgements} We are indebted to 
W.~Ochs,  M.~Pennington and G.~Wanders for many discussions and 
remarks concerning various aspects of our work. Also, we wish to
thank G.~Ecker for providing us with his notes on the problem 
that were very useful at an early phase of this
investigation. Moreover, we thank J.~Bijnens, P.~B\"uttiker, 
S.~Eidelman, F.~Jegerlehner, 
B.~Loiseau, B.~Moussallam, S.~Pislak, A.~Sarantsev, J.~Stern and B.~Zou
for informative comments, in particular also for detailed information 
on data and phase shift analyses. 
This work was supported by the Swiss National Science
Foundation, Contract No. 2000-55605.98, by TMR, BBW-Contract No. 97.0131  and  
EC-Contract
No. ERBFMRX-CT980169 (EURODA$\Phi$NE).

\appendix

\renewcommand{\theequation}{\thesection.\arabic{equation}}
\setcounter{equation}{0}
\section{Integral kernels}
\label{not}
Crossing symmetry, $A(s,u,t)=A(s,t,u)$, implies that the isospin
components $\vec{T}=(T^0,T^1,T^2)$ are subject to the constraints
$(u\equiv 4M_\pi^2-s-t)$
\bea \vec{T}(s,u)&=&C_{tu}\,\vec{T}(s,t)\;,\rule{3em}{0em}\no
     \vec{T}(t,s)&=&C_{st}\,\vec{T}(s,t)\;,\no
     \vec{T}(u,t)&=&C_{su}\,\vec{T}(s,t)\;,\nonumber\eea
where the
crossing matrices $C_{tu}=C_{ut},\,C_{su}=C_{us},\,C_{st}=C_{ts}$
are given by \bea &&\hspace{-1em}C_{tu}=
\left(\!\!\begin{tabular}{rrr}
1&0&0\\\rule{0em}{1em}0&--1&0\\\rule{0em}{1em}0&0&1\end{tabular}\!
\right)\hspace{0.8cm}
C_{su}=
\left(\!\!\!\!\begin{tabular}{rrr}
\mbox{$\frac{1}{3}$}&--1&\mbox{$\frac{5}{3}$}\\\rule{0em}{1em}
--\mbox{$\frac{1}{3}$}&\mbox{$\frac{1}{2}$}&\mbox{$\frac{5}{6}$}\\
\rule{0em}{1em}
\mbox{$\frac{1}{3}$}&\mbox{$\frac{1}{2}$}&\mbox{$\frac{1}{6}$}\end{tabular}
\!\right) \hspace{0.8cm}  C_{st}=
\left(\!\!\!\begin{tabular}{rrr}
\mbox{$\frac{1}{3}$}&1&\mbox{$\frac{5}{3}$}\\\rule{0em}{1em}
\mbox{$\frac{1}{3}$}&\mbox{$\frac{1}{2}$}&--\mbox{$\frac{5}{6}$}\\
\rule{0em}{1em}
\mbox{$\frac{1}{3}$}&--\mbox{$\frac{1}{2}$}&\mbox{$\frac{1}{6}$}\end{tabular}
\!\right)
\nonumber\eea
Their products obey the relations
\bea &&(C_{tu})^2= (C_{su})^2=(C_{st})^2={\bf 1}\;,\no
&&C_{st}\,C_{tu}=C_{tu}\,C_{us}=C_{us}\,C_{st}\;,\hspace{3em}
C_{su}\,C_{ut}=C_{ts}\,C_{su}=C_{ut}\,C_{ts}\;.
\nonumber\eea

The quantities $g_2(s,t,s')$, $g_3(s,t,s')$ occurring in the fixed-t dispersion
relation (\ref{fixedt}) represent $3\times3$ matrices built with $C_{st}$,
$C_{tu}$ and $C_{su}$,
\bea g_2(s,t,s')&=&-\frac{t}{\pi\, s'\,(s'-4M_\pi^2)}\,
(u\, C_{st} + s\, C_{st}\, C_{tu})\left(\frac{{\bf 1}}{s'-t}
+ \frac{C_{su}}{s'-u_0}\right)\co\no
g_3(s,t,s')&=&-\frac{s\,u}{\pi\,s'(s'-u_0)}\left(\frac{{\bf 1}}{s'-s}
+ \frac{C_{su}}{s'-u}\right)\co\eea
where $u=4M_\pi^2-s-t$ and $u_0=4M_\pi^2-t$.

The straightforward partial wave projection of the amplitude reads
\bea \label{pwp1}t_\ell^I(s)\al=\al\frac{1}{64\pi}\int_{-1}^1 dz  P_\ell(z)\,
T^I(s,t_z)\co\hspace{2em}
t_z=\mbox{$\frac{1}{2}$}(4M_\pi^2-s)(1-z)\fs
\eea
On account of crossing symmetry, the formula is equivalent to
\bea\label{pwp} t_\ell^I(s)\al=\al\frac{1}{32\pi}\int_{0}^1 dz  P_\ell(z)\,
T^I(s,t_z)\fs\eea
As pointed out by Roy \cite{Roy}, 
the second form of the projection is preferable in the present context, 
because it involves smaller values of $|t_z|$, so that
the domain of convergence of the partial wave series for the imaginary parts
on the r.h.s.~of the fixed-$t$ dispersion relation (\ref{fixedt}) 
becomes larger: Whereas for the 
straightforward projection, the large Lehmann-Martin ellipse is mapped into
$-4M_\pi^2<s<32M_\pi^2$, the one in eq.~(\ref{pwp}) corresponds to
$-4M_\pi^2<s<60M_\pi^2$.

The kernels $K_{\ell\ell'}^{II'}(s,s')$ that occur in eq.~(\ref{eq:req1})
are different from zero only if both $I\!+\ell$ and
$I'\!+\ell'$ are even. With the partial wave projection
(\ref{pwp}),
the explicit expression becomes\footnote{Note that the fixed-$t$ 
dispersion relation
(\ref{fixedt}) is not manifestly crossing symmetric -- for $\ell'\geq 2$,
the kernels do depend on the specific form used for the partial wave 
projection. In particular, the kernels 
occurring in the Roy equations for the waves with $\ell\geq 3$ 
are proportional to $(s-4M_\pi^2)^\ell$ only if the projection in 
eq.~(\ref{pwp1}) is used -- for the one we are using here,
the proper behaviour of the solutions
only results if the contributions from the imaginary 
parts of the different partial waves compensate one another near threshold
(see section \ref{B6}).
For a detailed discussion of these issues
we refer to \cite{Atkinson}.} 
\bea \label{intrep}
K_{\ell \ell'}^{I I'}(s,s')\al=\al(2\ell'+1)\,\!
\int_{0}^{1}
dz\, P_\ell(z)\,K_{\ell'}(s,t_z,s')^{II'}\co\no
t_z\al=\al\mbox{$\frac{1}{2}$}(4M_\pi^2-s)(1-z)\fs
\eea
The functions $K_{\ell'}(s,t,u)^{I I'}$ are the matrix elements of
\bea K_{\ell'}(s,t,s')=
g_2(s,t,s')+g_3(s,t,s')\,P_{\ell'}\!\left(1+
\frac{2t}{s'-4M_\pi^2}\right)\fs\eea

The kernels contain the usual pole at $s=s'$, generating the right hand cut of
the partial wave amplitudes, as well as
a piece $\bar{K}^{II'}_{\ell\ell'}(s,s')$ that is analytic in $\mbox{Re}\,s>0$,
but contains a logarithmic branch cut for $s\leq -(s'-4M_\pi^2)$:
\bea K_{\ell \ell'}^{I I'}(s,s')=\frac{1 }
{\pi(s'-s)}\,\delta^{II'}\,\delta_{\ell\ell'}
+\bar{K}^{II'}_{\ell\ell'}(s,s')\fs\nonumber\eea
To illustrate the structure of the second term, we give the explicit
expression for $I=I'=\ell=\ell'=0$:
\bdm\bar{K}^{00}_{00}(s,s')=\frac{2}{3\,\pi\,(s-4M_\pi^2)}\;\ell n\!\left(
\frac{s+s'-4M_\pi^2}{s'}\right)-\frac{2\,s+5\,s'-16M_\pi^2}
{3\pi\,s'\,(s'-4M_\pi^2)}\fs\edm
We do not need to list other components -- they may be generated from
the above formulae by means of standard integration routines.

\setcounter{equation}{0}
\section{Background amplitude}\label{background}
\subsection{Expansion of the background for small momenta}
\label{B1}
The background amplitude only contains very weak singularities 
at low energies. 
At small values of the arguments, $A(s,t,u)_d$ thus
represents a slowly varying function of $s,t,u$, which is adequately
approximated by a polynomial. We may, for instance, consider the Taylor series
expansion around the center of the Mandelstam triangle: Set
$s_0=\frac{4}{3}M_\pi^2, s=s_0+x,t=s_0-\frac{1}{2}(x-y)$, expand in powers of
$x$ and $y$ and truncate the series. Alternatively, we may
exploit the fact that, in view of the angular momentum barrier,
the dispersion integral over the imaginary parts of the higher partial waves
receives significant contributions only for $s'\gsim 1\,\mbox{GeV}^2$. For
small values of $s$ and $t$, we can therefore 
expand the kernels $g_2(s,t,s')$ and $g_3(s,t,s')$
in inverse powers of $s'$. The coefficients of this expansion are homogeneous
polynomials of $s,\,t$ and $M_\pi^2$, which may be ordered with the
standard chiral power counting. The corresponding expansion of the Legendre 
polynomial starts with
\bdm P_\ell\left(1+\frac{2\,t}{s'-4M_\pi^2}\right)=1 +\ell(\ell+1)\, \frac{t}
{s'}+O\left(p^4\right)\fs\edm
Truncating the expansion at order $p^6$, the background amplitude becomes
\bea\label{Taylor}\hspace{-1em} \al\al\hspace{-0.5em}
\vec{T}(s,t)_d= -32\,\pi\left\{(t\,u \,C_{st} + s\,
u\,C_{su} +s\,t\,C_{tu})\,({\bf 1}+C_{su})\,\vec{I}_0\right.\\
\al\al\;\;\;\;\;+\{s^2\,t\,C_{tu}+u^2\, s\,C_{su}+t^2\,u\,C_{st}+
 (t^2 \,s\, C_{tu}+s^2\,u\, C_{su}+ u^2\,t\,C_{st})\,
C_{su}\}\,\vec{I}_1 \no \al\al\;\;\;\;\;\left.+ s\,t\,u\,({\bf 1} +C_{su})\,
\vec{H}\right\}+O(p^8)\fs\nonumber\eea
The coefficients $\vec{I}_0$ and $\vec{I}_1$ represent moments\footnote{
The factor $1/(s-4M_\pi^2)$ could also be expanded in inverse powers of $s$,
but this would worsen the accuracy of the polynomial representation. Note 
that the same factor also occurs in the representation (\ref{W}) for the 
contributions generated by the imaginary part of the $S$- and $P$-waves below
$s_2$: The expansion of the functions $W^I(s)$ in powers of $s$ yields 
integrals of the same form. Hence the low energy expansion of the full
amplitude can be expressed in terms of moments of this type.}
of the imaginary part at $t=0$,
\bea
I^I_n=\frac{1}{32\,\pi^2}\int_{4M_\pi^2}^\infty
\frac{ds\;\mbox{Im}\,T^I(s,0)_d}{s^{n+2}(s-4M_\pi^2)}\fs\eea
In view of the optical theorem, these quantities are given by integrals over
the total cross section, except that the contributions from the $S$- and 
$P$-waves
below $s_2$ are to be removed. 
Equivalently, we may express these coefficients in
terms of the imaginary parts of the partial waves:
\bea\label{In}\hspace{-2em} I_n^I\al=\al
\sum_{\ell=2}^\infty \,\frac{(2l+1)}{\pi}\int_{4M_\pi^2}^{s_2}
\frac{ds\;\mbox{Im}\,t^I_\ell(s)}{s^{n+2}(s-4M_\pi^2)}+
\sum_{\ell=0}^\infty \,\frac{(2l+1)}{\pi}\int_{s_2}^\infty
\frac{ds\;\mbox{Im}\,t^I_\ell(s)}{s^{n+2}(s-4M_\pi^2)}\,.
\eea
Except for a contribution proportional to $I^1_1$, 
the last term in eq.~(\ref{Taylor}) may be expressed in 
terms of the derivative
of $ \mbox{Im}\,\vec{T}(s,t)_d$ with respect to $t$:
\bea H^I= -2\,I^1_1\,\delta^I_1+\frac{1}{32\,\pi^2}\int_{4M_\pi^2}^\infty
\frac{ds}{s^{3}}\;\frac{\partial\,\mbox{Im}\,T^I(s,t)_d}
{\partial t}\;\rule[-0.8em]{0.03em}{2.3em}_{\;t=0}\fs\eea
Here, only the higher partial waves contribute:
\bea  H^I\al=\al
\sum_{\ell=2}^\infty
(2l+1)\{\ell(\ell+1)-2\,\delta^I_1\}\,\frac{1}{\pi}\int_{4M_\pi^2}^\infty
\frac{ds\;\mbox{Im}\,t^I_\ell(s)}{s^{3}(s-4M_\pi^2)}\fs\label{H}\eea
The expression is similar to the one for $I^I_1$, except that the sum over 
the angular momenta picks up a factor of $\ell(\ell+1)$,
indicating that partial waves with higher values of $\ell$ are more significant
here. Note that all of the above moments are positive.

\subsection{Constraints due to crossing symmetry}
\label{B2}
The expansion of the background amplitude starts at order $p^4$, with a
mani\-festly crossing symmetric contribution determined by the moments
$\vec{I}_0$.
The term from $\vec{I}_1$ is also crossing symmetric, but the
one proportional to $s\,t\,u$ violates the condition
$\vec{T}(s,u)_d=C_{tu}\,\vec{T}(s,t)_d$,
unless the $I=1$ component of the vector $({\bf 1}+C_{su})\,\vec{H}$ vanishes,
i.e.
\be\label{SRL} 2 H^0=9H^1+5H^2\fs\ee
This sum rule is both
necessary and sufficient for the polynomial approximation to the
background amplitude to be crossing symmetric up to and including 
contributions of order $p^6$.

The sum rule illustrates the well-known fact that crossing symmetry
leads to stringent constraints on the
imaginary parts of the partial waves with $\ell\geq 2$
(for a thorough discussion, see \cite{Wanders 1971,Roy HPA}).
Crossing symmetry implies for instance that $\mbox{Im}\,t_2^0(s)$ can be
different from zero only if some of the higher partial waves
with $I=1$ or $I=2$ also possess an imaginary part --
in marked contrast to the situation for the $S$- and $P$-waves, where
crossing symmetry does not constrain the imaginary parts.

In the form given, the
sum rule only holds up to corrections of order $M_\pi^2$. We may, however,
establish an exact variant by expanding
the $I=1$ component of the relation
$\vec{T}(s,u)_d=C_{tu}\vec{T}(s,t)_d$ around threshold, for instance 
in powers of $t$
and $u$. In order for the term of order $t\,u$ occurring in the expansion of
the left hand side
to agree with the corresponding term on the right hand side, the imaginary
parts must obey the sum rule 
\bea \label{SRp6}\al\al\hspace{-1.5em}
\int_{4M_\pi^2}^\infty
\frac{ds}{s^{2}\,(s-4M_\pi^2)}\;\left\{2\,\mbox{Im}\,\dot{T}^{0}(s,0)
-5\,\mbox{Im}\,\dot{T}^{2}(s,0)\right\}=\no
\al\al 3\int_{4M_\pi^2}^\infty
\frac{ds\,(3\, s-4\,M_\pi^2)}{s^{2}\,(s-4M_\pi^2)^3}
\,\left\{(s-4\,M_\pi^2)\,\mbox{Im}\,
\dot{T}^{1}(s,0)-2\,\mbox{Im}\,T^1(s,0)\right\}\co\eea
where $\dot{T}^{I}(s,t)$ stands for the partial derivative of 
$T^I(s,t)$ with respect to $t$. Expressed in terms of the
imaginary parts of the partial waves, the relation reads
\bea\label{SRpw}\al\al\hspace{-1.5em}
\sum_{\ell=2,4,\,\ldots}(2\ell+1)\,\ell\,(\ell+1)\int_{4M_\pi^2}^\infty
\frac{ds}{s^{2}\,(s-4M_\pi^2)^2}\;\{2\,\mbox{Im}\,t_\ell^0(s)
-5\,\mbox{Im}\,t_\ell^2(s)\}=\no
\al\al
 \sum_{\ell=3,5,\,\ldots}(2\ell+1)\,\{\ell\,(\ell+1)-2\}\int_{4M_\pi^2}^\infty
\frac{ds\,(s-\frac{4}{3}M_\pi^2)}{s^{2}\,(s-4M_\pi^2)^3}\;
9\,\mbox{Im}\,t_\ell^1(s)\fs\eea
The approximate version (\ref{SRL}) differs from this exact result only through
terms of order $M_\pi^2$.

The constraints imposed by crossing symmetry show, in
particular, that the concept of tensor meson
dominance is subject to a limitation that does not occur in the case of
vector dominance:
The hypothesis that convergent dispersion integrals or sum rules are
saturated by the contributions from a spin 2 resonance leads to coherent
results only at leading order of the low energy expansion. The sum rule
(\ref{SRp6}) demonstrates that the hypothesis in general fails:
Crossing symmetry implies that singularities with $\ell\geq 2$
cannot be dealt with one by one.

Since the relation (\ref{SRL}) ensures crossing symmetry, the above low energy
expansion of the isospin components of the amplitude is equivalent
to a manifestly crossing symmetric representation of the 
background amplitude:
\be A(s,t,u)_d= p_1+p_2\,s+p_3\,s^2+p_4\,(t-u)^2+
p_5\,s^3+p_6\,s(t-u)^2+O(p^8)\,.\ee
By construction, $A(s,t,u)_d$ does not contribute to the
$S$-wave scattering lengths. This condition fixes
$p_1$ and $p_2$ in terms of the
remaining coefficients:
\be p_1= -16M_\pi^4\,p_4\,,\;\;\;p_2 =
4M_\pi^2\,(-p_3 + p_4 - 4M_\pi^2\,p_5)\,, \ee
The explicit expressions for the latter read 
\bea
\label{pn} p_3\al=\al\frac{8\,\pi}{3}\,(4I^0_0-9I^1_0-I^2_0)
      +\frac{16\,\pi}{3}\,M_\pi^2\,(-8\,I^0_1 - 21\, I^1_1 + 11\,I^2_1
        +12 \,H)\co\no
p_4\al=\al 8\,\pi \,(I^1_0+I^2_0)+
   16\,\pi\,M_\pi^2\,(I^1_1 +I^2_1)\co\\
p_5\al=\al \frac{4\,\pi}{3}\,(8\,I^0_1+9\,I^1_1-11\,I^2_1-6\,H)\co\no
p_6\al=\al 4\,\pi (I^1_1-3\,I^2_1+2\,H)
     \fs\nonumber\eea
In view of the sum rule (\ref{SRL}), only two of the components of $\vec{H}$
are independent. Moreover, the amplitude only involves a combination thereof:
\be\label{defH}  H\equiv \mbox{$\frac{2}{5}$}(H^0-2H^1)=
        \mbox{$\frac{2}{9}$}\,(H^0+2H^2)=H^1+H^2\fs\ee 

The above formulae show that the leading background contribution
is determined by the integrals $\vec{I}_0$, which yield
\bdm p_1=O(M_\pi^4)\co\hspace{1em}
p_2=O(M_\pi^2)\co\hspace{1em}p_3=O(1)\co\hspace{1em}p_4
=O(1)\fs\edm
The contributions from $\vec{I}_1$ and $\vec{H}$ 
modify the result by corrections that are suppressed by one power of $M_\pi^2$
and, in addition,
generate a polynomial of third degree in $s,\,t,\,u$, characterized 
by $p_5$ and $p_6$. 

\subsection{Background generated by the higher partial waves}
\label{B3}
Next, we turn to the numerical evaluation of the integrals $\vec{I}_0$, 
$\vec{I}_1$, $\vec{H}$
and first consider the contributions from the imaginary parts of the partial
waves with $\ell\geq 2$ in the region below 2 GeV. The integrals are
dominated by the resonances, which generate peaks in the imaginary parts.
In the vicinity of the peak, we may describe the phase shift with the
Breit-Wigner formula
\bdm e^{2i\delta(s)}=\frac{M^2_{r}+i\Gamma_r M_r-s}
{M_r^2-i\Gamma_r M_r-s}\co\edm
where $M_r$ and $\Gamma_r$ denote the mass and the width of the resonance,
respectively. 
To account for inelasticity (decays into states other
than $\pi\pi$), we multiply the corresponding expression for the 
imaginary part of the partial wave amplitude with the branching
fraction $\Gamma_{r\rightarrow \pi\pi}/\Gamma_r$. This leads to 
\bea \mbox{Im}\,t^{I_r}_{\ell_r}(s)=\sqrt{\frac{s}{s-4M_\pi^2}}
\,\frac{\Gamma_{r\rightarrow\pi\pi}\Gamma_r M_r^2}
{(s-M_r^2)^2+\Gamma_r^2M_r^2}\co\nonumber\eea
where $I_r$ and $\ell_r$ denote the isospin and 
the spin of the resonance, respectively.
In the narrow width approximation, the formula simplifies to
\bea \label{NW}\mbox{Im}\,t_{\ell_r}^{I_r}(s)=\pi\,\Gamma_{r\rightarrow\pi\pi} 
M_r(1-4M_\pi^2/M_r^2)^{-\frac{1}{2}}\delta(s-M_r^2)\fs\eea

Only four of the states listed in the particle data booklet \cite{PDG} 
below 2 GeV have spin $\ell\geq 2$ and carry the proper quantum numbers to be 
produced in $\pi\pi$ collisions: 
The spin 2 resonances $f_2(1275)$ and 
$f_2'(1525)$, the spin
3 state $\rho_3(1681)$ and the state $f_J(1710)$, whose spin is not 
firmly established, but must be even.
There is no evidence for exotic states: $f_2,f_2'$ and 
$f_J$ are isoscalars, while the $\rho_3$ is an isovector.

Very likely, the lightest spin 4 state is the $f_4(2044)$: 
A linear $\rho(770)-f_2(1275)-\rho_3(1691)$ Regge
trajectory calls for a spin 4 recurrence almost exactly there.
At any rate, if the spin of the state $f_J(1710)$ were equal to 4 or even
larger, it would sit above that trajectory and thus upset the standard 
Regge picture, which we will be making use of to estimate the asymptotic 
part of the driving terms. We take it for granted that $J=0$ or 2 and conclude
that only the $I=0$ $D$-wave and the $F$-wave contain resonances
below $2\,\mbox{GeV}$. 
In the following, we discuss the contributions generated by these states, 
comparing the result obtained from the narrow width formula
with the one found on the basis of two different phase shift analyses.

The most important contribution arises from the tensor meson $f_2(1275)$.
Inserting the values $M_{f_2}=1275\,\mbox{MeV}$, 
$\Gamma_{f_2\rightarrow\pi\pi}=157\,\mbox{MeV}$, the narrow width formula
gives $I^0_{0\,f_2}=.25\,\mbox{GeV}^{-4}$, 
$I^0_{1\,f_2}=.15\,\mbox{GeV}^{-6}$, $H^0_{f_2}=.93\,\mbox{GeV}^{-6}$, 
to be compared with
the results obtained with the parametrizations of the $D$-wave in
refs.~\cite{hyams} and \cite{bugg}, which yield
\bea 
\hspace{-3em}\cite{hyams}:\; I^0_{0\,D}\al=\al .25 \,\mbox{GeV}^{-4}\,, 
\hspace{0.3em}I^0_{1\,D}=.18
\,\mbox{GeV}^{-6}\,, \hspace{0.3em}
H^0_{D}=1.10\,\mbox{GeV}^{-6}\label{DHyams}\co\\
\cite{bugg}:\;I^0_{0\,D}\al=\al
.27\,\mbox{GeV}^{-4}\,, \hspace{0.3em}I^0_{1\,D}=.19
\,\mbox{GeV}^{-6}\,, \hspace{0.3em}
H^0_{D}=1.17\,\mbox{GeV}^{-6}\label{DZou}\fs\eea 
These numbers show that the contributions from the imaginary part of the
$D$-wave are dominated by the $f_2(1275)$.

We add a few remarks concerning the detailed behaviour of
$\mbox{Im}\,t^0_2(s)$ and first note that 
the $f_2'(1525)$ mainly decays into $K\bar{K}$. In the present context,
this state may be ignored, because the corresponding $\pi\pi$
partial width is tiny: $\Gamma_{f_2'\rightarrow\pi\pi}=
.62\pm .14\,\mbox{MeV}$. 
The phase shift analysis of ref.~\cite{bugg} does contain a
second resonance in the $D$-wave, which generates a small enhancement in the
integrands on the r.h.s.~of eqs.~(\ref{In}), (\ref{H})
towards the upper end of the range of integration. 
The numerical result in eq.~(\ref{DZou}) includes the tiny contribution
produced by this enhancement, but this effect only accounts for a small 
fraction of the difference in the values obtained 
with the two different phase shift analyses. The main reason for that
difference is that the two representations of the $D$-wave 
in refs.~\cite{hyams,bugg} do not agree very well
on the left wing of the $f_2(1275)$. 
In the context of the present paper, 
these details are not essential -- we use the difference between the
two phase shift analysis as a measure for the uncertainties to be 
attached to the moments. 

To estimate the significance of the remaining partial waves with $I=0$,
we consider the contribution generated by the $f_4(2044)$. 
This resonance also mostly decays into states other than $\pi\pi$. 
The relevant partial
width is $\Gamma_{f_4\rightarrow\pi\pi}=35\pm 4 \mbox{MeV}$. The narrow
width formula shows that the contribution from this
state is very small: $I^0_{0\,f_4}=.009\,\mbox{GeV}^{-4}$, 
$I^0_{1\,f_4}=.002\,\mbox{GeV}^{-6}$, $H^0_{f_4}=.04\,\mbox{GeV}^{-6}$.
Moreover, the center of the peak is outside our 
range of integration --  
more than half of the contribution from this level
is to be booked in the asymptotic part. We conclude
that the imaginary parts of the partial waves with $\ell\geq 4$ only matter 
at energies above 2 GeV. 

The $\rho_3(1681)$ shows up as a peak in the imaginary part of the $F$-wave. 
According to the particle
data tables \cite{PDG}, it mainly decays into $4\pi$. The partial width
of interest in our context is 
$\Gamma_{\rho_3\rightarrow \pi\pi}=38\pm 3 \,\mbox{MeV}$.
Inserting this in the narrow width formula, we obtain 
$I^1_{0\,\rho_3}=.020\,\mbox{GeV}^{-4}$,
$I^1_{1\,\rho_3}=.007\,\mbox{GeV}^{-6}$,  
$H^1_{\rho_3}=.07\,\mbox{GeV}^{-6}$, to be compared with 
the values found by performing the numerical integration 
with the representations
for the $F$-wave given in the two references quoted above: 
\bea 
\hspace{-3em}\cite{hyams}:\; I^1_{0\,F}\al=\al .028\,\mbox{GeV}^{-4}\,, 
\hspace{0.3em}I^1_{1\,F}=.012
\,\mbox{GeV}^{-6}\,, \hspace{0.3em}
H^1_{F}=.12\,\mbox{GeV}^{-6}\co\\
\hspace{-3em}\cite{bugg}:\;I^1_{0\,F}\al=\al
.030\,\mbox{GeV}^{-4}\,, \hspace{0.3em}I^1_{1\,F}=.016
\,\mbox{GeV}^{-6}\,, \hspace{0.3em}
H^1_{F}=.16\,\mbox{GeV}^{-6}\label{FZou}\fs\eea
In the present case, the narrow width
formula only accounts for about half of the result: 
The region below the resonance is equally
important. There, the difference between the two phase shift analyses 
is more pronounced than for the $D$-waves. Accordingly, the uncertainties in
the $F$-wave contributions to the moments are larger.  

The formula (\ref{NW}) predicts that the contribution 
generated by the imaginary part of the $I=2$
waves vanishes, because that channel does not contain any resonances. 
According to Martin, Morgan and Shaw
\cite{MMS}, the $D$-wave phase shift may be
approximated as $\delta^2_2(s)\simeq -0.003\,
(s/4M_\pi^2)\,(1-4M_\pi^2/s)^{\frac{5}{2}}$. The corresponding contributions
to the moments are indeed very small: $I^2_0=0.005\,\mbox{GeV}^{-4}$,
$I^2_1=0.006\,\mbox{GeV}^{-6}$, $H=0.04\,\mbox{GeV}^{-6}$. 
In the following, we assume that these estimates do hold to within 
a factor of two.

This completes our discussion of the contributions generated by
the higher partial waves in the region below 2 GeV. 
The net result is that these are
due almost exclusively to the $D$- and $F$-waves.
The numerical results are
listed in row L of table \ref{tab:moments}. For $I=0,1$, the values given
rely on the phase shift analyses of refs.~\cite{hyams,bugg}, 
while the estimates  
for $I=2$ correspond to the parametrization of ref.~\cite{MMS}.

\begin{table}[H]
\vspace*{-1em}
\begin{center}
\begin{tabular}{|c|c|c|c||c|c|c||c|c|c|}
\hline
&\multicolumn{3}{c||}{$I=0$}&\multicolumn{3}{|c||}{$I=1$}&
\multicolumn{3}{c|}{$I=2$}\\
\cline{2-10}
\rule{0em}{1em}&$I^0_0$ & $I^0_1$ & $H^0$ &$I^1_0$ & $I^1_1$ & $H^1$
&$I^2_0$
& $I^2_1$ & $H^2$\\
&\hspace*{-0.2em}$ \mbox{\footnotesize GeV}^{-4}$\hspace*{-0.3em} &
\hspace*{-0.2em}$ \mbox{\footnotesize GeV}^{-6}$\hspace*{-0.3em}&
\hspace*{-0.2em}$ \mbox{\footnotesize GeV}^{-6}$\hspace*{-0.3em}&
\hspace*{-0.2em}$ \mbox{\footnotesize GeV}^{-4}$\hspace*{-0.3em}&
\hspace*{-0.2em}$ \mbox{\footnotesize GeV}^{-6}$\hspace*{-0.3em}&
\hspace*{-0.2em}$ \mbox{\footnotesize GeV}^{-6}$\hspace*{-0.3em}&
\hspace*{-0.2em}$ \mbox{\footnotesize GeV}^{-4}$\hspace*{-0.3em}&
\hspace*{-0.2em}$ \mbox{\footnotesize GeV}^{-6}$\hspace*{-0.3em}&
\hspace*{-0.2em}$ \mbox{\footnotesize GeV}^{-6}$\hspace*{-0.3em}\\
\hline
L &.26   &  .19 &1.13 & .029 & .014& .14 &.005& .006 & .04\\
\hline
R & .03 &.004 & .11 & .018 & .003 & .07 & -- &
-- & --\\ 
\hline
P & .01 & .001 & .04 & .010& .001& .04 & .010 & .001& .04 \\
\hline\hline
total & .30 & .19 & 1.28& .058& .018 & .24 & .015 & .007 &.08\\
$\pm$ & .01&.01 &.05 &.007&.002&.03&.008&.006&.04\\
\hline
\end{tabular}
\caption{\label{tab:moments}Moments of the background amplitude.
The rows L, R and P indicate the
contributions from the region below $2\,\mbox{GeV}$, from the leading
Regge trajectory and from the Pomeron, respectively. The last two rows show
the result for the sum of these contributions and our estimate of the 
uncertainties, respectively.}
\vspace*{-1em}
\end{center}
\end{table}

\subsection{Asymptotic contributions}
\label{B4}

We now turn to the contributions from the 
high energy tail of the dispersion integrals.
The asymptotic behaviour of the scattering amplitude 
may be analyzed in terms of Regge
poles. A trajectory with isospin $I$ generates a
contribution $\propto s^{\alpha(t)}$ to the $t$-channel
isospin component $\mbox{Im}\,T^{(I)}(s,t)$, which is defined by
\bdm
\mbox{Im}\,T^{(I)}(s,t)=\sum_{I'}\,C_{st}^{I I'}\,\mbox{Im}\,T^{I'}(s,t)\fs
\edm 
The asymptotic behaviour of the amplitude with $I_t=1$ 
($s\rightarrow\infty$, $t$ fixed) is governed by
the $\rho\,$-trajectory,
\bea \mbox{Im}\,T^{(1)}(s,t)=\beta_\rho(t)\,s^{\alpha_\rho(t)}\fs\nonumber\eea
The Pomeron dominates the high energy behaviour of the $I_t=0$ amplitude.
Together with the contribution from the $f$-trajectory, the 
Regge representation of this component reads
\bdm \mbox{Im}\,T^{(0)}(s,t)=3\,P(s,t)
+\beta_f(t)\,s^{\alpha_f(t)}\fs\edm
In the absence of exotic trajectories, the component with $I_t=2$
rapidly tends to zero when $s$ becomes large.
The asymptotic behaviour of the $s$-channel isospin components thus
takes the form
\bea\label{Regge}
\mbox{Im}\,T^0(s,t)\al=\al
P(s,t)
+\mbox{$\frac{1}{3}$}\beta_f(t)\,s^{\alpha_f(t)}+
\beta_\rho(t)\,s^{\alpha_\rho(t)} + (t\leftrightarrow u)\co
\no
\mbox{Im}\,T^1(s,t)\al=\al
P(s,t)
+\mbox{$\frac{1}{3}$}\beta_f(t)\,s^{\alpha_f(t)} +
\mbox{$\frac{1}{2}$}\beta_\rho(t)\,s^{\alpha_\rho(t)} -(t\leftrightarrow u)\co
\\
\mbox{Im}\,T^2(s,t)\al=\al
P(s,t)+
\mbox{$\frac{1}{3}$}\beta_f(t)\,s^{\alpha_f(t)} -
\mbox{$\frac{1}{2}$}\beta_\rho(t)\,s^{\alpha_\rho(t)} +(t\leftrightarrow u)
\fs\nonumber\eea
If $t$ is kept fixed, the
terms with $P(s,t)$ and $\beta(t)\,s^{\alpha(t)}$ dominate, generating a
peak in the forward direction, while the analogous structure in the backward
direction (fixed $u$) is described by those with $P(s,u)$ and $\beta(u)\,
s^{\alpha(u)}$. At fixed $t$, the crossed terms drop off very rapidly with
$s$, so that their contribution disappears in the noise of the calculation and
may just as well be dropped. 

The Lovelace-Shapiro-Veneziano model \cite{Veneziano,Lovelace,Shapiro} 
provides a very instructive 
framework for understanding the interplay of the asymptotic contributions 
with the resonance structures seen at low energies (see appendix
\ref{veneziano}). In
that model, the $\rho\,$- and $f$-trajectories are linear and 
exchange degenerate,
\be\label{alpha1} \alpha_\rho(t)=\alpha_f(t)=\alpha_0+t\,\alpha_1\fs\ee
We fix the intercept with the Adler zero, $\alpha(M_\pi^2)=\frac{1}{2}$, and
choose the slope such that the spin 1 state on the leading
trajectory occurs at the proper mass:
\be\label{alpha2} \alpha_1=\mbox{$\frac{1}{2}$}\,(M_\rho^2-M_\pi^2)^{-1}\co
\hspace{3em}
\alpha_0=\mbox{$\frac{1}{2}$}-\alpha_1\,M_\pi^2  \fs\ee
The amplitude may be represented as a sum of narrow resonance contributions.
Since the model
does not contain exotic states, $\mbox{Im}\,T^2(s,t)$
vanishes, so that the residues $\beta_f(t)$ and $\beta_\rho(t)$ are in the
ratio 3:2. The explicit expression reads
\be\label{betaf} \beta_\rho(t)=
\mbox{$\frac{2}{3}$}\beta_f(t)=
\frac{\pi\,\lambda\, (\alpha_1)^{\alpha(t)}}{\Gamma[\alpha(t)]}\fs\ee
Finally, we fix the overall normalization constant $\lambda$ such that the
width of the $\rho$ agrees with what is observed. This requires
\bea \label{lambda}\lambda=
96\,\pi\,\Gamma_\rho\,M_\rho^2\,(M_\rho^2-4M_\pi^2)^{-\frac{3}{2}} \fs\eea

The model explicitly obeys crossing symmetry and yields a decent picture both 
for the masses and widths of the
resonances occurring on the leading trajectory and for the qualitative 
properties of the Regge residues $\beta_\rho(t)$, $\beta_f(t)$.
The main deficiency of the model is lack of unitarity: It does not
contain a Pomeron term, so that the total cross section tends to zero
at high energies. While the model yields quite decent values for the full
widths, it does not account for the fact that the 
resonances often decay into states
other than $\pi\pi$, particularly if the available phase space becomes
large -- in the model, the branching fraction
$\Gamma_{r\rightarrow\pi\pi}/\Gamma_r$ is equal to 1.
Consequently, the LSV-model overestimates the magnitude of the Regge 
residues -- a significant fraction thereof
should be transferred to the Pomeron term. For this reason, the model can
only serve as a semi-quantitative guide.

As discussed in section \ref{B2}, crossing symmetry
strongly correlates the asymptotic behaviour of the partial
waves with their properties at low energy. In particular, the parameters
occurring in the Regge representation of the scattering amplitude 
can be extracted from low energy
phenomenology. For a review of these calculations, we refer to 
the article by Pennington \cite{Pennington Annals}.
The value obtained for $\beta_\rho(0)$ is 
smaller\footnote{In ref.~\cite{Pennington Annals}, the residue is written as
$\beta_\rho(t)=\mbox{$\frac{16}{3}$} \pi \gamma_\rho(t)\, 
\alpha_1^{\alpha_\rho(t) - \frac{1}{2}} $. The result obtained for
the value at $t=0$ is $\gamma_\rho(0)=(0.6\pm0.1)M_\pi^{-1}$, to be compared
with the number $\gamma_\rho(0)=0.97 \,M_\pi^{-1}$ 
that follows from eqs.~(\ref{alpha1})-(\ref{lambda}).} 
than what follows from eqs.~(\ref{betaf}), (\ref{lambda}) by a factor 
of $0.6 \pm0.1$. Also, while the formula (\ref{betaf}) implies that the
residue contains a zero at 
$t_0=2M_\pi^2-M_\rho^2=-0.55\,\mbox{GeV}^2$ because $\alpha(t)$ vanishes
there, the calculation of
ref.~\cite{Pennington Annals} instead yields a zero at
$t_0=-0.44\pm0.05\,\mbox{GeV}^2$. This confirms the remarks made
above: The LSV-model describes the qualitative properties of the Regge 
residues quite decently, but overestimates their magnitude.

In the numerical evaluation, we use the linear $\rho\,$-trajectory 
specified above, $\alpha_\rho(t)=\alpha(t)$, and fix the corresponding 
residue with the results of 
ref.~\cite{Pennington Annals}, which are adequately described by a modified
version of the LSV-formula: 
\bea\label{betaPenn}\beta_\rho(t)=\frac{\pi\lambda_1\alpha_1^{\alpha(t)}}
{\Gamma[\,(t-t_0)\,\alpha_1]}\co\hspace{1.5em}t_0=-0.44\,\mbox{GeV}^2\co
\hspace{1.5em}\lambda_1=(.78\pm.13) \,\lambda\fs\eea
We determine the properties of the $f$-trajectory with exchange degeneracy, 
i.e.~set $\alpha_f(t)=\alpha(t)$ and
 $\beta_f(t)=\frac{3}{2}\beta_\rho(t)$. For the Pomeron, we use
the representation 
\bea P(s,t)=\sigma\, s\,e^{\frac{1}{2}b\,t}\fs\eea
While the parameter $b=8\,\mbox{GeV}^{-2}$ \cite{Pennington Annals} 
describes the width of the diffraction peak, 
the optical theorem implies that $\sigma$ represents the
asymptotic value of the total $\pi\pi$ cross section. Evidently, the above
parametrization can be adequate only in a limited range of
energies: The cross section does not tend to a constant, but grows
logarithmically. In the present context, however, 
the behaviour at very high energies is an academic issue, because the 
integrands of the moments
rapidly fall off with $s$. What counts is that the above representation
yields a decent approximation for c.m.~energies in the range between
2 and 3 GeV. There, the terms generated by the $\rho\,$-$f$-trajectory
are by no means negligible: The formula (\ref{Regge}) shows that at 2
GeV (3 GeV), these terms by themselves generate a contribution to 
$\mbox{Im}\,T^0(s,0)$ that corresponds to a total cross section of 
21 mb (14 mb) -- in the energy range relevant for the
moments, the Pomeron term does not represent the dominating contribution to
the total cross section. 
As discussed in detail in ref.~\cite{Pennington Annals},
crossing symmetry leads to the estimate  
$\sigma=(6 \pm 5)\,\mbox{mb}$. Although the error bar is large, the value
is significantly smaller than what is indicated by the
rule of thumb $\sigma_{tot}^{\pi\pi}\simeq\frac{2}{3}\,
\sigma_{tot}^{\pi N}\simeq\frac{4}{9}\,\sigma_{tot}^{NN}\simeq 20\,\mbox{mb}$.

Indeed, the sum rule (\ref{SRL}) confirms this result.
The numerical values obtained with the above representation for the
contributions from the $\rho\,$-$f$-trajectory 
are indicated in row R of table \ref{tab:moments}. If the high energy tail 
is omitted altogether, the l.h.s.~of the sum
rule (\ref{SRL}) becomes $(2H^0)_L=2.3\,\mbox{GeV}^{-6}$, while the 
r.h.s.~amounts to $(9H^1+5 H^2)_L=1.5\,\mbox{GeV}^{-6}$. Clearly, further
contributions are required to bring the sum rule into equilibrium. 
The Regge terms do contribute more to the right than to
the left and reduce the discrepancy by a factor of two. Since the Pomeron
affects the various isospin components almost equally, it contributes
about 7 times more to the right than to the left. For the sum rule
to be obeyed within the uncertainties of the remaining
contributions, the value of $\sigma$ must be in the range 
$\sigma=(5\pm 3)\,\mbox{mb}$. 

Let us compare our representation of the background with the model used for 
the asymptotic
behaviour in the early literature. Assume that, above an energy of
1.5 GeV, the imaginary parts can be described by a Pomeron term with
$\sigma_{tot}=20\,\mbox{mb}$ and a Regge term that corresponds to the
leading trajectory of the LSV-model. The l.h.s.~of the sum rule
(\ref{SRL}) then takes the value $2H^0=3.3$, while the r.h.s~yields
$9H^1+5H^2=6.1$ (to be compared with the value 2.6 
obtained for either one of the two sides 
with our representation of the background). Evidently, the
model is in conflict with crossing symmetry. 
In the region
relevant for the driving term integrals, 
the LSV-model overestimates the Regge residues by about 40\%  
\cite{Pennington Annals} and the sum rule (\ref{SRL}) then implies that
the value $\sigma=20\,\mbox{mb}$ is too large by about a factor of 4.

We repeat that our calculation has
no bearing on the asymptotic behaviour of the total cross section -- we
are merely observing that, unless the value of
$\sigma$ is in the range $5\pm3\,\mbox{mb}$, the representation used for the 
amplitude
violates crossing symmetry. The row P indicates the 
contributions to the moments generated by the Pomeron if $\sigma$ is taken in
the middle of this range. The net result of our calculation is contained
in the last two rows of table \ref{tab:moments}, which list the outcome for 
the moments
and for the error bars to be attached to these, respectively.
For the quantity $H$ defined in eq.~(\ref{defH}), we obtain
\be H=0.32\pm 0.02\,\mbox{GeV}^{-6}\fs\ee

\subsection{Driving terms}
\label{B5}
The polynomial approximation for the background amplitude can be used to
determine the low energy behaviour of the driving terms -- it suffices
to evaluate the partial wave projections of the polynomial 
$\vec{T}(s,t)_d$. The range of validity
of the resulting representation for the driving terms, however, only extends
to c.m.~energies of about 0.6 GeV. For our numerical work, we need a
representation that holds for higher energies.

The approximations for the imaginary parts discussed above 
yield the following representation of the driving terms:
\bea \al\al d_\ell^I(s)= d_\ell^I(s)_L+ d_\ell^I(s)_R+d_\ell^I(s)_P\co\no
\al\al d_\ell^I(s)_L =\sum_{I'=0}^2\sum_{\ell'=2}^3  \int_{4M_\pi^2}^{s_2}
ds'\,K_{\ell\ell'}^{I I'}(s,s')\,\mbox{Im} \, t_{\ell'}^{I'}(s')\co\no
\al\al d_\ell^I(s)_H= \frac{1}{32\, \pi}\int_{0}^1\!dz P_\ell(z)\,
T^I(s,t_z)_H\co\hspace{2em}\mbox{\footnotesize\it H}= \mbox{\footnotesize
  \it R,P} \no 
\al\al \vec{T}(s,t)_H=\int_{s_2}^\infty ds' g_2(s,t,s')\cdot
\mbox{Im}\,\vec{T}(s',0)_H+\int_{s_2}^\infty ds'
g_3(s,t,s')\cdot\mbox{Im}\,\vec{T}(s',t)_H \co\no
\al\al \mbox{Im}\,T^0(s,t)_R=\mbox{$\frac{3}{2}$}\,
\beta_\rho(t)\,s^{\alpha(t)}+\mbox{$\frac{3}{2}$}\,\beta_\rho(u)\,s^{\alpha(u)}
\co \no
\al\al \mbox{Im}\,T^1(s,t)_R=
\beta_\rho(t)\,s^{\alpha(t)}-\beta_\rho(u)\,s^{\alpha(u)}\co\no
\al\al \mbox{Im}\,T^2(s,t)_R= 0\co \no
\al\al \mbox{Im}\,T^0(s,t)_P= \mbox{Im}\,T^2(s,t)_P=P(s,t)+P(s,u)\co\no
\al\al \mbox{Im}\,T^1(s,t)_P= P(s,t)-P(s,u)\nonumber
\fs\eea
The result of the numerical evaluation of these integrals with the parameter
values specified above is given in eq.~(\ref{numerical driving
  terms}).

We use the difference between the results for $d^0_0(s)_L$ and $d^1_1(s)_L$
obtained with the two phase shift analyses quoted above 
as a measure for the uncertainties in these quantities. 
In the case of the $I=2$ $D$-wave,
we assume that the Martin-Morgan-Shaw formula does describe the behaviour of
the imaginary part to within a factor of 2. For the Regge-contributions,
we use the error estimate $\gamma_\rho(0)=(0.6\pm0.1)M_\pi^{-1}$ 
given in ref.~\cite{Pennington Annals}.
Finally, the uncertainties attached to the Pomeron term correspond to
those in the value $\sigma=5\pm 3\,\mbox{mb}$, obtained in section 
\ref{B4}. The result quoted in eq.~(\ref{errors in driving terms})
is obtained by adding the corresponding error bars quadratically and
fitting the outcome with a polynomial.

There is a neat and rather thorough check of the above 
calculation. The driving terms represent
the partial wave projections of the background amplitude. 
Since that amplitude must be crossing symmetric, we may equally well 
calculate the projections with the formula (\ref{pwp1}) instead of
using (\ref{pwp}) -- the result should be the same. The modification
of the partial wave projection changes the form of the 
kernels $K^{II'}_{\ell\ell'}(s,s')$ and  
the contributions from the imaginary parts of the higher partial waves below
2 GeV then
change, quite substantially. The contributions from the asymptotic region,
however, are also modified. In the sum, these changes indeed
cancel out, to a remarkable degree of accuracy. 
This corroborates the claim that our description of the background is 
approximately crossing symmetric. Evidently, the sum rule (\ref{SRL}) plays
an important role here, as it correlates the magnitude of the asymptotic
contributions with those from the low energy region.

\setcounter{equation}{0}
\section{Sum rules and asymptotic behaviour}\label{asymptotics} 
\subsection{Sum rules for the $P$-wave}
\label{C1}
As discussed in section \ref{Olsson sum rule}, the Olsson sum rule
ensures the correct asymptotic behaviour of the $t$-channel $I=1$ 
scattering amplitude $T^{(1)}(s,t)$ for $s\rightarrow\infty$, $t=0$. 
The requirement that this amplitude has the proper high energy behaviour
also for $t<0$ implies a further constraint, which is readily derived
from the fixed-$t$ dispersion relation (\ref{fixedt}). It suffices to evaluate
the coefficient of the term that grows linearly with $s$ and to subtract
the value at $t=0$. The result involves the following integrals over the 
imaginary parts of the amplitude ($t\leq 0$): 
\bea\label{SR1}  \al\al
S(t)\equiv
\int_{4M_\pi^2}^\infty\!ds\;\frac{
2\,\mbox{Im}\,\bar{T}^0(s,t)+3\,\mbox{Im}\,\bar{T}^1(s,t)-5\,
\mbox{Im}\,\bar{T}^2(s,t)}{12\,s\,(s+t-4M_\pi^2)}\\
\al\al\hspace{12em}-
\int_{4M_\pi^2}^\infty\!ds\;\frac{(s-2M_\pi^2)\,\mbox{Im}\,T^1(s,0)}
{s\, (s-4M_\pi^2)\,(s - t)\,(s+t-4M_\pi^2)}\fs\nonumber\eea
The barred quantities stand for $\mbox{Im}\,\bar{T}^I(s,t)=
\{\mbox{Im}\,T^I(s,t)-\mbox{Im}\,T^I(s,0)\}/t$. The amplitude $T^{(1)}(s,t)$
  has the proper asymptotic behaviour only if $S(t)$ vanishes for
space-like values of $t$.
Since the $S$-waves drop out, the condition amounts to a
family of sum rules that relate 
integrals over the imaginary part of the $P$-wave to the higher partial waves.
For $t=0$, for instance, the sum rule may be written in the 
form
\bea\label{SRt} \al\al\int_{4M_\pi^2}^\infty\!ds\;\frac{\mbox{Im}\,t_1^1(s)}
{s^2\,(s-4M_\pi^2)}
= \sum_{\ell=2,4,\,\ldots}(2\ell+1)\,\ell\,(\ell+1)
\int_{4M_\pi^2}^\infty\!ds\;\frac{
2\,\mbox{Im}\,t^0_\ell(s)-5\,
\mbox{Im}\,t^2_\ell(s)}
{18\,s\,(s-4M_\pi^2)^2}\no\al\al\hspace{4em}
+\sum_{\ell=3,5,\,\ldots}(2\ell+1)\,\int_{4M_\pi^2}^\infty\!ds\;\frac{
\{\ell(\ell+1)\,s-4\,(s-2\,M_\pi^2)\}\,\mbox{Im}\,t^1_\ell(s)}
{6\,s^2\,(s-4M_\pi^2)^2}
\fs\eea
The integrals over the individual partial waves converge more rapidly than in
the case of the Olsson sum rule, but the factor $\ell(\ell+1)$ gives
the higher partial waves more weight -- in fact, the contributions from the
asymptotic region are even more important here. The sum rule is of the same 
structure as the one that follows from crossing symmetry, eq.~(B.7), but
there are two differences: The integrals converge less 
rapidly by one power of $s$ and the $P$-wave does not drop out.

Since the sum rule (\ref{SRt}) offers a good opportunity to check the
representation used for the asymptotic region, we evaluate it
explicitly with our input for the imaginary
parts. We split the integration into one from threshold to 
$\Etwo=2\,\mbox{GeV}$
and one over the asymptotic region, $s>s_2$ (compare appendix B). 
Denoting the low energy
part of the integral over the $P$-wave by 
\bea S_{\ind P}=
\int_{4M_\pi^2}^{s_2}\!ds\;\frac{\mbox{Im}\,t_1^1(s)}
{s^2\,(s-4M_\pi^2)}\co\nonumber\eea
we write the sum rule in the form
\bea\label{SRE1} S_{\ind P}=S_{\ind D}+S_{\ind F}+S_{as}\co\eea
where $S_{\ind D}$ and $S_{\ind F}$ stand for the analogous integrals over
the $D$- and $F$-waves. While the low energy contributions 
from the waves with $\ell\geq 4$ are neglected, their high energy tails
are accounted for in the term $S_{as}$, which
collects all contributions from the region above $s_2$. 

The form (\ref{SRt}) of the sum rule
is suitable to calculate the contributions
from the interval $4M_\pi^2<s<s_2$. Numerically, we obtain
\bea S_{\ind P}= 1.93\pm0.08\co\hspace{2em} 
S_{\ind D}=0.55\pm 0.03\co\hspace{2em}
S_{\ind F}=0.13\pm 0.01\co\nonumber\eea
in units of $\mbox{GeV}^{-4}$.
To evaluate the asymptotic contributions, we instead use the form (\ref{SR1}):
The term $S_{as}$ coincides with the expression 
$S(0)/48 \pi$, except that the integration now only extends over the 
interval $s_2<s<\infty$. Inserting the representation specified
in appendix B.4, we find that the bulk  
stems from the leading Regge trajectory ($1.12\pm0.19$). The Pomeron does not 
contribute to the
first integral on the r.h.s.~of eq.~(\ref{SR1}), because that integral 
is of the same isospin
structure as the one occurring in the Olsson sum rule, but it does
generate a small negative term via the second integral ($-0.02\pm 0.01$). 
The net result for the asymptotic contributions,
\bea S_{as}=1.10\pm0.19\co\nonumber\eea 
leads to $S_{\ind D}+S_{\ind F}+S_{as}= 1.77\pm0.19$. Within the errors,
the outcome agrees with the numerical value $S_{\ind P}=1.93\pm0.08$
obtained for the l.h.s.~of the sum rule (\ref{SRE1}). Note that more than 
half of 
the r.h.s.~stems from the asymptotic region. We conclude that our asymptotic 
representation is valid within the estimated uncertainties, also for this sum 
rule, which converges more slowly than the moments considered in appendix B.
Since the Olsson sum rule belongs to the same convergence class as the one
above, we feel confident that our error estimates apply also in that case.

\subsection{Asymptotic behaviour of the Roy integrals}
\label{C2}
If the imaginary parts of the partial waves with $\ell>1$ are discarded, 
the Roy equations become a closed system for the $S$- and $P$-waves. 
The explicit expressions for the kernels show
that the dispersion integrals over the imaginary parts of these waves
grow linearly with $s$, like the subtraction polynomials. 
Except for the contributions from the higher partial waves, the
r.h.s.~of the Roy equations for the $S$- and $P$-waves thus grows in
 proportion to $s$:
\bea
\al\al \mbox{Re}\,t^0_0\rightarrow\frac{\Sigma\,s}{12M_\pi^2}\co\hspace{2em}
 \mbox{Re}\,t^1_1\rightarrow\frac{\Sigma\,s}{72M_\pi^2}\co\hspace{2em}
 \mbox{Re}\,t^2_0\rightarrow -\frac{\Sigma\,s}{24M_\pi^2}\co\no
\al\al
\Sigma=2\,a_0^0-5\,a_0^2-\frac{4 M_\pi^2}{\pi}\int_{4M_\pi^2}^\infty
\!\!ds\,\frac{2\,\mbox{Im}\,t^0_0(s)+27\,\mbox{Im}\,t^1_1(s)-5\,
\mbox{Im}\,t^2_0(s)}{s\,(s-4M_\pi^2)}\fs\rule{2em}{0em}\label{Sigma}\eea
So, if the coefficient $\Sigma$ vanishes, the contribution from
the dispersion integrals cancels the one from the subtraction polynomial,
simultaneously for all three partial waves 
\cite{Basdevant Guillou Navelet,MMS}. In fact, if the imaginary parts of the
higher partial waves are dropped and if $\Sigma$ is set equal to zero,
the Roy equations become identical to 
those proposed by Chew and Mandelstam \cite{Chew Mandelstam} 
(see ref.~\cite{Basdevant
  Guillou Navelet} for a detailed discussion).
The expression for $\Sigma$ resembles the Olsson sum rule, where the
contributions from the $S$- and $P$-wave read
\bea 2\,a_0^0-5\,a_0^2=\frac{4 M_\pi^2}{\pi}\int_{4M_\pi^2}^\infty
\!\!ds\,\frac{2\,\mbox{Im}\,t^0_0(s)+9\,\mbox{Im}\,t^1_1(s)-5\,
\mbox{Im}\,t^2_0(s)}{s\,(s-4M_\pi^2)}+ \ldots\nonumber\eea
If only the $S$-waves are retained, the Olsson sum rule
does imply that $\Sigma$ vanishes -- evidently, this sum rule is closely 
related to the observation that the linearly rising contribution from the
 subtraction terms must cancel the one from the dispersion integrals 
(section \ref{sec:math}). As is well-known, however, 
the coefficient of the $P$-wave term in $\Sigma$ differs from the one
in the Olsson sum rule. The implications of this
discrepancy for the Chew-Mandelstam framework are discussed in
the references quoted above. The family of sum rules derived 
in appendix \ref{C1} points in the same direction:  The imaginary part of 
the $P$-wave is tied together with those of the higher partial waves --
setting these equal to zero leads to inconsistencies \cite{Lovelace 1961}. 

For the above asymptotic formulae to apply at $E\sim 1\,\mbox{GeV}$, 
two conditions would have to be met: (a) the contributions from the higher
partial waves can be ignored at these energies 
and (b) the integrals over the imaginary parts of the $S$- and $P$-waves
are dominated by the contributions from low energies.
Unfortunately, neither of the two conditions is met. 
The solutions show a pronounced structure in the region  
above the matching point -- evidently, we are not dealing with
the asymptotic behaviour there.  The numerical
value of $\Sigma$ is negative: The integral in eq.~(\ref{Sigma}) 
over-compensates the term $2a_0^0-5a_0^2$. We may lay the blame
upon the contributions above the matching point -- if the integral were cut
off there, $\Sigma$ would approximately vanish. 

The situation is quite different for the Olsson
sum rule, which does not rely on low energy approximations but represents 
the exact version of the condition that must be obeyed by the two subtraction
constants for the scattering amplitude to have the proper asymptotic behaviour.
In that case, the coefficient of the $P$-wave is 
three times smaller -- the region above the matching point plays an
essential role in bringing the sum rule into balance. 
The numerical evaluation in section \ref{Olsson sum rule} shows that even 
those from the region above 2 GeV are significant. 
The rapid growth of the driving
terms indicates that the higher partial waves become increasingly
important as the energy rises -- it is clear that the asymptotic behaviour
of the partial wave amplitudes cannot be studied on the basis of the $S$- and
$P$-wave contributions to the r.h.s~of the Roy equations. 
 
We conclude that, at the quantitative level, the above simple mechanism cannot 
explain why, for suitable values of $a_0^0$ and $a_0^2$, our solutions 
remain within the bounds set by unitarity. For an analysis of the behaviour
above the matching point that neither
discards the higher partial waves, nor relies on low energy dominance,
we refer to sections \ref{sec:math} and \ref{Olsson sum rule}.

\setcounter{equation}{0}
\section{Explicit numerical solutions}\label{explicit numerical solutions}
\begin{table}
\begin{center}
\begin{tabular}{llllll}
$\rule[-0.8em]{0em}{0em}$
&\multicolumn{1}{c}{\hspace{-1.2em}$A_0^0$}&
\multicolumn{1}{c}{\hspace{-1.2em}$B_0^0$}&
\multicolumn{1}{c}{\hspace{-1.2em}$C_0^0$}&
\multicolumn{1}{c}{\hspace{-1.2em}$D_0^0$}&
\multicolumn{1}{c}{\hspace{-1.2em}$s_0^0$}\\ 
$z_1$&$\rule[0em]{0.8em}{0em}$.2250&  $\rule[0em]{0.8em}{0em}$.2463   
& $-.$1665$\cdot 10^{-1}$& $-.$6403$\cdot 10^{-3}$&
$\rule[0em]{0.8em}{0em}$.3672$
\cdot 10^2$\\
$z_2$&$\rule[0em]{0.8em}{0em}$.2250&  $\rule[0em]{0.8em}{0em}$.1985   & 
$\rule[0em]{0.8em}{0em}$.3283$\cdot 10^{-2}$& $-.$4136$\cdot 10^{-2}$&
$\rule[0em]{0.8em}{0em}$.1339$\cdot 10$\\
$z_3$&$\rule[0em]{0.8em}{0em}$.0000&  $\rule[0em]{0.8em}{0em}$.1289   & 
$\rule[0em]{0.8em}{0em}$.1142$\cdot 10^{-1}$& $-.$3699$\cdot 10^{-2}$&
$\rule[0em]{0.8em}{0em}$.6504\\
$z_4$&$\rule[0em]{0.8em}{0em}$.0000&  $\rule[0em]{0.8em}{0em}$.1426$
\cdot 10^{-1}$& $\rule[0em]{0.8em}{0em}$.1400$\cdot 10^{-1}$& $-.$3980$
\cdot 10^{-2}$&$-.$3211$\cdot 10$\\
$z_5$&$\rule[0em]{0.8em}{0em}$.0000&  $\rule[0em]{0.8em}{0em}$.8717$
\cdot 10^{-2}$& $\rule[0em]{0.8em}{0em}$.1613$\cdot 10^{-1}$& $-.$3152$
\cdot 10^{-2}$&$-.$1396$\cdot 10$\\
$z_6$&$\rule[0em]{0.8em}{0em}$.0000&  $\rule[0em]{0.8em}{0em}$.5058$
\cdot 10^{-1}$& $\rule[0em]{0.8em}{0em}$.3000$\cdot 10^{-1}$& $-.$7354$
\cdot 10^{-2}$&$-.$4114$\cdot 10$\\
$z_7$&$\rule[0em]{0.8em}{0em}$.0000&  $-.$4266$\cdot 10^{-2}$& $-.$4045$
\cdot 10^{-2}$& $-.$1212$\cdot 10^{-2}$&$-.$3447$\cdot 10$\\
$z_8$&$\rule[0em]{0.8em}{0em}$.0000&  $-.$4658$\cdot 10^{-2}$& $
\rule[0em]{0.8em}{0em}$.2110$\cdot 10^{-2}$& $-.$4544$\cdot 10^{-2}$&
$-.$8428$\cdot 10$\\
$z_9$&$\rule[0em]{0.8em}{0em}$.0000&  $-.$5358$\cdot 10^{-2}$& 
$\rule[0em]{0.8em}{0em}$.1095$\cdot 10^{-1}$& $-.$4558$\cdot 10^{-2}$&
$-.$6350$\cdot 10$\\
$z_{10}$&$\rule[0em]{0.8em}{0em}$.0000&$-.$2555$\cdot 10^{-2}$& 
$\rule[0em]{0.8em}{0em}$.4249$\cdot 10^{-2}$& $-.$1271$\cdot 10^{-2}$&
$-.$1486$\cdot 10$\\
\multicolumn{5}{c}{}\\
$\rule[-0.8em]{0em}{0em}$
&\multicolumn{1}{c}{\hspace{-1.2em}$A_1^1$}&
\multicolumn{1}{c}{\hspace{-1.2em}$B_1^1$}&
\multicolumn{1}{c}{\hspace{-1.2em}$C_1^1$}&
\multicolumn{1}{c}{\hspace{-1.2em}$D_1^1$}&
\multicolumn{1}{c}{\hspace{-1.2em}$s_1^1$}\\ 
$z_1$&    $\rule[0em]{0.8em}{0em}$.3626$\cdot 10^{-1}$&
$\rule[0em]{0.8em}{0em}$.1337$\cdot 10^{-3}$&$-.$6976$\cdot 10^{-4}$&
$\rule[0em]{0.8em}{0em}$.1408$\cdot 10^{-5}$&
$\rule[0em]{0.8em}{0em}$.3074$\cdot 10^2$\\
$z_2$&    $\rule[0em]{0.8em}{0em}$.1834$\cdot 10^{-1}$&$-.$2336$
\cdot 10^{-2}$&$\rule[0em]{0.8em}{0em}$.1965$\cdot 10^{-3}$&$-.$1974$
\cdot 10^{-4}$&$-.$2459\\
$z_3$&    $\rule[0em]{0.8em}{0em}$.1081$\cdot 10^{-1}$&$-.$8563$
\cdot 10^{-3}$&$\rule[0em]{0.8em}{0em}$.3268$\cdot 10^{-4}$&$-.$8821$
\cdot 10^{-5}$&$-.$1733\\
$z_4$&    $-.$3195$\cdot 10^{-2}$&$\rule[0em]{0.8em}{0em}$.1678$
\cdot 10^{-3}$&$\rule[0em]{0.8em}{0em}$.2173$\cdot 10^{-4}$&$-.$6047$
\cdot 10^{-6}$&$\rule[0em]{0.8em}{0em}$.6323$\cdot 10^{-1}$\\
$z_5$&    $\rule[0em]{0.8em}{0em}$.1670$\cdot 10^{-3}$&$
\rule[0em]{0.8em}{0em}$.4147$\cdot 10^{-4}$&$\rule[0em]{0.8em}{0em}$.3267$
\cdot 10^{-5}$&$-.$1617$\cdot 10^{-5}$&$-.$1090$\cdot 10^{-2}$\\
$z_6$&    $-.$9543$\cdot 10^{-3}$&$\rule[0em]{0.8em}{0em}$.8402$
\cdot 10^{-4}$&$\rule[0em]{0.8em}{0em}$.2059$\cdot 10^{-4}$&$-.$3125$
\cdot 10^{-5}$&$\rule[0em]{0.8em}{0em}$.2724$\cdot 10^{-1}$\\
$z_7$&    $\rule[0em]{0.8em}{0em}$.5049$\cdot 10^{-3}$&$-.$9308$
\cdot 10^{-4}$&$\rule[0em]{0.8em}{0em}$.1070$\cdot 10^{-4}$&$-.$1257$
\cdot 10^{-5}$&$-.$7218$\cdot 10^{-2}$\\
$z_8$&    $\rule[0em]{0.8em}{0em}$.4595$\cdot 10^{-4}$&$-.$2755$
\cdot 10^{-3}$&$\rule[0em]{0.8em}{0em}$.5554$\cdot 10^{-4}$&$-.$4432$
\cdot 10^{-5}$&$\rule[0em]{0.8em}{0em}$.1483$\cdot 10^{-1}$\\
$z_9$&    $-.$9000$\cdot 10^{-4}$&$-.$2308$\cdot 10^{-3}$&$
\rule[0em]{0.8em}{0em}
$.5307$\cdot 10^{-4}$&$-.$4415$\cdot 10^{-5}$&$\rule[0em]{0.8em}{0em}$.1813$
\cdot 10^{-1}$\\
$z_{10}$&$-.$1198$\cdot 10^{-4}$&$-.$6120$\cdot 10^{-4}$&$
\rule[0em]{0.8em}{0em}
$.1519$\cdot
10^{-4}$&$-.$1344$\cdot 10^{-5}$&$\rule[0em]{0.8em}{0em}$.5016$\cdot 10^{-2}$\\
\multicolumn{5}{c}{}\\
$\rule[-0.8em]{0em}{0em}$
&\multicolumn{1}{c}{\hspace{-1.2em}$A_0^2$}&
\multicolumn{1}{c}{\hspace{-1.2em}$B_0^2$}&
\multicolumn{1}{c}{\hspace{-1.2em}$C_0^2$}&
\multicolumn{1}{c}{\hspace{-1.2em}$D_0^2$}&
\multicolumn{1}{c}{\hspace{-1.2em}$s_0^2$}\\ 
$z_1$&  $-.$3706$\cdot 10^{-1}$&$-.$8553$\cdot 10^{-1}$&$-.$7542$
\cdot 10^{-2}$&
$\rule[0em]{0.8em}{0em}$.1987$\cdot 10^{-3}$&$-.$1192$\cdot 10^2$\\
$z_2$&  $\rule[0em]{0.8em}{0em}$.0000 &$-.$1236$\cdot 10^{-1}$&$
\rule[0em]{0.8em}{0em}$.3466$\cdot 10^{-1}$&$-.$2524$\cdot 10^{-2}$&$-.$4040$
\cdot 10^2$\\
$z_3$&  $-.$3706$\cdot 10^{-1}$&$-.$6673$\cdot 10^{-2}$&$\rule[0em]{0.8em}{0em}
$.2857$\cdot 10^{-1}$&$-.$1993$\cdot 10^{-2}$&$-.$3457$\cdot 10^2$\\
$z_4$&  $\rule[0em]{0.8em}{0em}$.0000 &$\rule[0em]{0.8em}{0em}$.4901$
\cdot 10^{-2}$&$\rule[0em]{0.8em}{0em}$.2674$\cdot 10^{-2}$&$
\rule[0em]{0.8em}{0em}$.1506$\cdot 10^{-2}$&$-.$9879$\cdot 10^2$\\
$z_5$&  $\rule[0em]{0.8em}{0em}$.0000 &$\rule[0em]{0.8em}{0em}$.2810$
\cdot 10^{-1}$&$\rule[0em]{0.8em}{0em}$.1477$\cdot 10^{-1}$&$
\rule[0em]{0.8em}{0em}$.2915$\cdot 10^{-3}$&$-.$9856$\cdot 10^2$\\
$z_6$&  $\rule[0em]{0.8em}{0em}$.0000 &$\rule[0em]{0.8em}{0em}$.4010$
\cdot 10^{-1}$&$\rule[0em]{0.8em}{0em}$.2458$\cdot 10^{-1}$&$
\rule[0em]{0.8em}{0em}$.1325$\cdot 10^{-2}$&$-.$2072$\cdot 10^3$\\
$z_7$&  $\rule[0em]{0.8em}{0em}$.0000 &$-.$1663$\cdot 10^{-1}$&$-.$3030$
\cdot 10^{-1}$&$\rule[0em]{0.8em}{0em}$.8759$\cdot 10^{-3}$&$-.$1589$
\cdot 10^3$\\
$z_8$&  $\rule[0em]{0.8em}{0em}$.0000 &$-.$6784$\cdot 10^{-1}$&$-.$9512$
\cdot 10^{-1}$&$\rule[0em]{0.8em}{0em}$.4713$\cdot 10^{-2}$&$-.$5259$
\cdot 10^3$\\
$z_9$&  $\rule[0em]{0.8em}{0em}$.0000 &$-.$5429$\cdot 10^{-1}$&$-.$8744$
\cdot 10^{-1}$&$\rule[0em]{0.8em}{0em}$.5313$\cdot 10^{-2}$&$-.$5366$
\cdot 10^3$\\
$z_{10}$&$\rule[0em]{0.8em}{0em}$.0000&$-.$1178$\cdot 10^{-1}$&$-.$2535$
\cdot 10^{-1}$&$\rule[0em]{0.8em}{0em}$.1730$\cdot 10^{-2}$&$-.$1723$
\cdot 10^3$\\
\end{tabular}
\caption{Polynomial coefficients for Roy solutions.\label{tab:coeff}}
\end{center}
\end{table}

In this appendix, we make available our explicit numerical solutions of the
Roy integral equations.
 We proceed as follows. For a few tens of pairs
  $(a_0^0,a_0^2)$ in the universal
 band (see fig.~\ref{fig:UB}), we have constructed the three lowest
partial waves at
$2M_\pi \leq\sqrt{s} \leq 0.8$ GeV. 
As we explain in the main text, we parametrize the phase
shifts $\delta_\ell^I$ of the solutions  in the manner proposed by 
Schenk~\cite{schenk},
\begin{equation}
\tan \delta_\ell^I = \sqrt{1-{4 M_\pi^2 \over s}}\; q^{2 \ell} \left\{A^I_\ell
+ B^I_\ell q^2 + C^I_\ell q^4 + D^I_\ell q^6 \right\} \left({4
  M_\pi^2 - s^I_\ell \over s-s^I_\ell} \right) \; \; ,
\end{equation}
Each solution of the Roy equations corresponds to a specific value of
the $3\times 5$ coefficients in this  expansion,
\bea
A_\ell^I=A_\ell^I(a_0^0,a_0^2),\,\ldots\,,\, 
 s_\ell^I=s_\ell^I(a_0^0,a_0^2)\per\nonumber\eea
We approximate  these   by
 a polynomial of third degree in the scattering 
lengths $a_0^0$ and $a_0^2$. In terms of the variables
\bea u=\frac{a_0^0}{p_0}-1\scs
      v=\frac{a_0^2}{p_2}-1\scs
        p_0=0.225\scs
      p_2=-0.03706 \scs\nonumber\eea
the numerical representation for the coefficient $B_0^0$, for instance, reads 
\bea
B_0^0\al=\al z_1
    +z_2\,u
    +z_3\,v
    +z_4\,u^2 
    +z_5\,v^2
     +z_6\,u\, v
    +z_7\,u^3
     +z_8\, u^2\, v 
    +z_9\,u\, v^2
    +z_{10}\,v^3\fs \nonumber\eea
The $15 \times 10$ numbers $z_1,\ldots,z_{10}$ for the coefficients 
$A_0^0,B_0^0,\ldots  ,s_0^2$ are displayed in table \ref{tab:coeff}, in
units of $M_\pi^2$. 

\setcounter{equation}{0}
\section{Lovelace-Shapiro-Veneziano model}
\label{veneziano}
In this appendix, we describe the model used to illustrate the basic properties
of the Regge poles occurring in the asymptotic representation of the
scattering amplitude \cite{Veneziano,Lovelace,Shapiro}.
In this model, the $\pi\pi$ scattering amplitude is taken to be of the form
\bea A(s,t,u)_V=\lambda_1\, \Phi(\alpha_s,\alpha_t)+ \lambda_1\,
\Phi(\alpha_s,\alpha_u)             + \lambda_2\, \Phi(\alpha_t,\alpha_u)\,,
\nonumber\eea
where $\Phi(\alpha,\beta)$ is closely related to the Beta-function,
\bea \Phi(\alpha,\beta)=\frac{\Gamma(1-\alpha)\Gamma(1-\beta)}
            {\Gamma(1-\alpha-\beta)}\fs\nonumber\eea
and $\alpha_s$ represents a linear Regge trajectory,
\bea \alpha_s=\alpha_0+\alpha_1 s\,.\nonumber\eea
At fixed $t$, the function $\Phi(\alpha_s,\alpha_t)$ shows Regge behaviour
when $s$ tends to infinity:
\bea \Phi(\alpha_s,\alpha_t)\rightarrow (-\alpha_s)^{\alpha_t}
\Gamma(1-\alpha_t)\,.\nonumber\eea
At the same time, the representation ($1-\alpha_t>0$)
\bea\label{poles}
\Phi(\alpha_s,\alpha_t)\al=\al(1-\alpha_s-\alpha_t)\,B(1-\alpha_s,1-\alpha_t)\\
\al=\al \rule{0em}{2em}(1-\alpha_s-\alpha_t)\,\left\{\frac{1}{1-\alpha_s}
+\sum_{n=1}^\infty
\frac{\alpha_t(\alpha_t+1)\cdots(\alpha_t+n-1)}{n!\,(n+1-\alpha_s)}
\right\}\nonumber
\eea
shows that the amplitude may be expressed as a sum of narrow
resonance contributions, with mass
\bea M_n^2=(\alpha_1)^{-1}(n-\alpha_0)\co\hspace{2em}n=1,2,\ldots\nonumber\eea

The coupling constants $\lambda_1,\lambda_2$ may be chosen such that the
amplitude does not contain resonances with $I=2$. For this condition to be
satisfied, the corresponding $s$-channel isospin component
\bea T^2(s,t)_V=
 2\, \lambda_1\, \Phi(\alpha_t,\alpha_u)+(\lambda_1+\lambda_2)\,
\left(\Phi(\alpha_s,\alpha_t)+\Phi(\alpha_s,\alpha_u)\right)\nonumber\eea
should be free of poles in the physical region of the $s$-channel.
This requires
\bdm \lambda_2=-\lambda_1\equiv\mbox{$\frac{1}{2}$}\lambda \,,\edm
so that the amplitude takes the form
\bea A(s,t,u)_V\al=\al- \mbox{$\frac{1}{2}$}\lambda\left\{
   \Phi(\alpha_s,\alpha_t) +\Phi(\alpha_s,\alpha_u) - \Phi(\alpha_t,\alpha_u)
            \right\}\co\no
T^0(s,t)_V\al=\al-\mbox{$\frac{1}{2}$}\lambda\left\{
3\,\Phi(\alpha_s,\alpha_t)+3\,\Phi(\alpha_s,\alpha_u)
-\Phi(\alpha_t,\alpha_u)
\right\}\co\label{LSV2}\\
T^1(s,t)_V\al=\al-\lambda\left\{\Phi(\alpha_s,\alpha_t)
-\Phi(\alpha_s,\alpha_u)\right\}\co\no
T^2(s,t)_V\al=\al-\lambda\,\Phi(\alpha_t,\alpha_u)\fs\nonumber\eea

In the chiral limit, where the Mandelstam triangle shrinks to the point
$s=t=u=0$, the amplitude must contain an Adler zero there. Indeed, the
factor $1-\alpha_s-\alpha_t$ generates such a zero if
$\alpha_0=\frac{1}{2}$. Hence
the deviation of $\alpha_0$ from
$\frac{1}{2}$ must be of order $M_\pi^2$, so that
$\alpha_s-\frac{1}{2}$ represents a quantity of order $p^2$. At leading
order of the low energy expansion, the behaviour
of the amplitude therefore represents the first term in the expansion around
the point $\alpha_s=\alpha_t=\alpha_u=\frac{1}{2}$, which in view of
$\Gamma(\frac{1}{2})=\sqrt{\pi}$ yields
\be \label{le}A(s,t,u)_V=\pi\,\lambda\,(\alpha_s -\mbox{$\frac{1}{2}$})+
O(p^4)\,,\ee
This does have the structure of the Weinberg formula, provided
the intercept $\alpha_0$ is chosen such that $\alpha_s$ passes through the
value $\frac{1}{2}$ at $s=M_\pi^2$, i.e. \cite{Lovelace}
\bdm \alpha_0=\mbox{$\frac{1}{2}$}-\alpha_1\,M_\pi^2\,.\edm
The lowest levels of spin 1, 2, 3, 4 indeed occur on an approximately linear
trajectory with this intercept:
Fixing the value of the slope with $M_\rho$,
\bdm \alpha_1=\mbox{$\frac{1}{2}$}\,(M_\rho^2-M_\pi^2)^{-1}\,,\edm the model
predicts $$M_{\!f_2}=1319\, (1275)\,\mbox{MeV},\;
M_{\rho_3}=1699\,
(1691)\,\mbox{MeV},\;
M_{f_4}=2008\,(2044)\,\mbox{MeV},$$
 where the numbers in brackets are
those in the data tables \cite{PDG}.

The representation (\ref{poles}) shows that for $s>4\,M_\pi^2,\,t<0$, the
imaginary part of $\Phi(\alpha_s,\alpha_t)$ consists of a sequence of
$\delta$-functions:
\bea
\mbox{Im}\,\Phi(\alpha_s,\alpha_t)\al=\al -\pi\sum_{n=1}^\infty
R_n(\alpha_t)\,\delta(\alpha_s - n )\co\no
R_n(\alpha)\al=\al\frac{\Gamma(\alpha_t+n)}{\Gamma(n)\Gamma(\alpha_t)}=
\frac{1}{(n-1)!}\;
\alpha_t\,(\alpha_t+1)\,\cdots\,(\alpha_t+n-1)\fs \nonumber\eea
For the imaginary part of the $s$-channel isospin components, we thus obtain
\bea
\mbox{Im}\,T^0(s,t)_V\al=\al\frac{3\,\lambda\,\pi}
{2\,\alpha_1}\sum_{n=1}^\infty
\{R_n(\alpha_t)+R_n(\alpha_u)\}\,\delta(s-M_n^2)\co\no
    \mbox{Im}\,T^1(s,t)_V\al=\al\frac{\lambda\,\pi}{\alpha_1}\sum_{n=1}^\infty
\{R_n(\alpha_t)-R_n(\alpha_u)\}\,\delta(s-M_n^2)\co \no
    \mbox{Im}\,T^2(s,t)_V\al=\al 0\,,\nonumber\eea
with $u=4\,M_\pi^2-t-M_n^2$.

We may then read off the imaginary parts of the partial wave amplitudes
by decomposing the polynomial $R_n(\alpha)$ into a Legendre series:\footnote{In
the case of $t^0_0(s)$, the sum over $n$ only starts at $n=1$.}
\bea R_n(\alpha_t)\al=\al \sum_{\ell=0}^n
(2\,\ell+1)\,P_\ell\left(1+\frac{2\,t} {M_n^2-4\,M_\pi^2}\right)
\,r_{n\,\ell}\co\no
\mbox{Im}\,t^0_\ell(s)_V\al=\al\frac{3\,\lambda}{64\,\alpha_1}
\{1+(-1)^\ell\}\sum_{n=\ell}^\infty\, r_{n\,\ell}\,\delta(s-M_n^2)\co\no
\mbox{Im}\,t^1_\ell(s)_V\al=\al\frac{\lambda}{32\,\alpha_1}
\{1-(-1)^\ell\}\sum_{n=\ell}^\infty\,r_{n\,\ell} \,\delta(s-M_n^2)\co\no
\rule{0em}{1.5em}\mbox{Im}\,t^2_\ell(s)_V\al=\al 0\fs
\nonumber\eea
On the leading trajectory, the coefficients are
\bdm r_{n\,n}=\frac{n}{2^n\,(2\,n+1)!!}\;
\alpha_1^n\,(M_n^2-4\,M_\pi^2)^n\,.\edm

Comparison with the narrow width formula (\ref{NW}) shows that
the model predicts the width of the various levels as\footnote{
The formula reproduces the numerical results in Table I of 
ref.~\cite{Shapiro}, if the parameter values are adapted accordingly
($\alpha_0=0.48,\,\alpha_1=0.9\, {\rm  GeV}^{-2},\,  
\Gamma_\rho=112\,  {\rm MeV}, M_\rho=764\, {\rm MeV}$).}
\be\label{width} \Gamma_{n\,\ell}^{\pi\pi}=\frac{\lambda\,\omega^I r_{n\,\ell}}
{32\,\pi\,\alpha_1\,M_n^2}(M_n^2-4\,M_\pi^2)^{\frac{1}{2}}\,,\ee
where $\omega^I$ depends on the isospin of the particle: $\omega^0=3$,
$\omega^1=2$, $\omega^2=0$. In particular, the result for the width of the 
$\rho$ reads
\be \Gamma_\rho=\frac{\lambda}{96\,\pi\,M_\rho^2}\,
(M_\rho^2-4\,M_\pi^2)^{\frac{3}{2}}\,.\ee
Fixing the coupling constant with the experimental value
$\Gamma_\rho=151.2\,\mbox{MeV}$, we obtain $\lambda/32\pi=0.728$. The 
formula (\ref{width}) then predicts
\bdm\Gamma_{\!f_2}^{\pi\pi}=130\, (157)\,\mbox{MeV}\,,\hspace{2em}
\Gamma_{\rho_3}^{\pi\pi}=51\, (51)\,\mbox{MeV}\,,\hspace{2em}
\Gamma_{\!f_4}^{\pi\pi}=46\, (35)\,\mbox{MeV}\,,\edm
 where the numbers in brackets
are again taken from the data tables \cite{PDG}. This shows that the model 
does yield a decent picture, not only for the masses but also for the widths of
the particles on the leading trajectory.

In addition to the levels on the leading trajectory, the model, however, also
contains plenty of daughters, with a rather strong coupling to the
$\pi\pi$-channel. For the states on the first daughter trajectory, for
instance, equation (\ref{width}) yields
$\Gamma_{10}^{\pi\pi}=783\,\mbox{MeV}$,
$\Gamma_{21}^{\pi\pi}=154\,\mbox{MeV}$,
$\Gamma_{32}^{\pi\pi}=113\,\mbox{MeV}$, $\Gamma_{43}^{\pi\pi}=42 \,\mbox{MeV}$,
etc. The scalar daughter of the $\rho$ is particularly fat. 

It is clear that an amplitude that describes
all of the levels as narrow resonances fails here. Unitarity implies
the bound \bdm \int_{4\,M_\pi^2}^{M^2}
ds\,\mbox{Im}\,t_0^0(s)\,\sqrt{1-4M_\pi^2/s} \leq M^2-4 M_\pi^2\,.\edm
This condition is violated for $M<1.3 \,\mbox{GeV}$. Also, if the
intercept of the leading trajectory is fixed with the Adler condition as
above, the scalar daughter of the
$f_2$ is a ghost: The formula (\ref{width}) yields a negative decay width
\cite{Shapiro}. In this respect, the model is
deficient --  as witnessed by the life of even royal families, the decency of
a mother does not ensure that her daughters
behave.

The problem also shows up in the S-wave
scattering lengths: Chiral symmetry relates the coefficient of the leading term
in the low energy expansion (\ref{le}) to the pion decay constant,
\be \pi\,\lambda\,\alpha_1=\frac{1}{F_{\pi}^2}\fs\ee
If the coupling constant $\lambda$ is fixed such that the model yields
the proper width for the $\rho$, the l.h.s. of this relation exceeds the
r.h.s. by a factor of 1.7. Accordingly, the prediction of the model for $a_0^0$
exceeds the current algebra result by about this factor.
In the
vicinity of threshold, the behaviour of the amplitude is determined by the
properties of the function $\phi(\alpha,\beta)$ for
$\alpha\simeq\beta\simeq\frac{1}{2} $. There,
the first term in the series (\ref{poles}) accounts for about two thirds of
the sum. The spin 1 component of this term is due to $\rho\,$-exchange,
while the spin 0 part arises from the scalar daughter of the $\rho$.
By construction, the former does have the proper magnitude. The S-wave
scattering lengths are too large
because the scalar daughter of the $\rho$ is
too fat.

As was noted from the start \cite{Shapiro}, the
model is not unique. To arrive at a more realistic model, we could add
extra terms that domesticate the
daughters and leave the leading trajectory and the position of the Adler
zero untouched. Note, however, that the number of states
occurring in the
Veneziano model corresponds to the degrees of freedom of a string, while the
spectrum of bound states in QCD is the one of a local field theory, where
the number of independent states grows much less
rapidly with the mass. Modifications of the type just mentioned
can at best provide a partial cure. In particular, these do not remove 
the main deficiency of the model, lack of unitarity.


\begin{thebibliography}{99}

\bibitem{Roy}
S.~M.~Roy,
Phys.\ Lett.\  {\bf B36} (1971) 353.


\bibitem{Basdevant Guillou Navelet}
J.~L.~Basdevant, J.~C.~Le Guillou and H.~Navelet,
Nuovo Cim.\  {\bf A7} (1972) 363.

\bibitem{PP1}
M.R.~Pennington and S.D.~Protopopescu,
Phys.\ Rev.\  {\bf D7} (1973) 1429.

\bibitem{PP2}
M.R.~Pennington and S.D.~Protopopescu,
Phys.\ Rev.\  {\bf D7} (1973) 2591.

\bibitem{BFP1}
J.~L.~Basdevant, C.~D.~Froggatt and J.~L.~Petersen,
Phys.\ Lett.\  {\bf B41} (1972) 173; ibid. 178.

\bibitem{BFP2}            
J.~L.~Basdevant, C.~D.~Froggatt and J.~L.~Petersen,
Nucl.\ Phys.\  {\bf B72} (1974) 413.

\bibitem{Petersen} J.~L.~Petersen, 
Acta Phys.\ Austriaca Suppl.\  {\bf 13} (1974) 291;
Yellow report CERN 77-04 (1977).

\bibitem{Froggatt Petersen}
C.~D.~Froggatt and J.~L.~Petersen,
Nucl.\ Phys.\  {\bf B91} (1975) 454;
ibid. {\bf B104} (1976) 186 (E); ibid. {\bf B129} (1977) 89.

\bibitem{Morgan Pennington handbook}
D.~Morgan and M.R.~Pennington, in ref.~\cite{handbook}, p.~193.

\bibitem{handbook}
L.~Maiani, G.~Pancheri and N.~Paver,
\newblock {\it The Second} DA{$\Phi$}NE {\it Physics Handbook}
  (INFN-LNF-Divisione Ricerca, SIS-Ufficio Publicazioni, Frascati, 1995).

\bibitem{mahoux}          
G.~Mahoux, S.~M.~Roy and G.~Wanders,
Nucl.\ Phys.\  {\bf B70} (1974) 297.   

\bibitem{Anant}
B.~Ananthanarayan,
Phys.\ Rev.\  {\bf D58} (1998) 036002
[hep-ph/9802338].

\bibitem{Auberson Epele}
G.~Auberson and L.~Epele,
Nuovo Cim.\  {\bf A25} (1975) 453.

\bibitem{pomponiuw}         
C.~Pomponiu and G.~Wanders,
Nucl.\ Phys.\  {\bf B103} (1976) 172.

\bibitem{n/d}                
D.~Atkinson and R.~L.~Warnock,
Phys.\ Rev.\  {\bf D16} (1977) 1948.


\bibitem{slim}              
D.~Atkinson, T.~P.~Pool and H.~Slim,
J.\ Math.\ Phys.\  {\bf 18} (1977) 2407.

\bibitem{epelew}        
L.~Epele and G.~Wanders,
Phys.\ Lett.\  {\bf B72} (1978) 390;
Nucl. Phys.\ {\bf B137} (1978) 521.

\bibitem{proofpool}   
T.~P.~Pool,
Nuovo Cim.\  {\bf A45} (1978) 207.

\bibitem{proofheemskerk}
A.C. Heemskerk,
\newblock {\it Application of the N/D method to the {R}oy equations},
\newblock PhD thesis, University of Groningen, 1978.

\bibitem{heemskerk}
A.~C.~Heemskerk and T.~P.~Pool,
Nuovo Cim.\  {\bf A49} (1979) 393.

\bibitem{lovelace}
C.~Lovelace, 
Comm.\ Math.\ Phys. {\bf 4} (1967) 261.

\bibitem{brander}
O.~Brander, 
Comm.\ Math.\ Phys. {\bf 40} (1975) 97.

\bibitem{buettiker}
P.~B\"uttiker,
\newblock {\it Comparison of Chiral Perturbation Theory with a Dispersive
  Analysis of} $\pi\pi$ {\it Scattering},
\newblock PhD thesis, Universit\"at Bern, 1996;

B.~Ananthanarayan and P.~B\"uttiker,
Phys.\ Rev.\  {\bf D54} (1996) 1125
[hep-ph/9601285];
Phys.\ Rev.\  {\bf D54} (1996) 5501
[hep-ph/9604217];
Phys.\ Lett.\  {\bf B415} (1997) 402
[hep-ph/9707305] 
and in  ref.~\cite{mainz}, p.~370.

\bibitem{patarakin}
O.~O.~Patarakin, V.~N.~Tikhonov and K.~N.~Mukhin,
Nucl.\ Phys.\  {\bf A598} (1996) 335;

O.~O.~Patarakin (for the CHAOS collaboration),
in  ref.~\cite{mainz}, p.~376, and hep-ph/9711361;
$\pi N$ Newsletter 13 (1997) 27;

M.~Kermani {\it et al.}  [CHAOS Collaboration],
Phys.\ Rev.\  {\bf C58} (1998) 3431.

\bibitem{mainz}
A.M. Bernstein, D. Drechsel and T. Walcher, editors,
\newblock {\it Chiral Dynamics: Theory and Experiment}, Workshop held in Mainz,
  Germany, 1-5 Sept.~1997, Lecture Notes in Physics Vol. 513, Springer, 1997.

\bibitem{hadatom99}
J.~Gasser, A.~Rusetsky, and J.~Schacher, Miniproceedings of the
Workshop {\it HadAtom99}, held at the University of  Bern, Oct. 14-15, 1999,
[hep-ph/9911339].

\bibitem{e865}
J.~Lowe, in  ref.~\cite{mainz}, p.~375, 
and hep-ph/9711361;
\newblock talk at {\it Workshop on Physics and Detectors for} {DA$\Phi$NE},
  Frascati, Nov.~16-19, 1999,
\newblock to appear in the Proceedings
  [http://wwwsis.lnf.infn.it/talkshow/dafne99.htm];

S.~Pislak, in ref.~\cite{hadatom99}, p.~25.

\bibitem{kloe}
P.~de Simone, in ref.~\cite{hadatom99}, p.~24.

\bibitem{dirac}B. Adeva et al., Proposal to the SPSLC, CERN/SPSLC 95-1
(1995).

\bibitem{Gasser Wanders}
J.~Gasser and G.~Wanders,
Eur.\ Phys.\ J.\  {\bf C10} (1999) 159
[hep-ph/9903443].

\bibitem{wamultich}
G.~Wanders, 
hep-ph/0005042.

\bibitem{Weinberg 1966}
S.~Weinberg,
Phys.\ Rev.\ Lett.\  {\bf 17} (1966) 616.

\bibitem{GL 1983}          
J.~Gasser and H.~Leutwyler,
Phys.\ Lett.\  {\bf B125} (1983) 325.

\bibitem{gchpt}          
N.~H.~Fuchs, H.~Sazdjian and J.~Stern,
Phys.\ Lett.\  {\bf B269} (1991) 183;

J.~Stern, H.~Sazdjian and N.~H.~Fuchs,
Phys.\ Rev.\  {\bf D47} (1993) 3814
[hep-ph/9301244];

M.~Knecht and J.~Stern, in ref.~\cite{handbook}, p.~169, and references
cited therein; 

J. Stern, in ref.~\cite{mainz}, p.~26, 
and  hep-ph/9712438.

\bibitem{Borges}   J.\ S\'{a} Borges, 
Nucl.\ Phys.\ {\bf B51} (1973) 189;
Phys.\ Lett.\ {\bf B149} (1984) 21; 
ibid.{\bf B262} (1991) 320;     
Phys.\ Lett.\  {\bf B149} (1984) 21;

J.~Sa Borges and F.~R.~Simao,
Phys.\ Rev.\  {\bf D53} (1996) 4806;

J.~Sa Borges, J.~Soares Barbosa and V.~Oguri,
Phys.\ Lett.\  {\bf B393} (1997) 413;

J.~Sa Borges, J.~Soares Barbosa and M.~D.~Tonasse,
Phys.\ Rev.\  {\bf D57} (1998) 4108
[hep-ph/9707394].

\bibitem{pipigchpt}     
M.~Knecht {\it et al.},
Nucl.\ Phys.\  {\bf B457} (1995) 513
[hep-ph/9507319];
ibid. {\bf B471} (1996) 445 [hep-ph/9512404].

\bibitem{Girlanda Moussallam Stern Knecht}
L.~Girlanda {\it et al.},
Phys.\ Lett.\  {\bf B409} (1997) 461
[hep-ph/9703448].

\bibitem{Olsson}        
M.~G.~Olsson,
Phys.\ Lett.\  {\bf B410} (1997) 311
[hep-ph/9703247].

\bibitem{pocanic}
D. Po\v{c}ani\'c,  in ref.~\cite{mainz}, p. 352, 
and hep-ph/9801366;
\newblock in {\it Proc.~International Workshop on Hadronic Atoms and
  Positronium in the Standard Model}, Dubna, Russia, edited by {M.~A.~Ivanov
  {\it et al.}}, p.~33, 1998,
\newblock and hep-ph/9809455.

\bibitem{pipi6}            
J.~Bijnens {\it et al.},
Phys.\ Lett.\  {\bf B374} (1996) 210
[hep-ph/9511397];
Nucl.\ Phys.\  {\bf B508} (1997) 263
[hep-ph/9707291];
ibid. {\bf B517} (1998) 639 (E).

\bibitem{Martin}
A.~Martin, Nuovo Cim. {\bf 42} (1966) 930;
ibid. {\bf 44} (1966) 1219;
{\it Scattering Theory: Unitarity,
Analyticity and Crossing}, Lecture Notes in Physics, vol. 3, 
Springer-Verlag, Berlin-Heidelberg-New York,
1969.

\bibitem{Roy HPA}    
S.~M.~Roy,
Helv.\ Phys.\ Acta {\bf 63} (1990) 627.

\bibitem{Pennington Annals}
M.~R.~Pennington,
Annals Phys.\  {\bf 92} (1975) 164.

\bibitem{Morgan Shaw}
D.~Morgan and G.~Shaw, 
Nucl.\ Phys.\ {\bf B10} (1969) 261; Phys.\ Rev.\ {\bf D2} (1970) 520.

\bibitem{saclay}
J.P. Baton {\it et al.}, 
Phys.\ Lett.\ {\bf B25} (1967) 419; 
Nucl.\ Phys. {\bf B3} (1967) 349;
Phys.\ Lett.\ {\bf B33} (1970) 525; 
ibid. 528.

\bibitem{ochs_phd}
W.~Ochs,
\newblock {\it Die Bestimmung von} $\pi\pi$-{\it Streuphasen auf der Grundlage
  einer Amplitudenanalyse der Reaktion} $\pi^-$p$ \rightarrow\pi^-\pi^+$n {\it
  bei} 17 GeV/c {\it Prim\"arimpuls},
\newblock PhD thesis, Ludwig-Maximilians-Universit\"at, M\"unchen, 1973.

\bibitem{hyams}       
B.~Hyams {\it et al.},
Nucl.\ Phys.\  {\bf B64} (1973) 134.

\bibitem{Protopopescu}
S.~D.~Protopopescu {\it et al.},
Phys.\ Rev.\  {\bf D7} (1973) 1279.

\bibitem{Grayer}G.~Grayer {\it et al.}, Nucl.Phys. {\bf B75} (1974) 189.
 
\bibitem{EM}                  
P.~Estabrooks and A.~D.~Martin,
Nucl.\ Phys.\  {\bf B79} (1974) 301.

\bibitem{losty}        
M.~J.~Losty {\it et al.},
Nucl.\ Phys.\  {\bf B69} (1974) 185.

\bibitem{hyams1}B.~Hyams et al., Nucl.\ Phys.\ {\bf B100} (1975) 205.

\bibitem{hoogland}       
W.~Hoogland {\it et al.},
Nucl.\ Phys.\  {\bf B126} (1977) 109.

\bibitem{Becker}
H.~Becker {\it et al.}  [CERN-Cracow-Munich Collaboration],
Nucl.\ Phys.\  {\bf B150} (1979) 301.

\bibitem{au}             
K.~L.~Au, D.~Morgan and M.~R.~Pennington,
Phys.\ Rev.\  {\bf D35} (1987) 1633.

\bibitem{Bugg1}      
B.~S.~Zou and D.~V.~Bugg,
Phys.\ Rev.\  {\bf D48} (1993) 3948;
ibid. {\bf D50} (1994) 591;

V.~V.~Anisovich, D.~V.~Bugg, A.~V.~Sarantsev and B.~S.~Zou,
Phys.\ Rev.\  {\bf D50} (1994) 1972; 
ibid. 4412.

\bibitem{bugg}         
D.~V.~Bugg, B.~S.~Zou and A.~V.~Sarantsev,
Nucl.\ Phys.\  {\bf B471} (1996) 59.
We thank B.S.~Zou and A.V.~Sarantsev for providing us with the 
corresponding Fortran codes.

\bibitem{Kaminski}

R.~Kaminski, L.~Lesniak and K.~Rybicki,
Z.\ Phys.\  {\bf C74} (1997) 79
[hep-ph/9606362];

R.~Kaminski, L.~Lesniak and B.~Loiseau,
Eur.\ Phys.\ J.\  {\bf C9} (1999) 141
[hep-ph/9810386];

R.~Kaminski, L.~Lesniak and K.~Rybicki,
Acta Phys.\ Polon.\  {\bf B31} (2000) 895
[hep-ph/9912354].

\bibitem{E852} J.Gunter {\it et al.} [E852 Collaboration], 
hep-ex/0001038.

\bibitem{Ochs Newsletter}
W.~Ochs,
$\pi N$ Newslett. {\bf 3} (1991)  25. 

\bibitem{Eidelman Jegerlehner}
S.~Eidelman and F.~Jegerlehner,
Z.\ Phys.\  {\bf C67} (1995) 585
[hep-ph/9502298].

\bibitem{aleph}        
R.~Barate {\it et al.}  [ALEPH Collaboration],
Z.\ Phys.\  {\bf C76} (1997) 15.

\bibitem{Lukaszuk}L.~{\L}ukaszuk, Phys. Lett. {\bf B47} (1973) 51.


\bibitem{cleo}
S.~Anderson {\it et al.}  [CLEO Collaboration],
hep-ex/9910046.

\bibitem{schenk}       
A.~Schenk,
Nucl.\ Phys.\  {\bf B363} (1991) 97.

\bibitem{NAG}
The NAG library, The Numerical Algorithms Group Ltd, Oxford UK. 

\bibitem{olsson sum rule}
M.~G.~Olsson, 
Phys.\ Rev.\ {\bf 162} (1967) 1338.

\bibitem{Chew Mandelstam}
G.~F.~Chew and S.~Mandelstam, 
Phys.\ Rev.\ {\bf 119} (1960) 467.

\bibitem{Rosselet}        
L.~Rosselet {\it et al.},
Phys.\ Rev.\  {\bf D15} (1977) 574.

\bibitem{PDG} C.~Caso {\it et al.} [Particle Data Group], 
Eur.\ Phys.\ J.\  {\bf C3} (1998) 1.

\bibitem{Pisut Roos}
J.~Pi\v{s}\'ut and M.~Roos, 
Nucl.\ Phys. {\bf B6} (1968) 325.

\bibitem{Lang}
C.~B.~Lang and A.~Mas-Parareda,
Phys.\ Rev.\  {\bf D19} (1979) 956.
 
\bibitem{Bohacik}
J.~Bohacik and H.~Kuhnelt,
Phys.\ Rev.\  {\bf D21} (1980) 1342.

\bibitem{Wanders sum rules}
G.~Wanders, Helv.\ Phys.\ Acta {\bf 39} (1966) 228.

\bibitem{Nagels}          
M.~M.~Nagels {\it et al.},
Nucl.\ Phys.\  {\bf B147} (1979) 189.

\bibitem{Leutwyler Frascati}
H.~Leutwyler,
Nucl.\ Phys.\  {\bf A623} (1997) 169C
[hep-ph/9709406].

\bibitem{Atkinson}       
D.~Atkinson and T.~P.~Pool,
Nucl.\ Phys.\  {\bf B81} (1974) 502.

\bibitem{donoghue}
J.~F.~Donoghue, C.~Ramirez and G.~Valencia,
Phys.\ Rev.\  {\bf D38} (1988) 2195.

\bibitem{gassermeissner}
J.~Gasser and U.~G.~Meissner,
Phys.\ Lett.\  {\bf B258} (1991) 219.

\bibitem{Roy_chiral} G.~Colangelo, J.~Gasser and H.~Leutwyler, work in
  progress. 

\bibitem{Wanders 1971}
G.~Wanders,
{\it Springer Tracts in Modern Physics} 57 (1971) 22.

\bibitem{MMS}
B.R. Martin, D. Morgan and G. Shaw,
\newblock {\it Pion-Pion Interactions in Particle Physics} (Academic Press,
  London, 1976).

\bibitem{Veneziano}G.~Veneziano, 
Nuovo Cim. {\bf 57} (1968) 190.

\bibitem{Lovelace} C.~Lovelace, 
Phys.\ Lett. {\bf B28} (1968) 264.

\bibitem{Shapiro} J.~A.~Shapiro, Phys.\ Rev. {\bf 179} (1969) 1345.

\bibitem{Lovelace 1961}
C.~Lovelace, 
Nuovo Cim. {\bf 21} (1961) 305.


\end{thebibliography}
\end{document}